\documentclass[iop]{emulateapj}
\newcommand{\CIV}{\ion{C}{4}}
\newcommand{\HI}{\ion{H}{1}}
\newcommand{\NV}{\ion{N}{5}}
\newcommand{\OI}{\ion{O}{1}}
\newcommand{\OVI}{\ion{O}{6}}
\newcommand{\OVII}{\ion{O}{7}}
\newcommand{\OVIII}{\ion{O}{8}}
\newcommand{\SiIV}{\ion{Si}{4}}

\newcommand{\Halpha}{H$\alpha$}

\newcommand{\Msol}{\ensuremath{M_{\odot}}}

\newcommand{\nH}{\ensuremath{n_{\mathrm{H}}}}
\newcommand{\nHe}{\ensuremath{n_{\mathrm{He}}}}

\newcommand{\NHI}{\ensuremath{N(\mbox{\HI})}}

\newcommand{\vLSR}{\ensuremath{v_\mathrm{LSR}}}

\newcommand{\nm}{\ensuremath{\mbox{\nm}}}
\newcommand{\cm}{\ensuremath{\mbox{cm}}}

\newcommand{\km}{\ensuremath{\mbox{km}}}

\newcommand{\pc}{\ensuremath{\mbox{pc}}}
\newcommand{\kpc}{\ensuremath{\mbox{kpc}}}

\newcommand{\s}{\ensuremath{\mbox{s}}}

\newcommand{\Myr}{\ensuremath{\mbox{Myr}}}

\newcommand{\erg}{\ensuremath{\mbox{erg}}}

\newcommand{\K}{\ensuremath{\mbox{K}}}

\newcommand{\cmsq}{\ensuremath{\cm^2}}

\newcommand{\pcc}{\ensuremath{\cm^{-3}}}
\newcommand{\pcmsq}{\ensuremath{\cm^{-2}}}

\newcommand{\ps}{\ensuremath{\s^{-1}}}

\newcommand{\kmps}{\km\ \ps}

\newcommand{\fuse}{\textit{FUSE}}

\newcommand{\eqref}[1]{equation~(\ref{#1})}

\newcommand{\betaall}{\ensuremath{\beta_\mathrm{All}}}
\newcommand{\betaHVC}{\ensuremath{\beta_\mathrm{HVC}}}
\newcommand{\Ksp}{\ensuremath{\kappa_\mathrm{Sp}}}
\newcommand{\Kturb}{\ensuremath{\kappa_\mathrm{turb}}}
\newcommand{\Linj}{\ensuremath{L_\mathrm{inj}}}
\newcommand{\MinitT}{\ensuremath{M_{\mathrm{init,}T}}}
\newcommand{\Minitv}{\ensuremath{M_{\mathrm{init,}v}}}
\newcommand{\MHIinit}{\ensuremath{M_\mathrm{H\,I,init}}}
\newcommand{\nHI}{\ensuremath{\mathcal{N}_\mathrm{H\,I}}}
\newcommand{\nHIinit}{\ensuremath{\mathcal{N}_\mathrm{H\,I,init}}}
\newcommand{\nHcl}{\ensuremath{n_\mathrm{H,cl}}}
\newcommand{\nHISM}{\ensuremath{n_\mathrm{H,ISM}}}
\newcommand{\NCIV}{\ensuremath{N(\mbox{\CIV})}}
\newcommand{\NNV}{\ensuremath{N(\mbox{\NV})}}
\newcommand{\NOVI}{\ensuremath{N(\mbox{\OVI})}}
\newcommand{\Ncutoff}{\ensuremath{N_\mathrm{cut}}}
\newcommand{\Tcl}{\ensuremath{T_\mathrm{cl}}}
\newcommand{\TISM}{\ensuremath{T_\mathrm{ISM}}}
\newcommand{\vzcl}{\ensuremath{v_\mathrm{z,cl}}}
\newcommand{\vturb}{\ensuremath{v_\mathrm{turb}}}
\newcommand{\vz}{\ensuremath{v_\mathrm{z}}}

\newcommand{\vzsim}{\ensuremath{v_\mathrm{z,sim}}}

\shorttitle{SIMULATIONS OF HVCS}
\shorttitle{KWAK ET AL.}

\begin{document}

\title{Simulations of High-Velocity Clouds. I. Hydrodynamics and High-Velocity High Ions}
\author{Kyujin Kwak, David B. Henley, and Robin L. Shelton}
\affil{Department of Physics and Astronomy, University of Georgia, Athens, GA 30602\\
       \texttt{kkwak@physast.uga.edu}, \texttt{dbh@physast.uga.edu}, \texttt{rls@physast.uga.edu}}

\begin{abstract}
We present hydrodynamic simulations of high-velocity clouds (HVCs) traveling through the hot,
tenuous medium in the Galactic halo.  A suite of models was created using the FLASH hydrodynamics
code, sampling various cloud sizes, densities, and velocities.  In all cases, the cloud-halo
interaction ablates material from the clouds.  The ablated material falls behind the clouds, where
it mixes with the ambient medium to produce intermediate-temperature gas, some of which radiatively
cools to less than 10,000~K.  Using a non-equilibrium ionization (NEI) algorithm, we track the
ionization levels of carbon, nitrogen, and oxygen in the gas throughout the simulation period. We
present observation-related predictions, including the expected \HI\ and high ion (\CIV, \NV, and
\OVI) column densities on sight lines through the clouds as functions of evolutionary time and
off-center distance.  The predicted column densities overlap those observed for Complex~C.  The
observations are best matched by clouds that have interacted with the Galactic environment for tens
to hundreds of megayears.  Given the large distances across which the clouds would travel during
such time, our results are consistent with Complex~C having an extragalactic origin.  The
destruction of HVCs is also of interest; the smallest cloud ($\mbox{initial mass} \approx
120~\Msol$) lost most of its mass during the simulation period (60~Myr), while the largest cloud
($\mbox{initial mass} \approx 4\times10^5~\Msol$) remained largely intact, although deformed, during
its simulation period (240~Myr).
\end{abstract}


\section{INTRODUCTION}
\label{sec:Introduction}

The halo of the Milky Way contains clouds of neutral hydrogen (\HI) with $|\vLSR| \ga 90~\kmps$,
known as high-velocity clouds (HVCs; \citealt{muller63,wakker97}). Although some HVCs are relatively
close to the Galactic disk (e.g., $|z| \la 4~\kpc$ for Complex~M; \citealt{danly93,keenan95}), other
HVCs are known to be in the upper Galactic halo, where the Galaxy interacts with its surroundings
(e.g., $d = 10 \pm 2.5~\kpc$ for Complex~C; \citealt{thom08}). The distant HVCs are thought to be
gas stripped off satellite galaxies \citep[e.g.,][]{gardiner96,putman04}, extragalactic gas falling
into the Galaxy \citep{oort66,blitz99}, or gas left over from the formation of the Galaxy
\citep{maller04}.

Observations of interstellar absorption of light from AGN show that the halo also contains highly
ionized gas (including high ions such as \CIV, \NV, and \OVI) moving with velocities comparable to
those of the \HI\ HVCs \citep{sembach03,fox04,fox05,fox06,collins07}. These high ions trace gas with
$T \sim (\mbox{1--3}) \times 10^5~\K$, and enable us to probe the connection between the hottest
Galactic gas ($T \ga 10^6~\K$) and cooler phases of the interstellar medium (ISM). These ions also
help us investigate the interaction between HVCs and the ambient halo gas, and, by extension, the
properties of the clouds and the halo gas. Some highly ionized HVCs are seen in the same directions
as \HI\ HVCs, but there are also highly ionized HVCs without corresponding high-velocity
\HI\ \citep{sembach03}. The sky covering fraction of \OVI\ HVCs ($\sim$60\%; \citealt{sembach03}) is
larger than that of \HI\ HVCs ($\sim$37\%; \citealt{lockman02}), suggesting the possibility that in
the halo, warm-hot (a few times $10^5$~K) high-velocity gas traced by high ions could be more common
than cool high-velocity gas traced by \HI.

High ions in HVCs could hypothetically result from several physical processes, including mixing of
cool neutral HVC gas with hot ($T \ga 10^6~K$) highly ionized ambient gas, thermal conduction
between cool HVC gas and hot ambient gas, radiative cooling of hot high-velocity gas, and shock
heating. For example, when a cool cloud is embedded in hot gas, thermal conduction can form a layer
of intermediate temperature gas ($T \sim \mbox{a few times}~10^5$~K) in which high ions are abundant
(e.g., \citealt{borkowski90}, for stationary clouds).  Also, simple radiative cooling of hot gas
will result in it passing through the temperature range optimal for high ions. Collisional
ionization equilibrium (CIE) and non-equilibrium ionization (NEI) calculations of this cooling, for
stationary gas at least, have been carried out by many authors (see \citealt{gnat07}, and references
therein). High ions could also be produced in the hot gas behind the shocks that form as HVCs travel
supersonically through the ambient medium. The observed ratios of high ions' column densities or
line intensities can be compared with the results of these models to determine how high ions are
produced; e.g., see Figure~7 in KS10. Although this particular comparison was made for low-velocity
ions, one could test models for high-velocity ions in a similar way.

Some of these hypothetical processes require that hot gas is plentiful in the halo.  Observations of
the diffuse soft X-ray background (SXRB;
\citealt{burrows91,snowden91,snowden98,kuntz00,smith07a,galeazzi07,henley08a,lei09,yoshino09,henley10b})
and of X-ray absorption lines, such as \OVII\ and \OVIII\ (e.g.,
\citealt{yao05,yao07a,fang06,bregman07,yao09}) show that hot gas with $T \sim 1$--$3 \times 10^6~\K$
is indeed present in the halo.  Although the distance and origin of this hot gas are uncertain at
present, it is likely that some of this hot gas exists in the same region as the \HI\ HVCs. If so,
then it is plausible that high ions are produced in the turbulent mixing layers (TMLs) between the
cool HVCs and the hot ambient ISM. This turbulent mixing of cool and hot gas arises from the
Kelvin-Helmholtz or shear instability induced by the velocity difference across the interface
between the two types of gas.  The TML model was first suggested by \citet{begelman90} and later
developed by \citet{slavin93a}.  Column densities of various high ions (including \CIV, \NV, and
\OVI) have been predicted for the TML model using a variety of techniques: analytical calculations
\citep{slavin93a}, 3D magnetohydrodynamic (MHD) simulations \citep{esquivel06}, and 2D hydrodynamic
simulations incorporating NEI calculations \citep[hereafter KS10]{kwak10}. Note that these studies
all used a plane-parallel geometry, rather than a HVC-like geometry. TML column density predictions
have been compared with observations of HVCs (\citealt{sembach03,fox04,collins07}; KS10), and should
be of value for studying other astrophysical situations in which cool or warm material slides past
hot material.

In this paper, we present the results of simulations that trace the interaction between an \HI\ HVC
and the hot ambient medium in a more realistic geometry than previous TML studies, that is, a
spherical cloud falling through the hot ISM. As in our previous study (KS10), we trace the
ionization states of carbon, nitrogen, and oxygen with non-equilibrium calculations which allow us
to estimate the amounts of interesting ions (\CIV, \NV\ and \OVI) more accurately than CIE
calculations. The new simulations enable us to investigate how the cold gas ablates from a
spherically shaped cloud due to shear instabilities, how it mixes with the hot ambient gas, and
where the high ions that are produced in the process of ablation and mixing reside. In addition, we
can examine the velocities of the high ions: there are high-velocity high ions that move with
similar velocities as the \HI\ HVCs and low-velocity high-ions that move with similar velocities as
the ISM.  Using the results of our simulations, we will answer the following questions: (1) How many
high ions are produced by \HI\ HVCs traveling through hot ambient ISM? (2) How many high-velocity
high ions result? (3) How do the ratios of high ions and \HI\ compare with observations? (4) Are
HVCs likely to reach the Galactic disk intact?  In this paper we will concentrate on the
high-velocity ions, deferring the discussion of low-velocity ions to Paper~II (D.~B. Henley
et~al. 2011, in preparation).

This paper is organized as follows. In the next section, we briefly summarize our numerical methods
and present the parameters for our suite of seven simulational models. In Section~\ref{sec:Results}
we present the results of our simulations. Specifically, Section~\ref{subsec:CloudEvolution}
describes in detail the hydrodynamical evolution of one of our model clouds,
Section~\ref{subsec:LossOfHI} describes the evolution of the amount of neutral material,
Section~\ref{subsec:CloudEvolutionDifferences} describes the effect of the different model
parameters on the cloud evolution, Section~\ref{subsec:FateOfHVCs} discusses whether or not HVCs can
reach the Galactic disk, and Section~\ref{subsec:HighStageIons} describes the production of
high ions in our simulations. In Section~\ref{sec:ComparisonWithObservations} we compare the
column densities and column density ratios predicted by our simulations with observed Complex~C
values. In Section~\ref{sec:NeglectedPhysics} we discuss the effects that the various assumptions in
our simulations have on our results. We summarize our results in Section~\ref{sec:Summary}.

\section{NUMERICAL METHODS AND MODEL PARAMETERS}
\label{sec:Method}

\begin{deluxetable*}{ccccccc}
\tablewidth{0pt}
\tablecaption{Model Parameters\label{tab:ModelParameters}}
\tablehead{
\colhead{Model}         & \colhead{Radius}      & \colhead{\vzcl}   & \colhead{\nHcl}       & \colhead{\MinitT}     & \colhead{\Minitv}     & \colhead{\nHISM} \\
\colhead{}              & \colhead{(\pc)}       & \colhead{(\kmps)}     & \colhead{(\pcc)}      & \colhead{(\Msol)}     & \colhead{(\Msol)}     & \colhead{(\pcc)} \\
\colhead{(1)}           & \colhead{(2)}         & \colhead{(3)}         & \colhead{(4)}         & \colhead{(5)}         & \colhead{(6)}         & \colhead{(7)}
}

\startdata
A &  20 & $-$100 & 0.1  & 120              & 130             & $1.0\times10^{-4}$ \\
B & 150 & $-$100 & 0.1  & $4.9\times 10^4$ & $5.1\times 10^4$ & $1.0\times10^{-4}$ \\
C & 150 & $-$150 & 0.1  & $4.9\times 10^4$ & $5.1\times 10^4$ & $1.0\times10^{-4}$ \\
D & 150 & $-$300 & 0.1  & $4.9\times 10^4$ & $5.1\times 10^4$ & $1.0\times10^{-4}$ \\
E & 150 & $-$150 & 0.1  & $4.9\times 10^4$ & $4.9\times 10^4$ & $1.0\times10^{-4}$ \\
F & 300 & $-$100 & 0.1  & $4.0\times 10^5$ & $4.2\times 10^5$ & $1.0\times10^{-4}$ \\
G & 150 & $-$100 & 0.01 & $4.9\times 10^3$ & $5.1\times 10^3$ & $1.0\times10^{-5}$
\enddata

\tablecomments{
  Col.~(1): Model identifiers.
  Col.~(2): Approximate radius of the model cloud (except for Model~E). See Figure~\ref{fig:CloudProfile} for the
  detailed density profile of each model cloud. Model~E is a uniform-density cloud with an exact radius.
  Col.~(3): Initial velocity of the cloud along the $z$-direction measured in the observer's frame.
  Col.~(4): Initial hydrogen number density of the cloud at its center.
  Col.~(5): Initial mass of cloud having a temperature $T < 10^4~\K$.
  Col.~(6): Initial mass of cloud moving with \vzcl. Note that all cloud material with a hydrogen number density
  greater than $5 \nHISM$, where \nHISM\ is the hydrogen number density of the ISM (Col.~7), moves initially at
  speed \vzcl\ relative to the ISM.
  All models have the same cloud temperatures ($\Tcl=10^3$~K at cloud centers) and ISM temperatures ($\TISM=10^6$~K).
}

\end{deluxetable*}

We carried out our simulations using the same code as KS10, namely FLASH version 2.5
\citep{fryxell00}. We used the FLASH NEI module to track the ionization evolution of carbon,
nitrogen, and oxygen.\footnote{\SiIV\ is another high ion that has been observed in the halo of the
  Milky Way via its UV lines. However, this ion is more susceptible to photoionization than \CIV,
  \NV, or \OVI, due to its lower ionization potential. Modeling photoionization is beyond the scope
  of this paper, so we do not include silicon in our NEI calculations. Note that the ionization and
  recombination rates in the FLASH NEI module include only the effects of collisional ionization,
  auto-ionization, radiative recombination, and dielectronic recombination.} The simulations include
radiative cooling, although the cooling curve was calculated assuming CIE. See KS10 for more details
of the code, including some discussion of CIE versus NEI cooling rates. As in KS10, we used the
abundances from \citet{allen73}, which are the default abundances in FLASH. Our results can be
rescaled to give results for different abundance tables, assuming that the radiative cooling curve
does not change significantly with the assumed abundances (an assumption which may not always be
valid).

In KS10 the simulations were run in 2D Cartesian coordinates to study TMLs in a plane-parallel
geometry. Here, we use 2D cylindrical coordinates to study initially spherical clouds as they fall
through the hot ISM. As in KS10, we use 2D simulations in order to minimize the unaffordably large
computing resources (particularly memory) needed to track the ionization of the three elements in
our NEI simulations.  In addition, instead of the cloud moving relative to a stationary ISM, in our
simulations the ambient medium moves relative to an initially stationary cloud. In this way, we can
trace the cloud evolution for a long time without requiring a large computational domain, reducing
the amount of memory and CPU time needed to run the simulations.  However, although the cloud is
initially stationary in the computational domain, all velocities stated in this paper are in the
observer's frame, in which the ISM is stationary. In the computational grid's reference frame, the
ISM is initially moving in the $+z$ direction with velocity $|\vzcl|$, where $-|\vzcl|$ is the
initial velocity of the cloud in the observer's frame.  Therefore, the $z$-velocity of gas in a
given grid cell in the observer's frame is $\vz = \vzsim - |\vzcl|$, where \vzsim\ is the
$z$-velocity of the gas in that cell obtained directly from the simulation. Note that, although we
present results for HVCs with negative velocities (i.e., moving toward us), our results are equally
applicable to HVCs with positive velocities if the signs on the velocities are changed.

\begin{figure}
\plotone{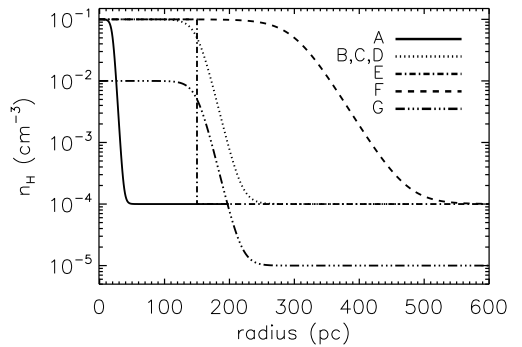}
\caption{Initial hydrogen number density profiles for our various models (see also
  Table~\ref{tab:ModelParameters}).
  \label{fig:CloudProfile}}
\end{figure}

The parameters of our various model clouds are presented in Table~\ref{tab:ModelParameters} and
Figure~\ref{fig:CloudProfile}. We ran seven different models in which we varied the cloud's initial
radius (Table~\ref{tab:ModelParameters}, Col.~2), radial profile (Figure~\ref{fig:CloudProfile}),
$z$-velocity, \vzcl\ (measured in the observer's frame; Table~\ref{tab:ModelParameters}, Col.~3),
and number density expressed in terms of the hydrogen number density, where $\nH / \nHe = 10$
(Table~\ref{tab:ModelParameters}, Col.~4). The ambient number density is $10^3$ times smaller than
the initial cloud number density (Table~\ref{tab:ModelParameters}, Col.~7).  All cloud material with
a density $\ge5$ times the ambient density initially moves at \vzcl\ relative to the ambient medium,
while the less dense outskirts of the cloud are initially set to the ISM's velocity.  As in KS10, we
required that the cloud and the ISM were initially in pressure balance. We achieved this by varying
the temperature from $\Tcl = 10^3~\K$ at the cloud center to $\TISM = 10^6~\K$ in the ambient
medium, such that the pressure was constant.

Because the clouds generally do not have sharp edges, the cloud mass is not well defined. For this
reason, we have calculated two masses for each cloud: the mass of material with $T < 10^4~\K$
(Table~\ref{tab:ModelParameters}, Col.~5), and the mass of material initially moving at
\vzcl\ relative to the ambient medium (Table~\ref{tab:ModelParameters}, Col.~6). These two masses
are generally similar to each other.  The cloud in Model~E has a sharp edge, and so the two masses
are identical in this model.

By running Models~B, C, and D, we can see the effects of varying the cloud's initial velocity
(Section~\ref{subsubsec:EffectOfVelocity}), because these models otherwise have the same initial
parameters. Apart from the radial density profile, Models~C and E are identical, and so these two
models show the effect of the cloud's initial density profile
(Section~\ref{subsubsec:EffectOfDensity}). The cloud and ISM densities in Model~G are an order of
magnitude smaller than those in the other models, and so comparing Models~B and G reveals the effect
of varying the cloud's density (Section~\ref{subsubsec:EffectOfDensity}).  Finally, Models~A, B, and
F reveal the effect of varying the cloud size (Section~\ref{subsubsec:EffectOfSize}).

Although the number densities of HVCs and of the upper halo are not well constrained, our chosen
model parameters are consistent with existing observational and theoretical constraints. The
\HI\ column densities on sightlines passing through the centers of our model clouds range from
$9.3 \times 10^{18}~\mbox{cm}^{-2}$ (Model~G) to $1.9 \times 10^{20}~\mbox{cm}^{-2} $ (Model~F),
and are consistent with measured column densities of \HI\ HVCs \citep[e.g.,][]{hulsbosch88}. The
density of the hot gas in the upper halo is more uncertain, with observational estimates including
$\nH < 6.3 \times 10^{-4}~\pcc$ from pulsar dispersion measures \citep{gaensler08} and $\nH = 7.5
\times 10^{-4}~\pcc$ from \OVII\ column density measurements (assuming a uniform spherical halo of
radius 20~\kpc; \citealt{bregman07}).\footnote{These estimates of the density of the upper halo were
  given as electron number densities. We have converted them to hydrogen number densities assuming
  $\nH / \nHe = 10$.} The halo density has also been constrained by combining observations with
theoretical models. \citet{peek07} obtained an upper halo density of $\nH \sim 2 \times
10^{-4}~\pcc$ by considering the drag force on an HVC complex observed in \HI. \citet{grcevich09}
estimated that the number density in the upper halo is $> (\mbox{2--3}) \times 10^{-4}~\pcc$, assuming
that the Milky Way's satellite galaxies lost their gas through ram-pressure stripping as they passed
through the upper halo.  The halo density may not decrease dramatically with distance --
\citet{weiner96} estimated that the density near the Magellanic Stream ($d \sim 50~\kpc$) is $\nH
\sim 1 \times 10^{-4}~\pcc$, assuming that the observed \Halpha\ emission arises from an interaction
between the Magellanic Stream and the ambient gas.  Note that most of these estimated halo densities
are slightly larger than the halo density used in most of our simulations ($\nH = 1 \times
10^{-4}~\pcc$ for Models A through F). However, \citet{sembach03} found that their observations of
\OVI\ HVCs favored a low-density extended hot halo ($\nH \la 10^{-4}$--$10^{-5}~\pcc$, $r \ga
70~\kpc$). The lower halo density in Model~G ($\nH = 10^{-5}~\pcc$) was chosen to investigate the
lower range of halo densities, as well as a lower cloud density.

\section{RESULTS}
\label{sec:Results}

\subsection{Hydrodynamical Evolution of High-Velocity Clouds}
\label{subsec:CloudEvolution}

\begin{figure*}
\centering
\hspace*{0.25in}
\includegraphics[scale=0.25]{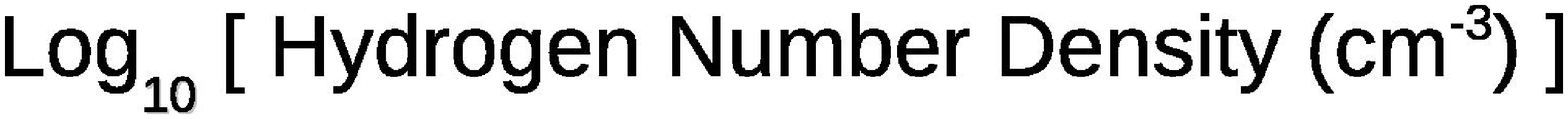} \\
\includegraphics[scale=0.18]{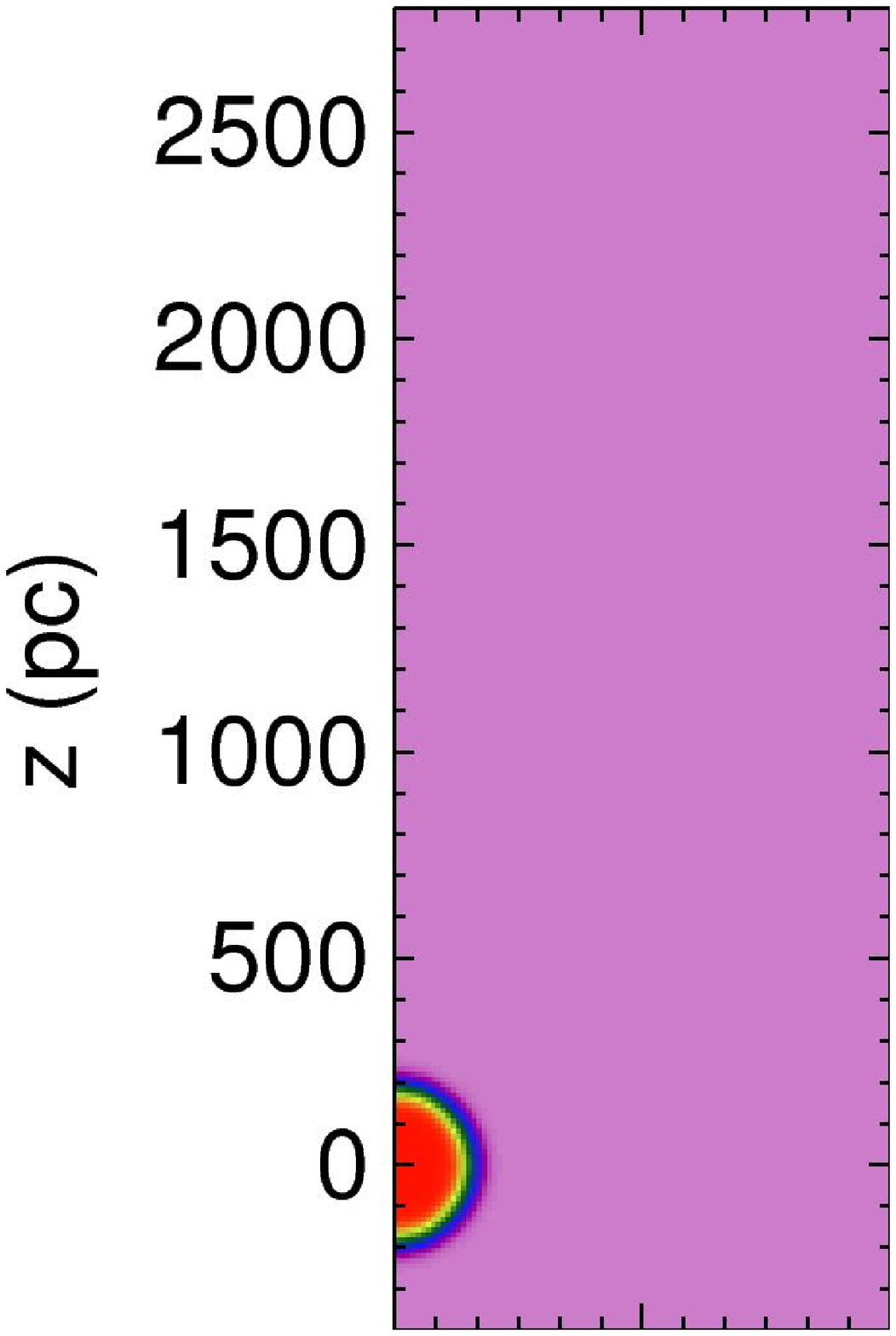}
\hspace{0.070in}
\includegraphics[scale=0.18]{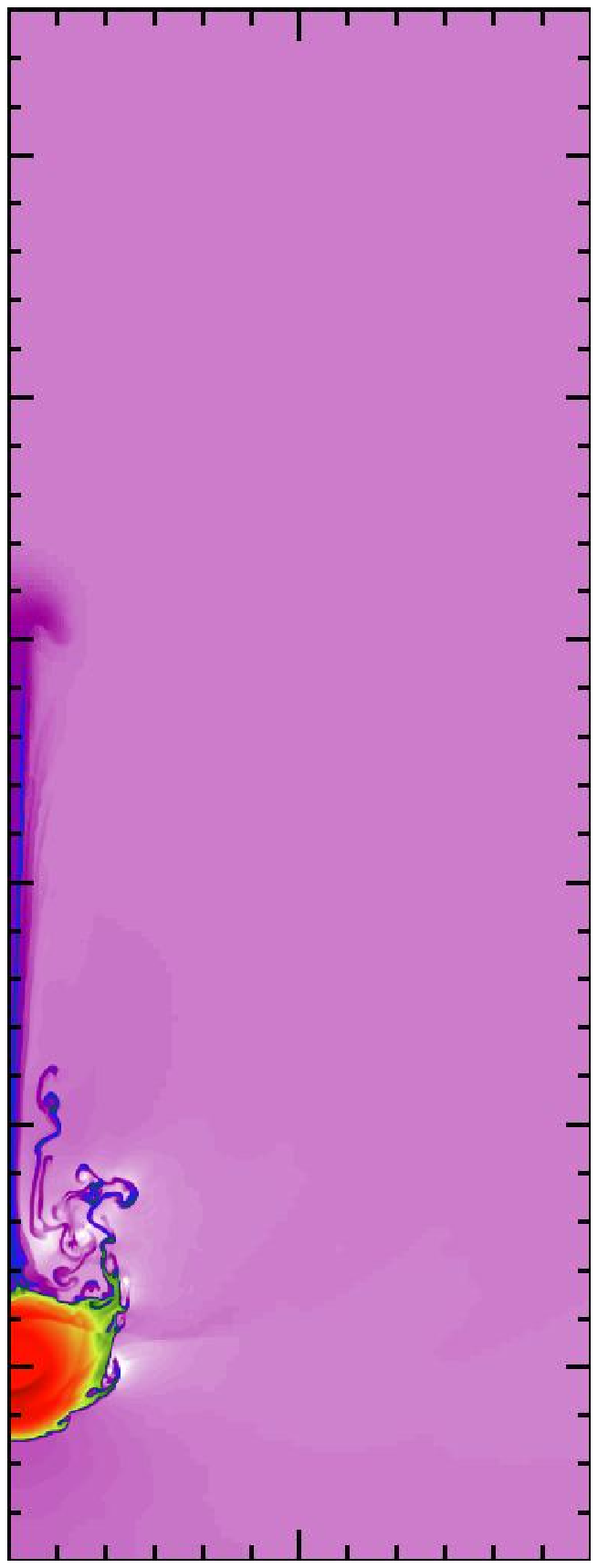}
\hspace{0.070in}
\includegraphics[scale=0.18]{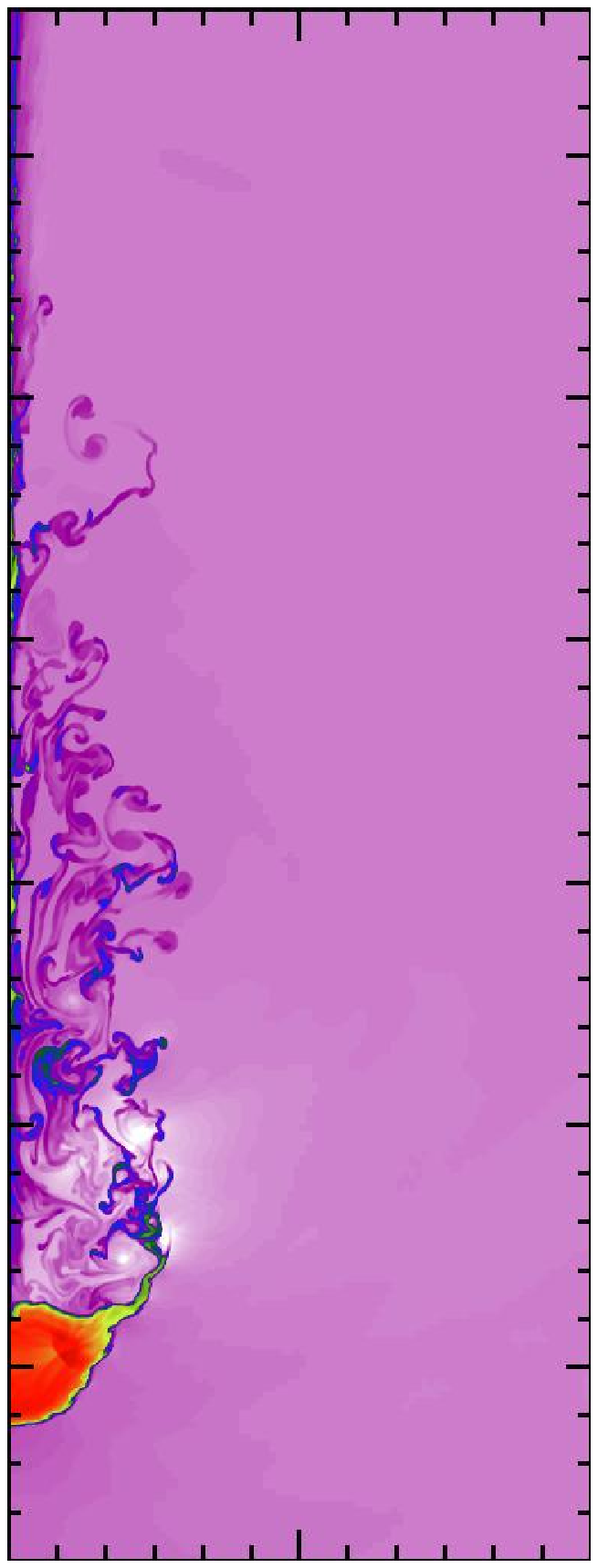}
\hspace{0.070in}
\includegraphics[scale=0.18]{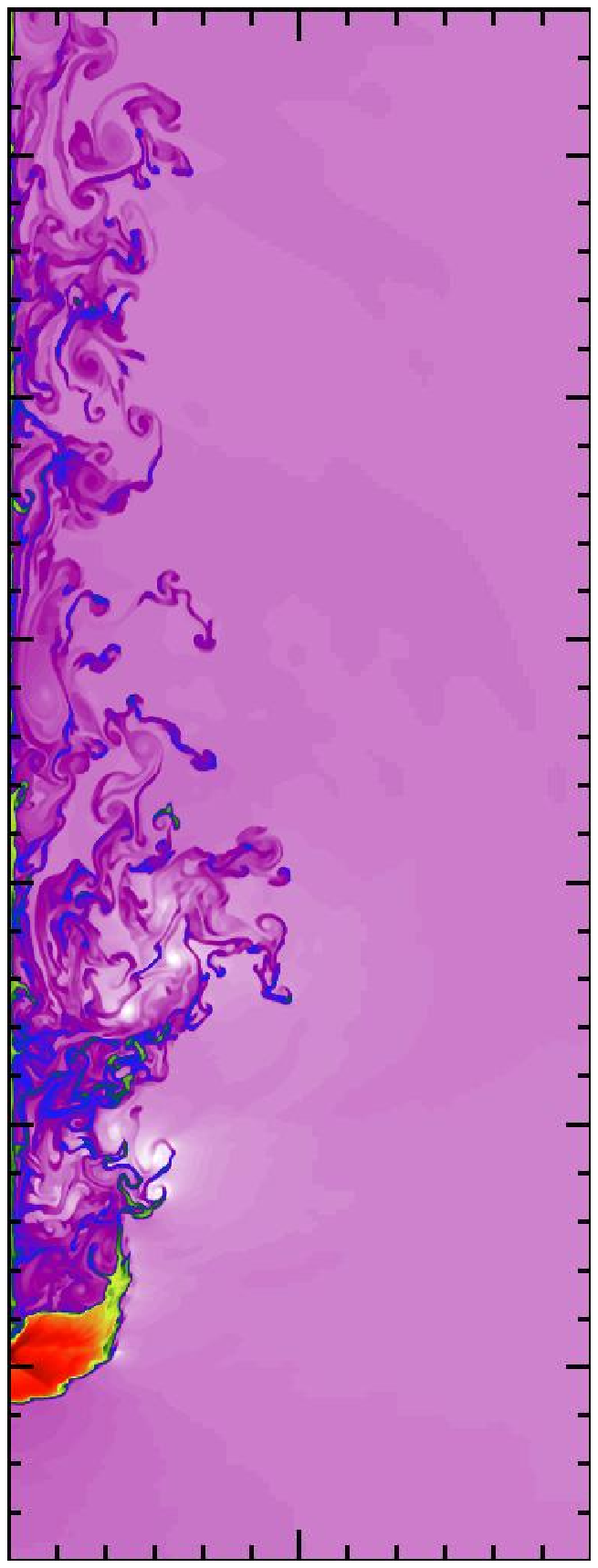}
\hspace{0.070in}
\includegraphics[scale=0.18]{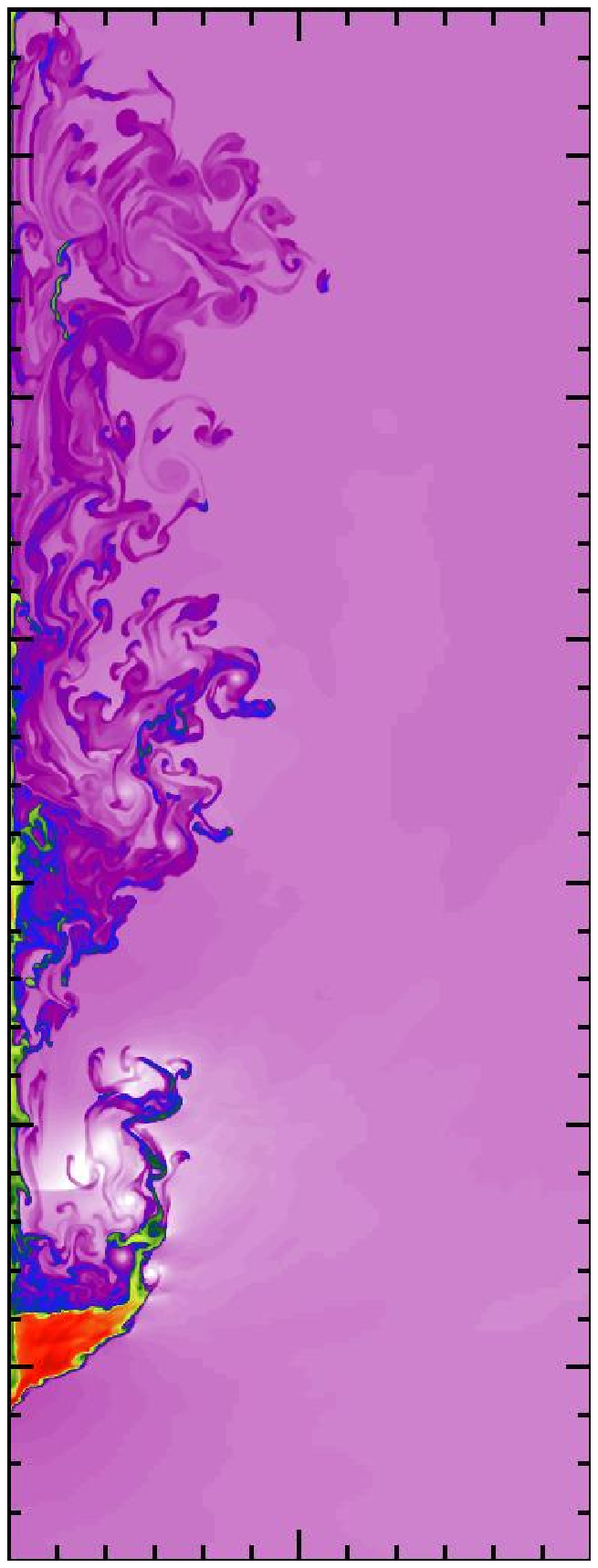}
\hspace{0.070in}
\includegraphics[scale=0.18]{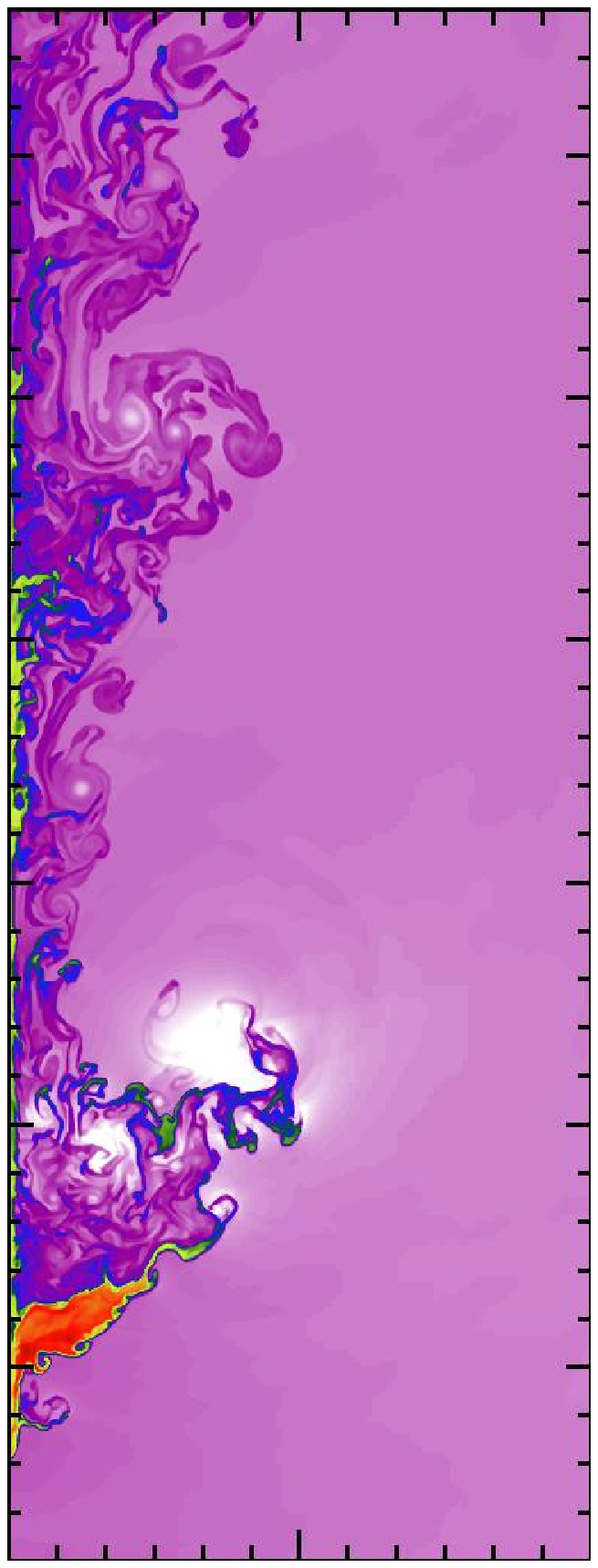}
\hspace{0.070in}
\includegraphics[scale=0.18]{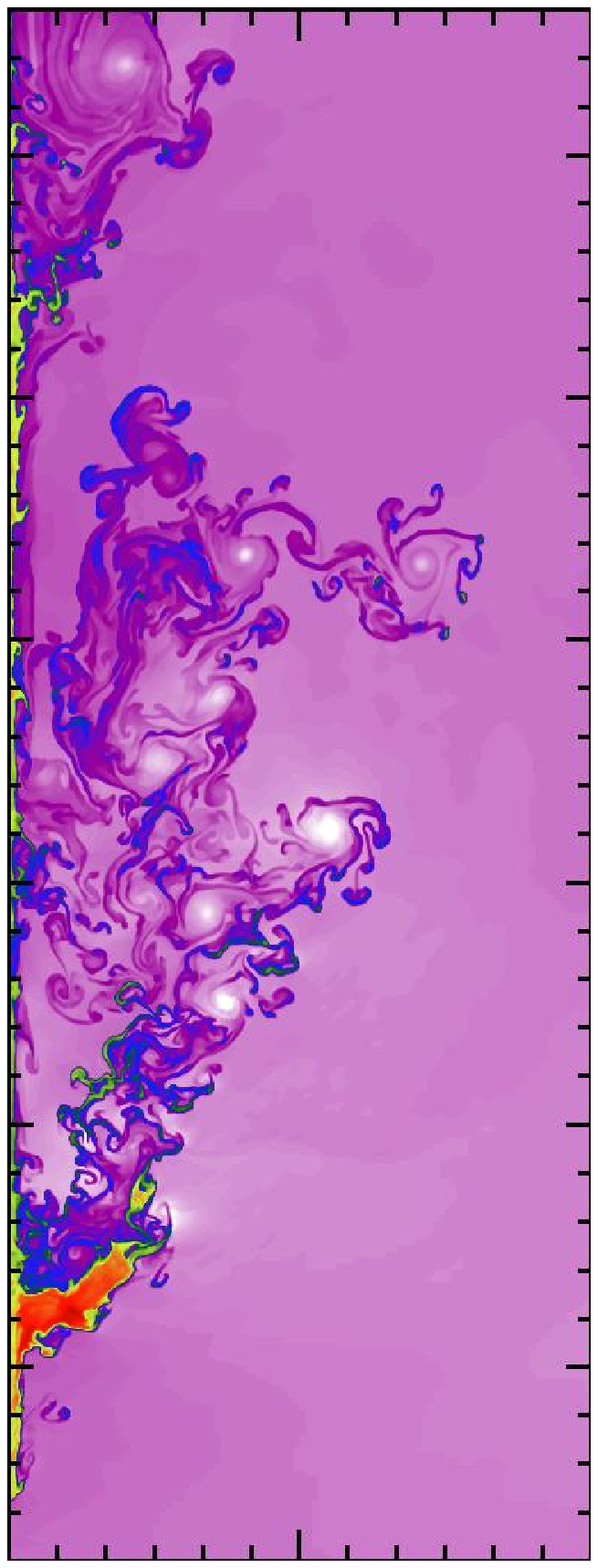}
\hspace{0.070in}
\includegraphics[scale=0.18]{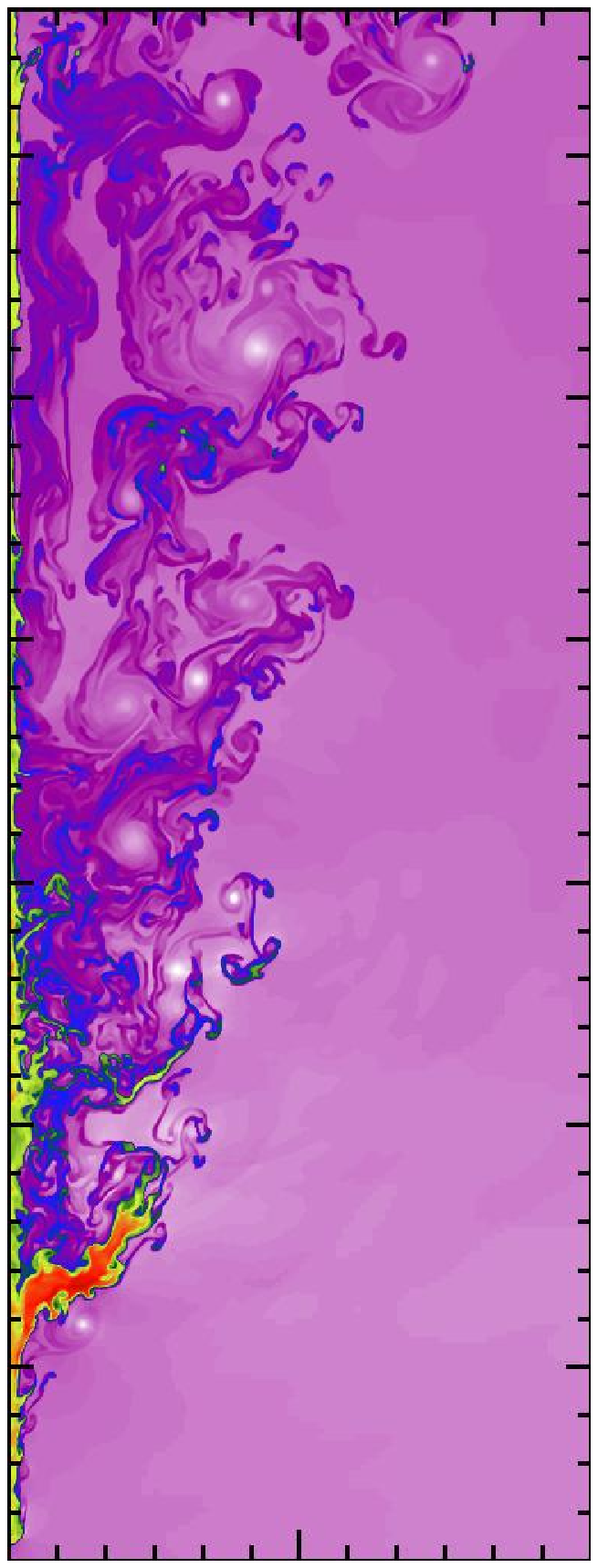}
\hspace{0.070in}
\includegraphics[scale=0.18]{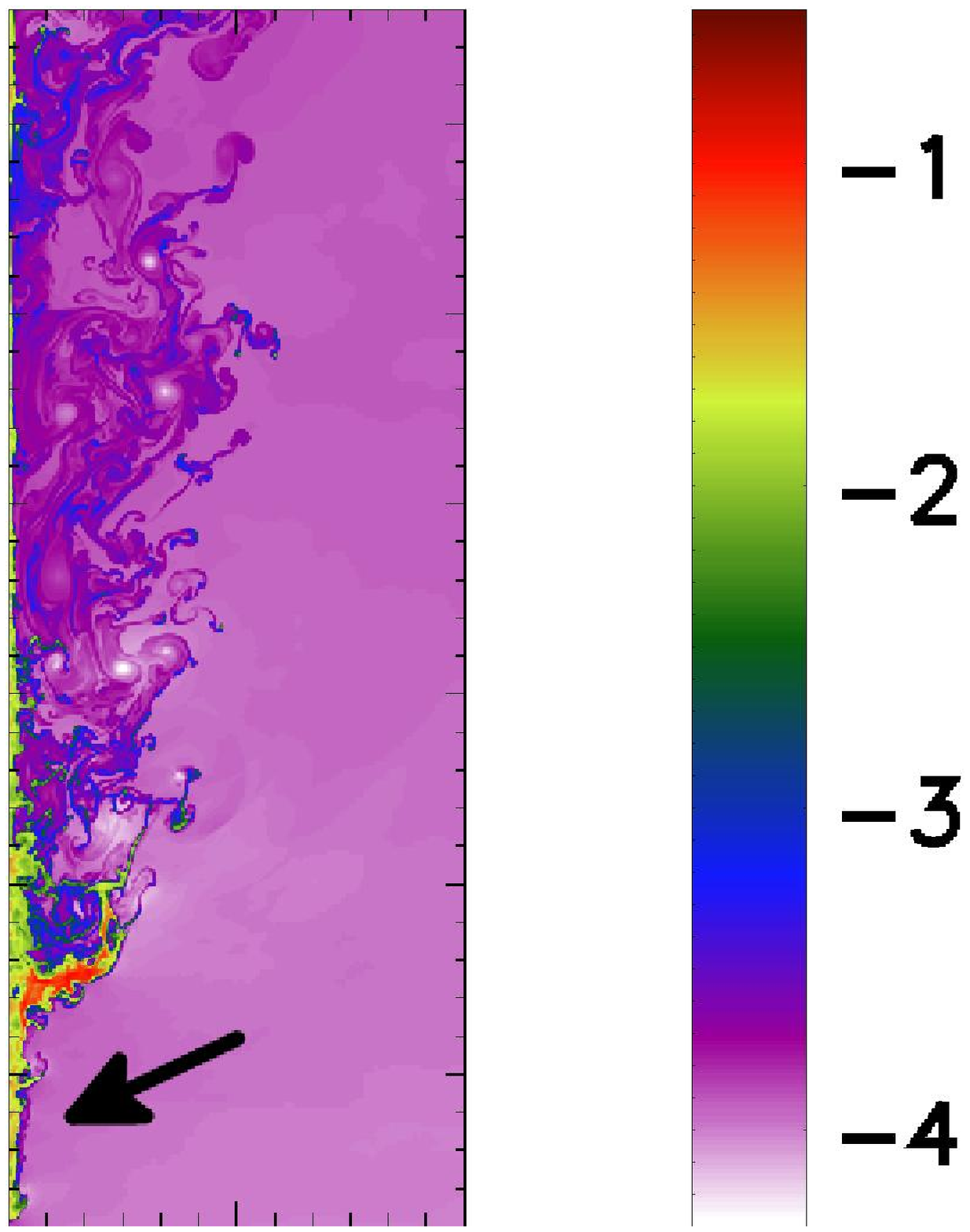} \\
\vspace{0.15in}
\hspace*{0.25in}
\includegraphics[scale=0.25]{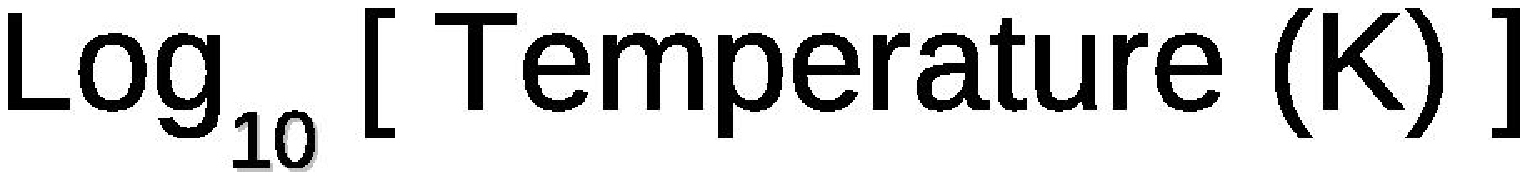} \\
\includegraphics[scale=0.18]{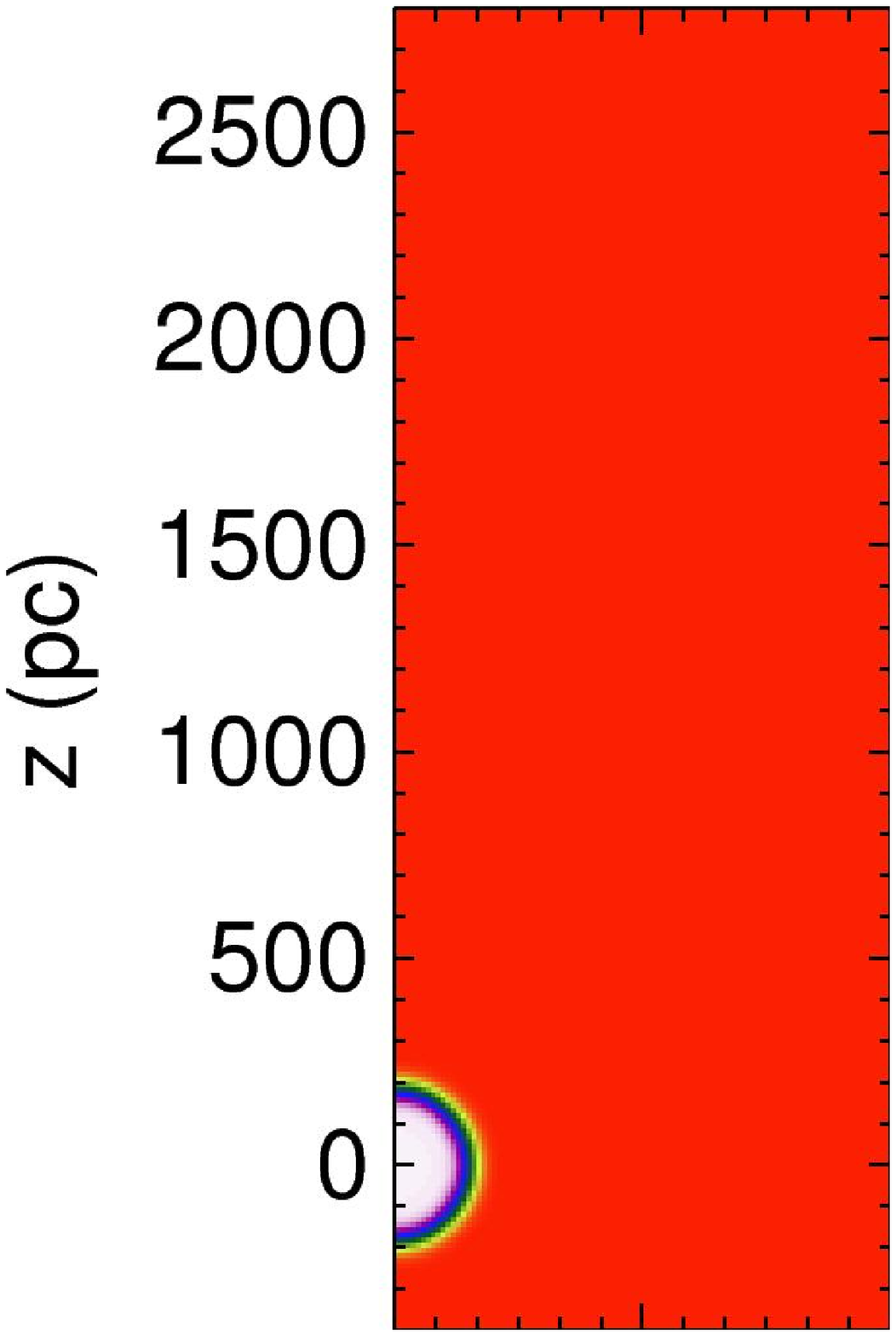}
\hspace{0.070in}
\includegraphics[scale=0.18]{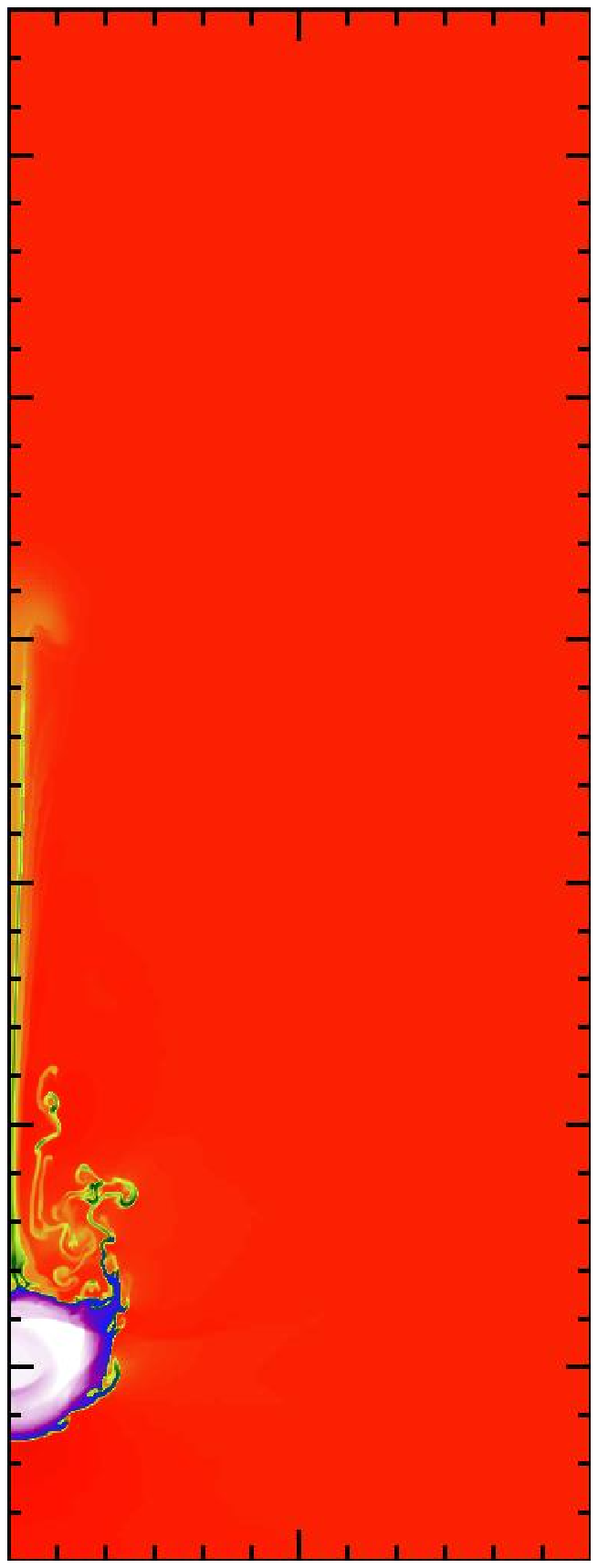}
\hspace{0.070in}
\includegraphics[scale=0.18]{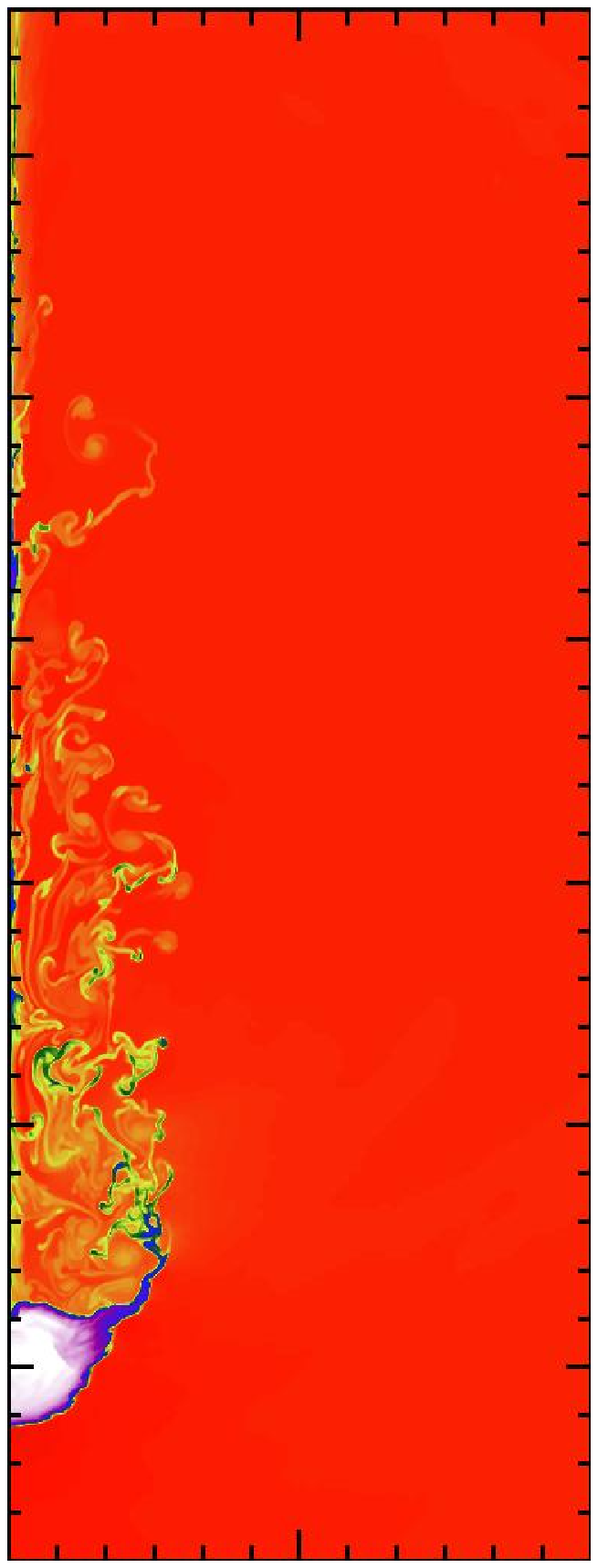}
\hspace{0.070in}
\includegraphics[scale=0.18]{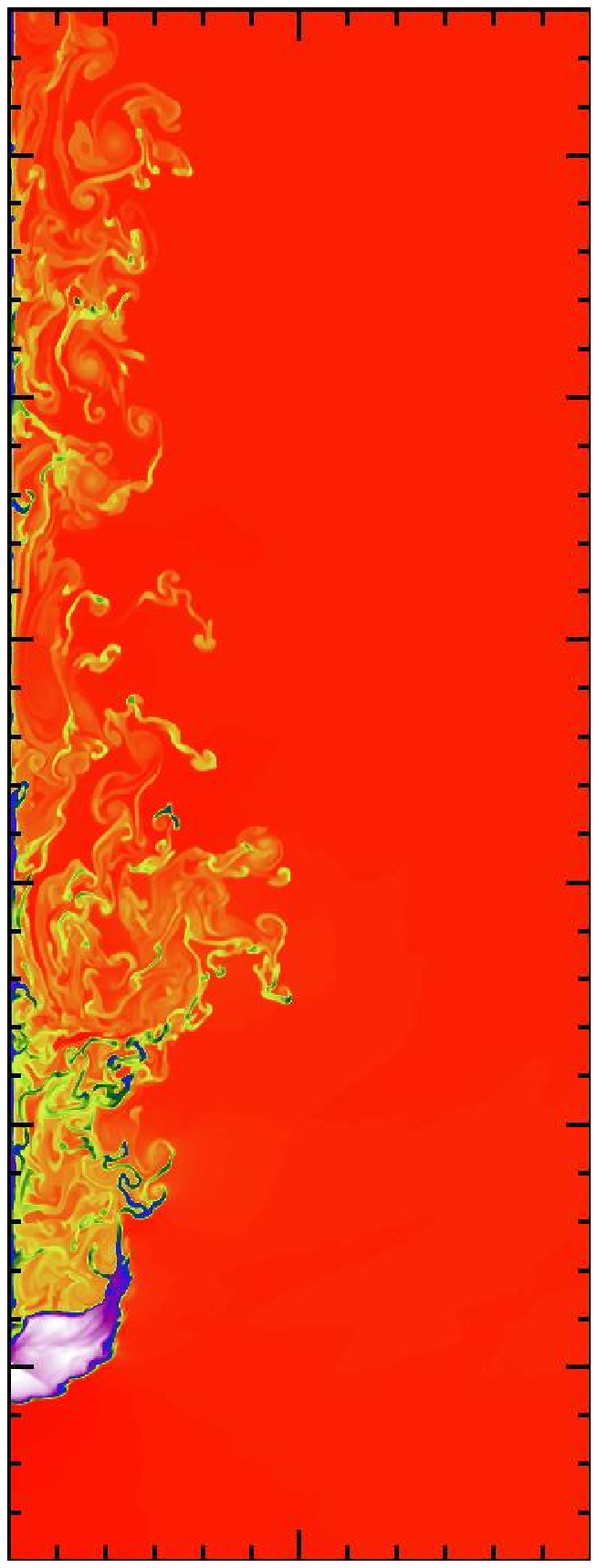}
\hspace{0.070in}
\includegraphics[scale=0.18]{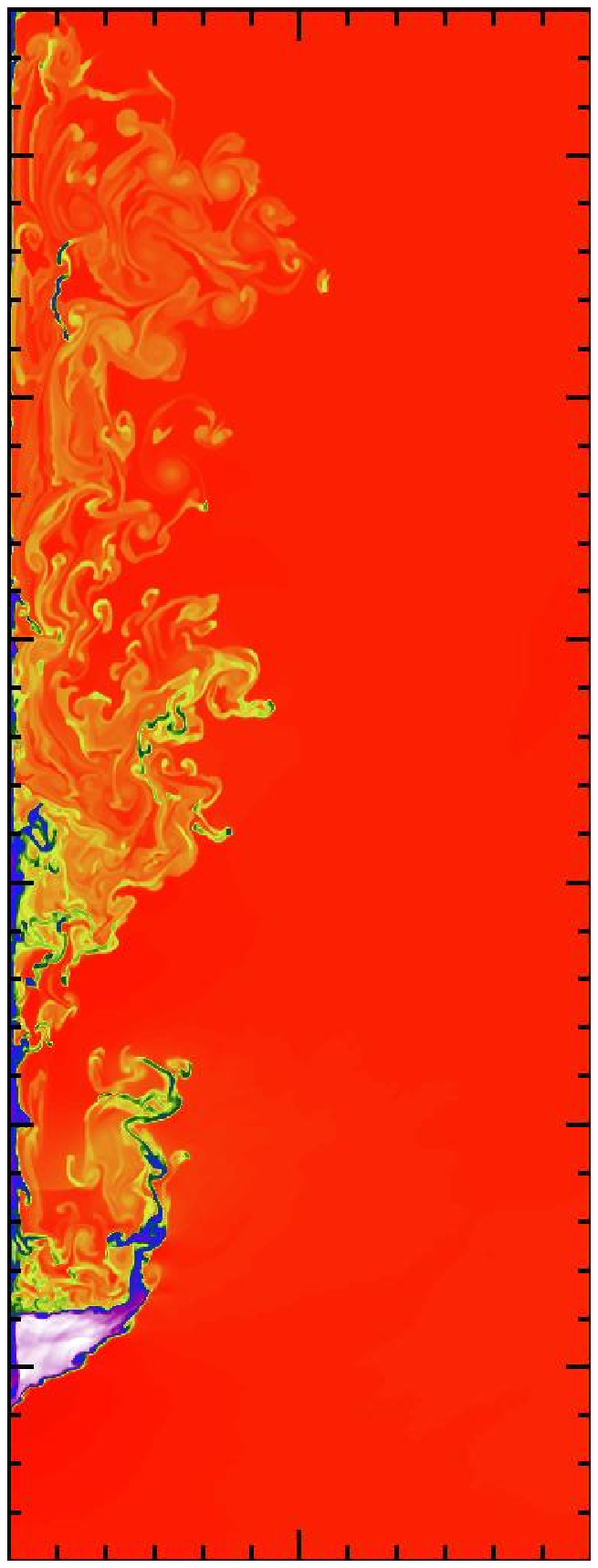}
\hspace{0.070in}
\includegraphics[scale=0.18]{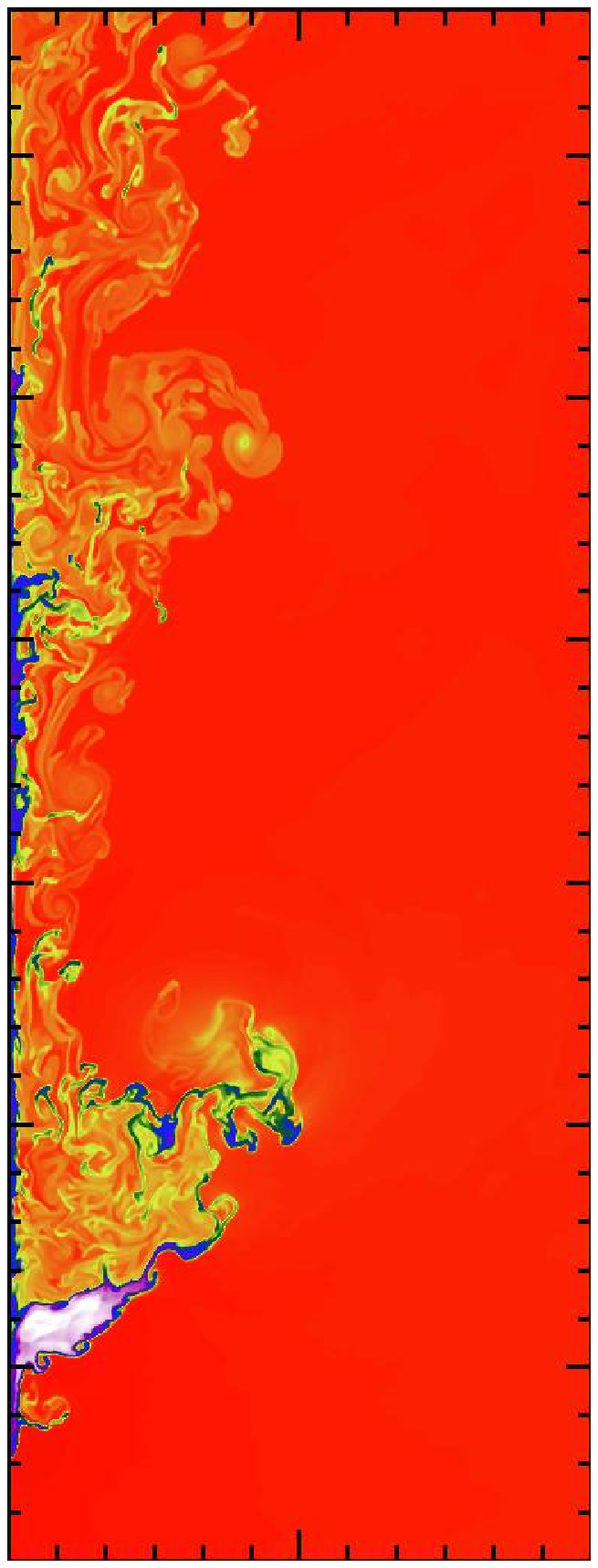}
\hspace{0.070in}
\includegraphics[scale=0.18]{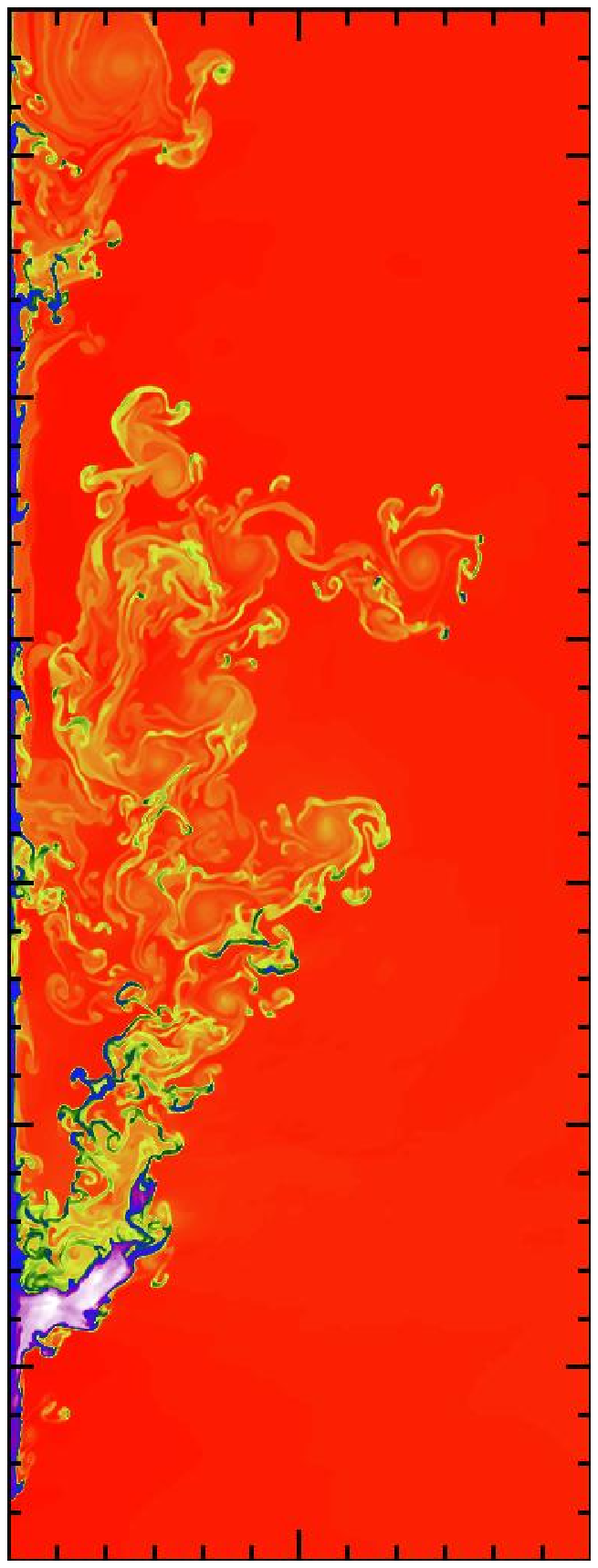}
\hspace{0.070in}
\includegraphics[scale=0.18]{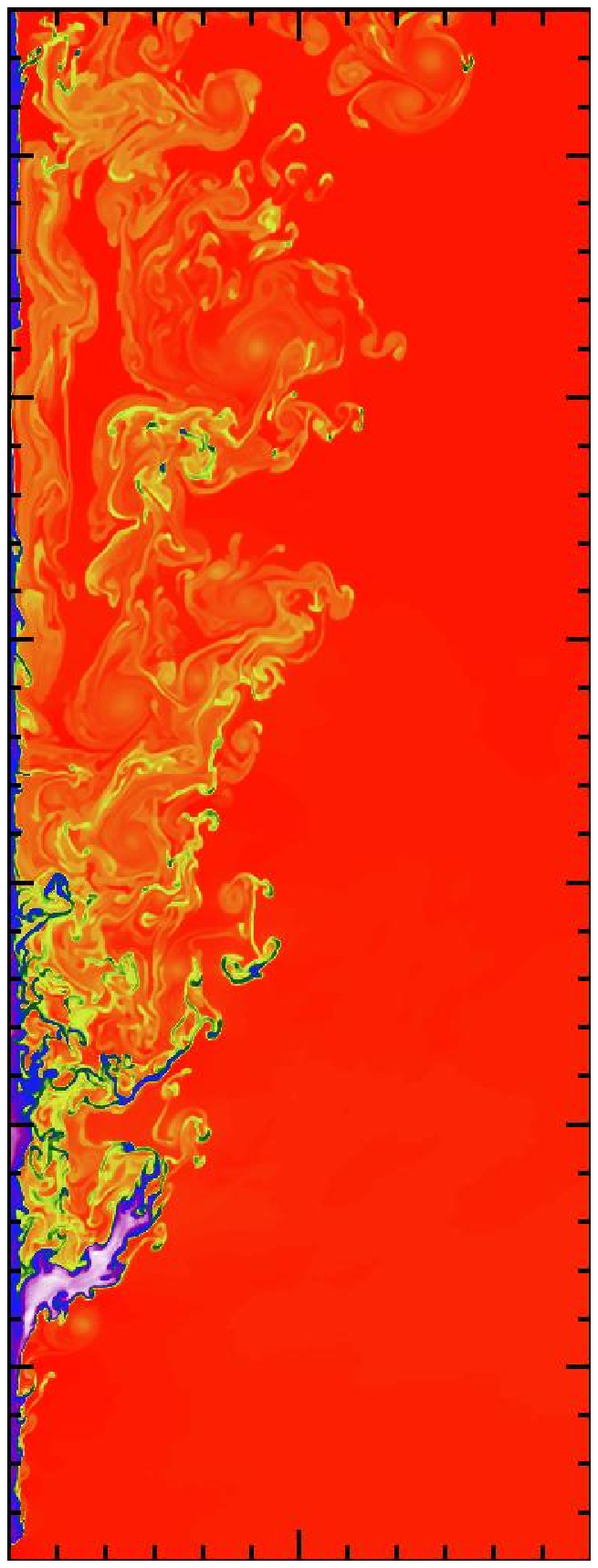}
\hspace{0.070in}
\includegraphics[scale=0.18]{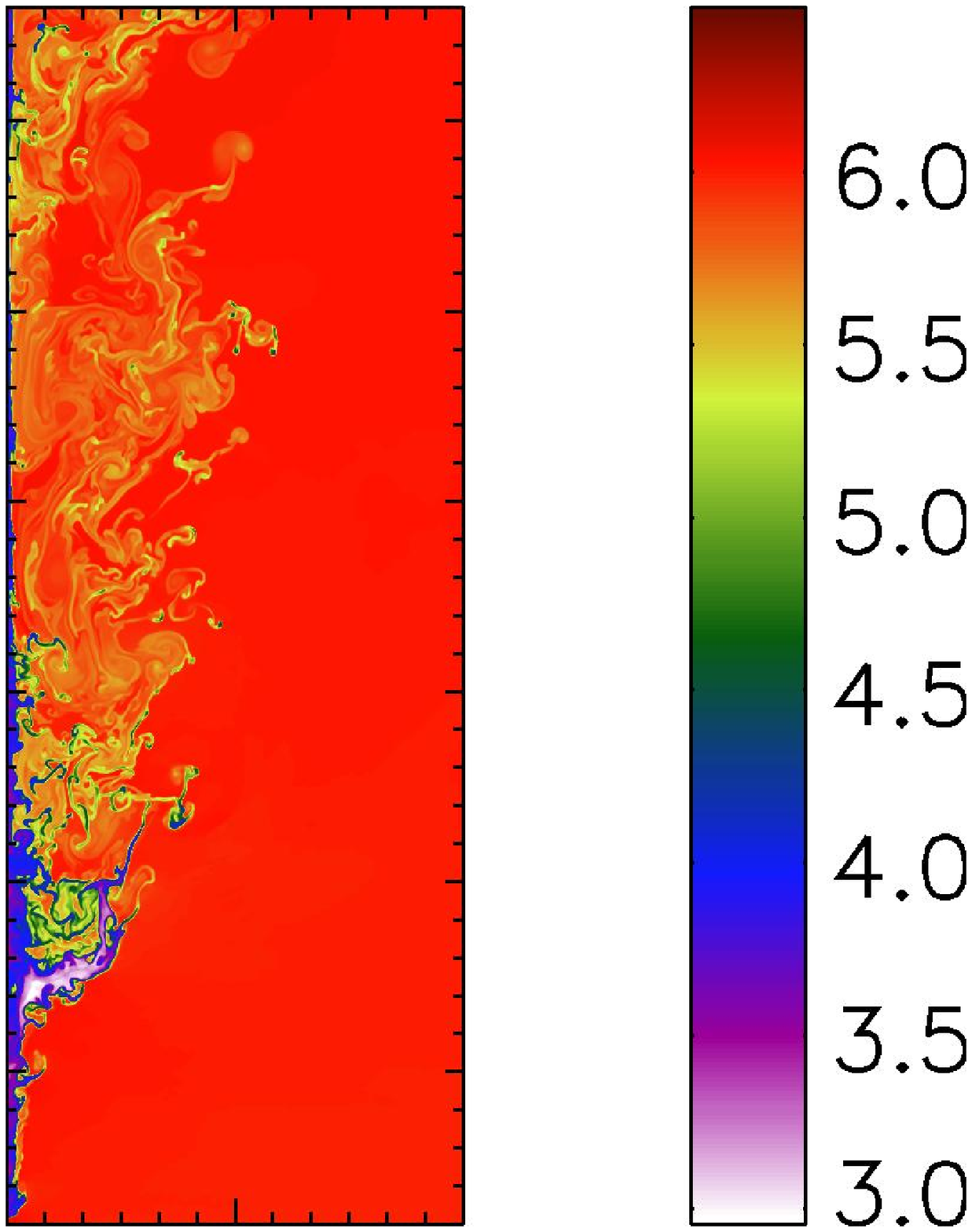} \\
\vspace{0.15in}
\hspace*{0.25in}
\includegraphics[scale=0.25]{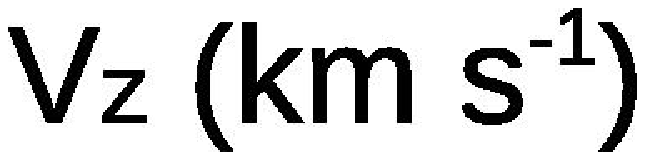} \\
\includegraphics[scale=0.18]{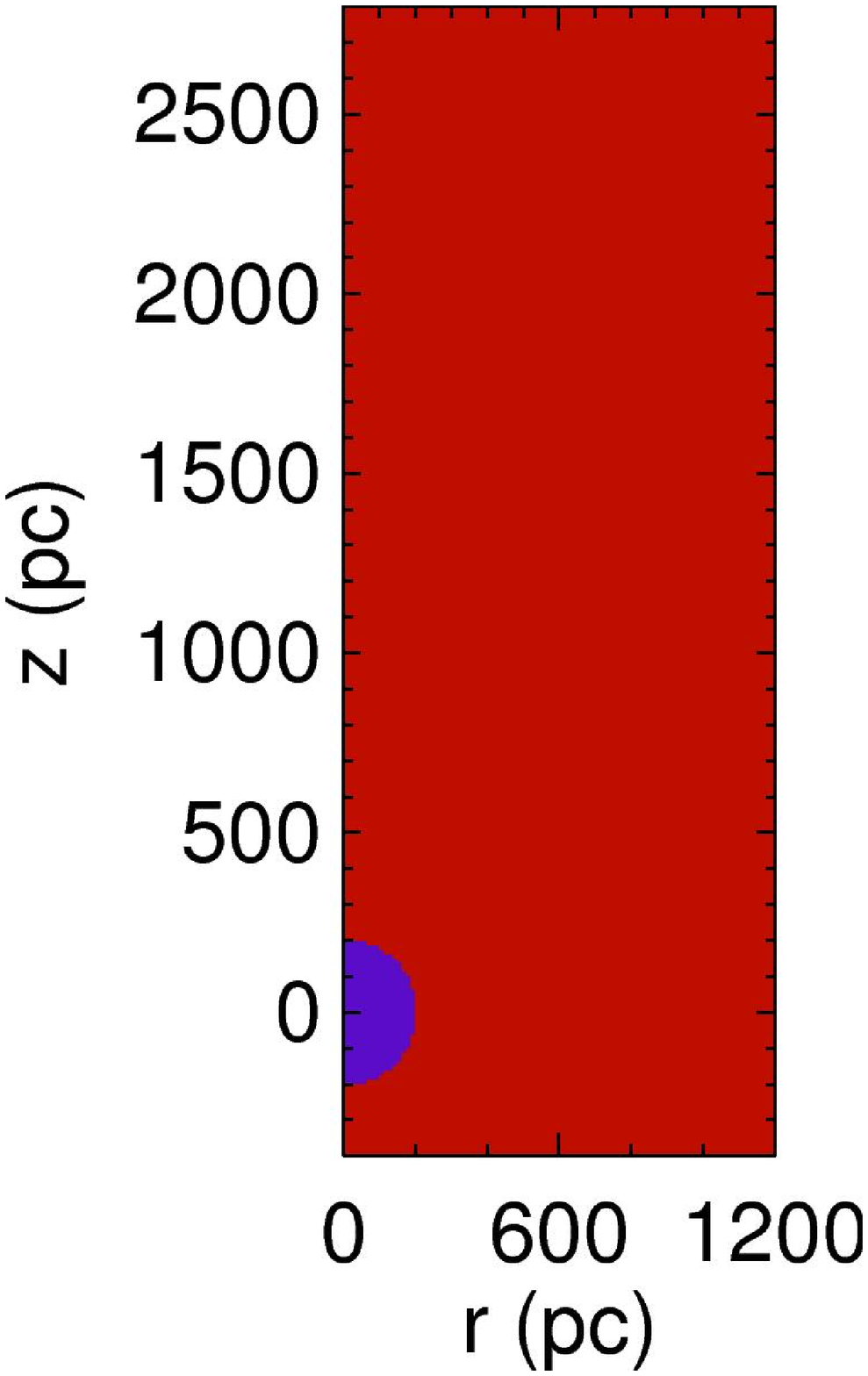}
\includegraphics[scale=0.18]{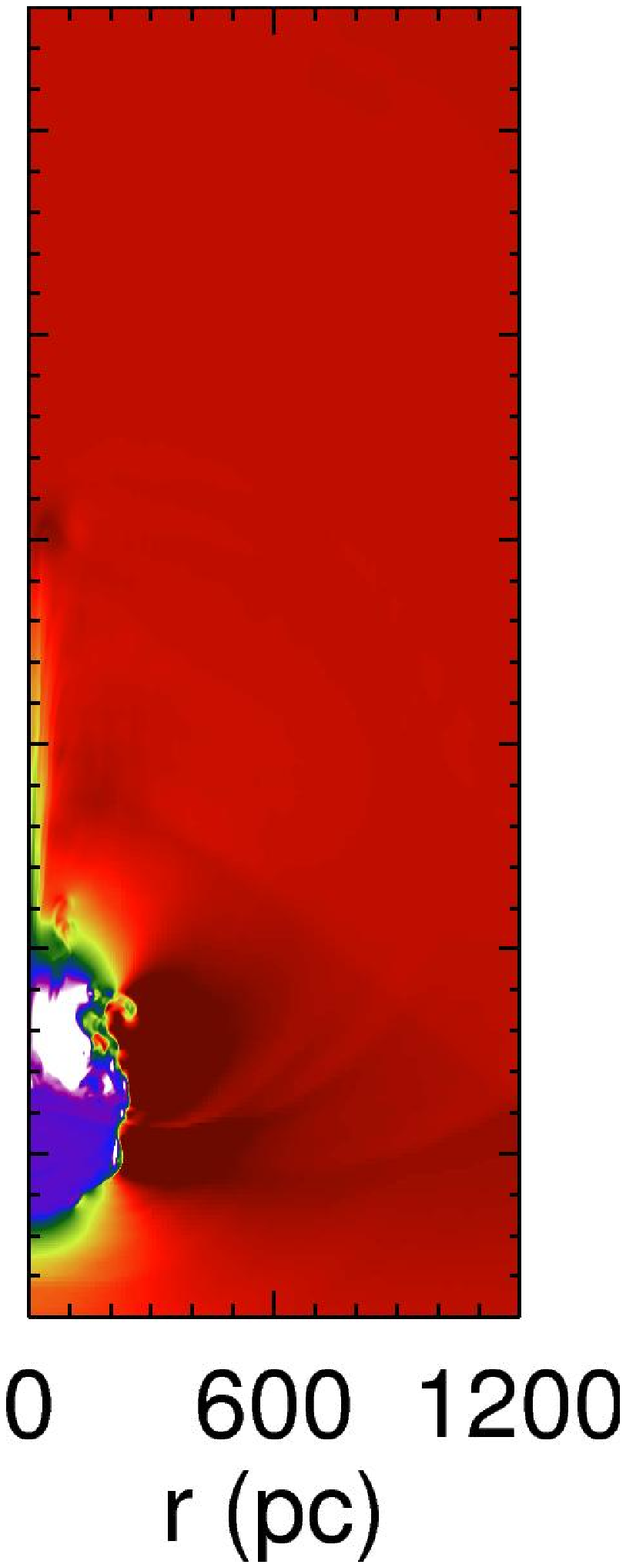}
\includegraphics[scale=0.18]{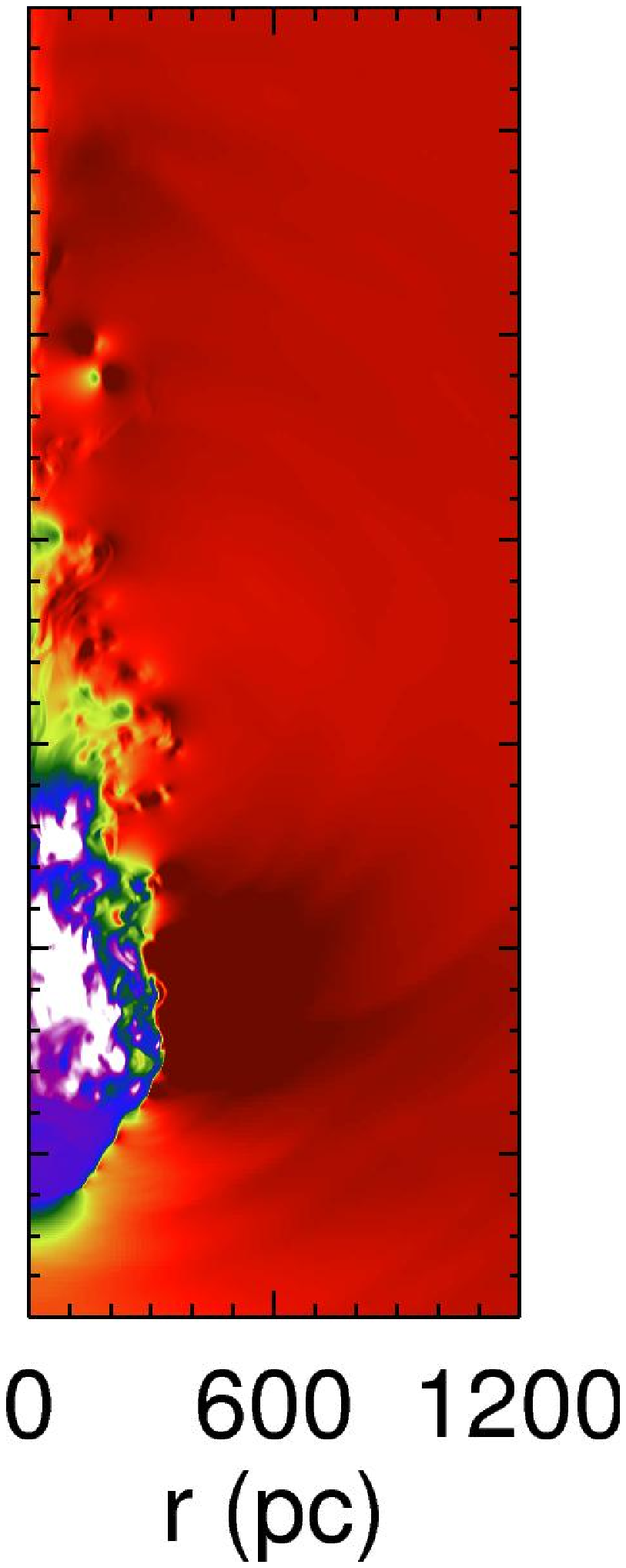}
\includegraphics[scale=0.18]{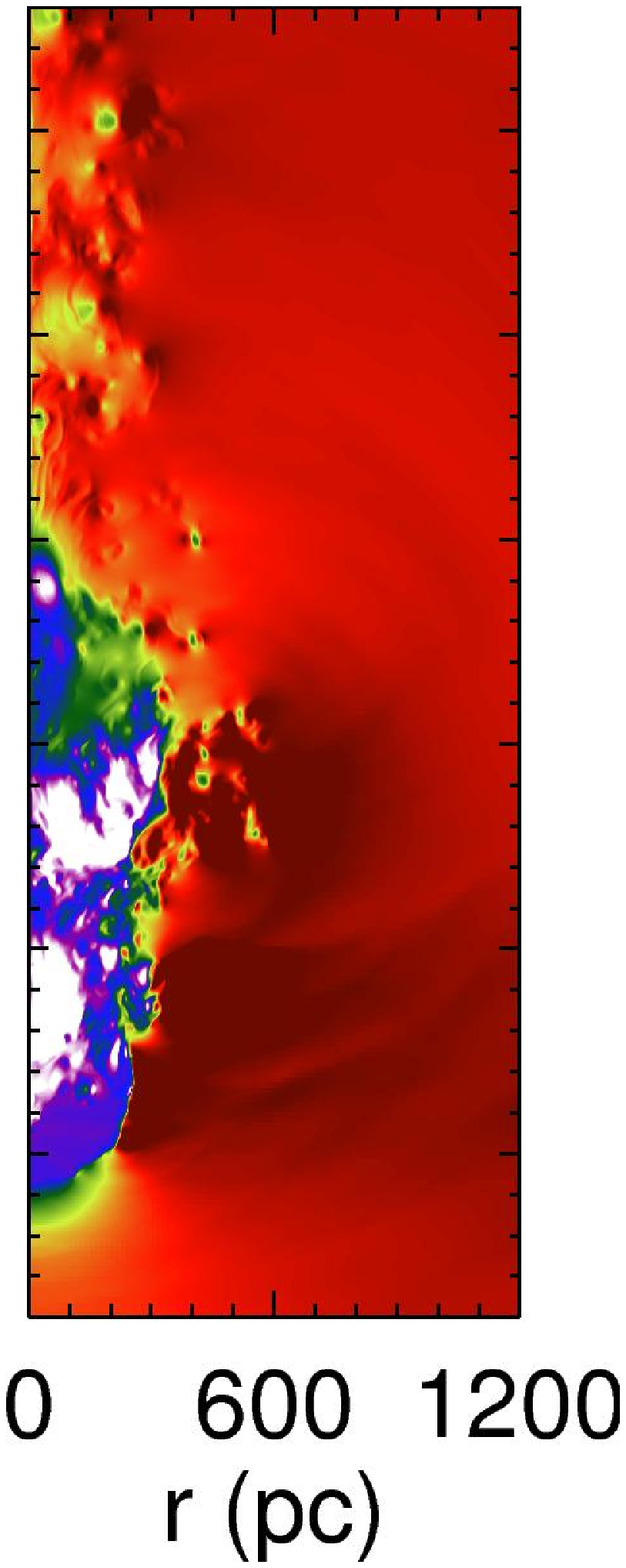}
\includegraphics[scale=0.18]{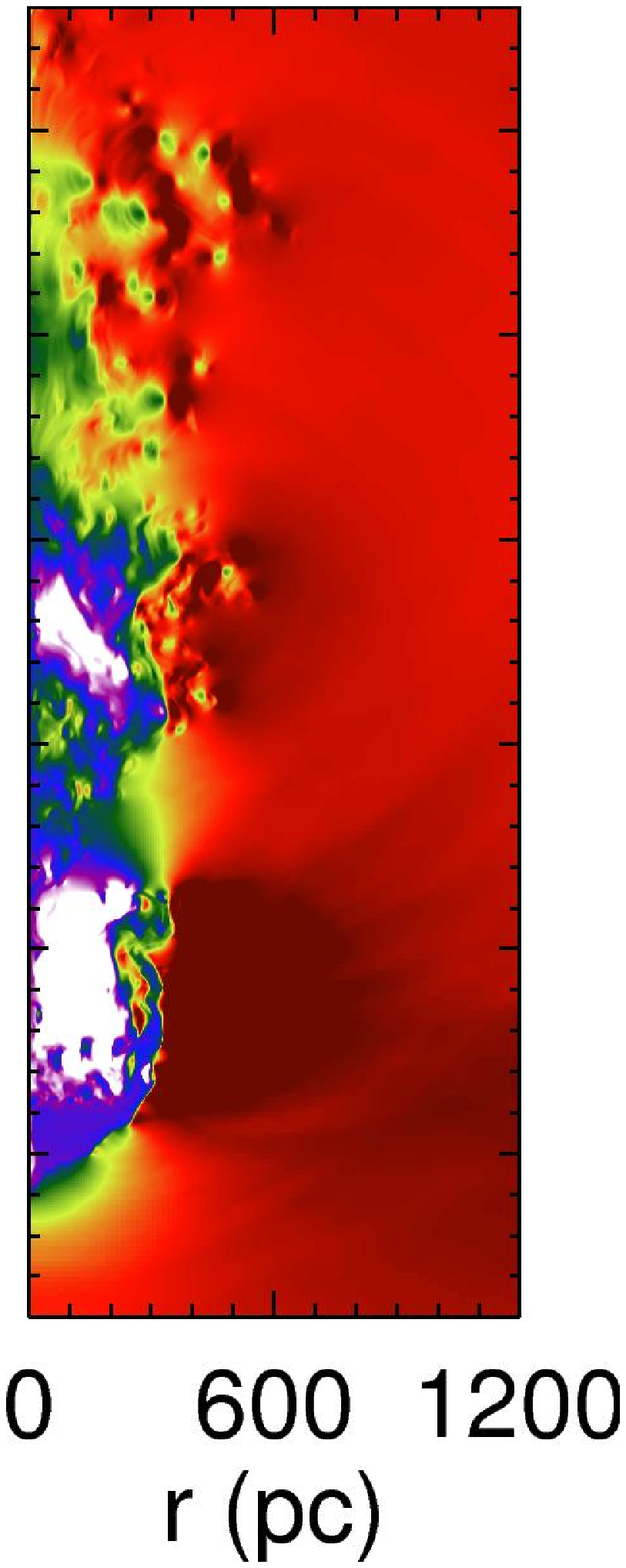}
\includegraphics[scale=0.18]{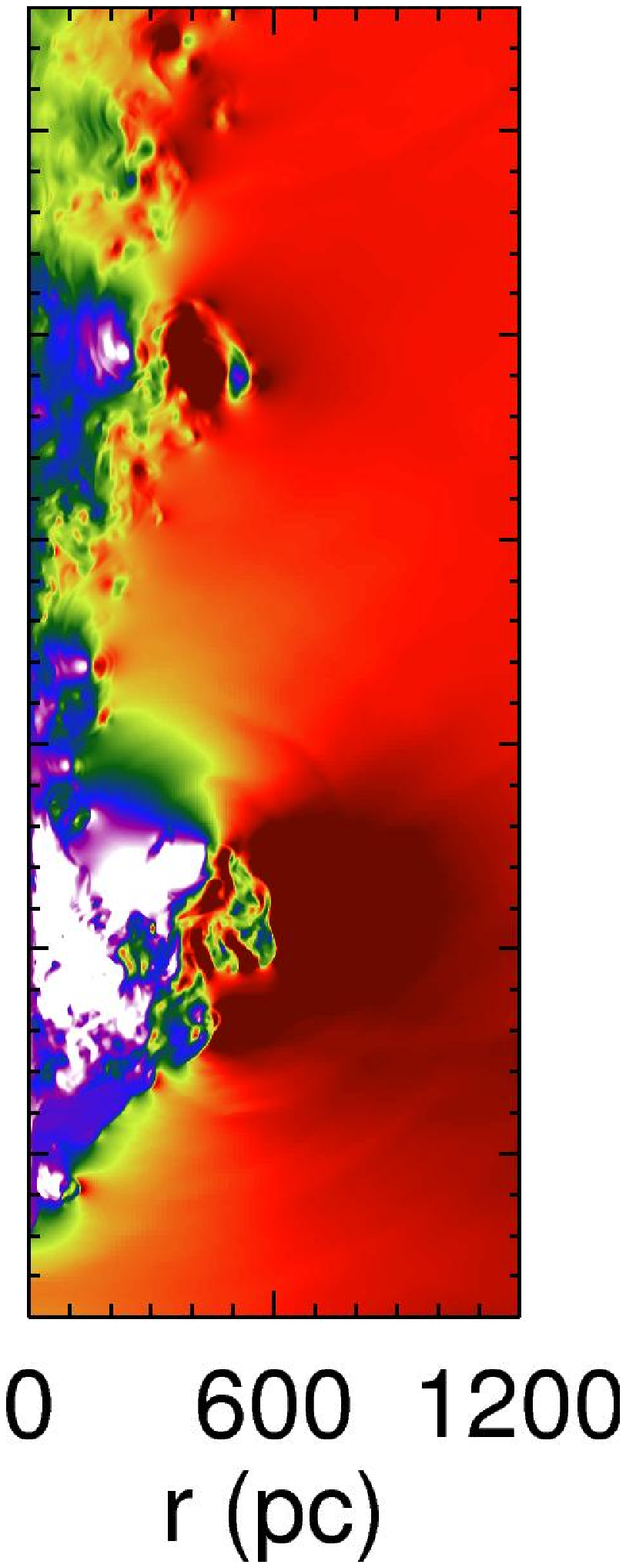}
\includegraphics[scale=0.18]{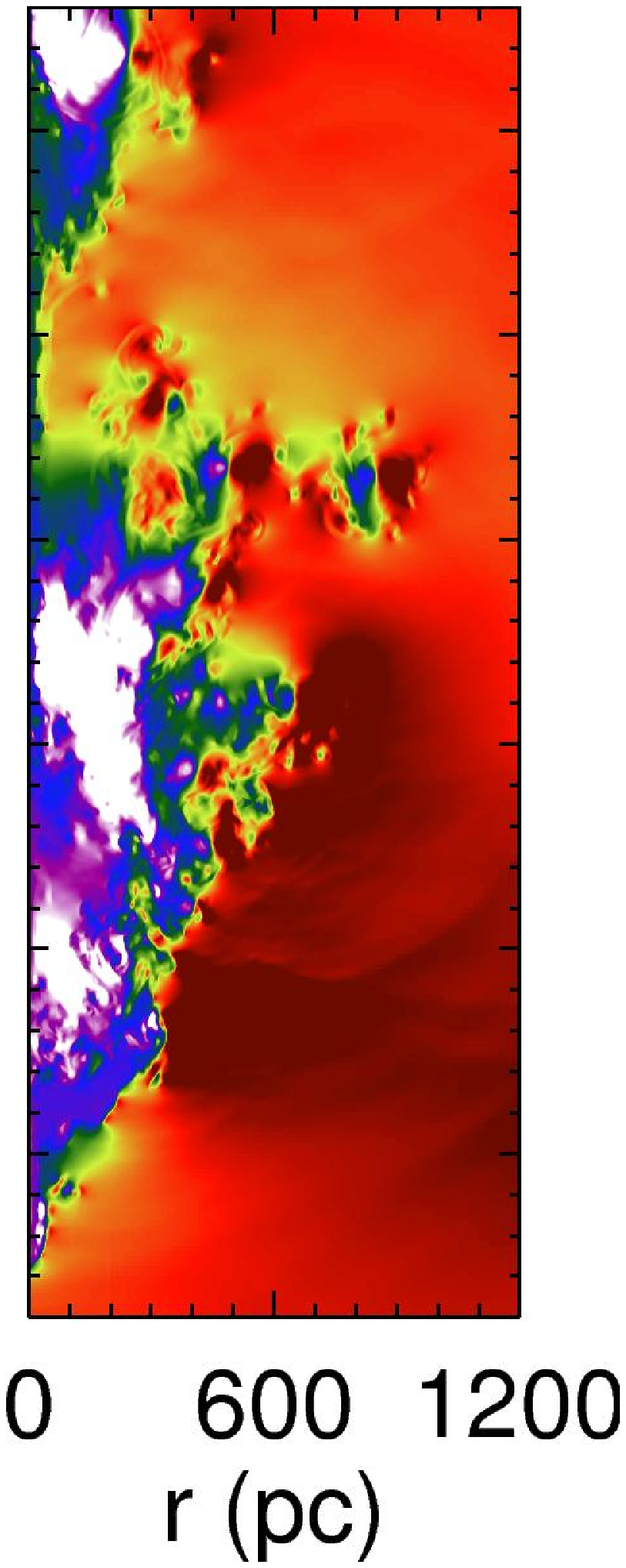}
\includegraphics[scale=0.18]{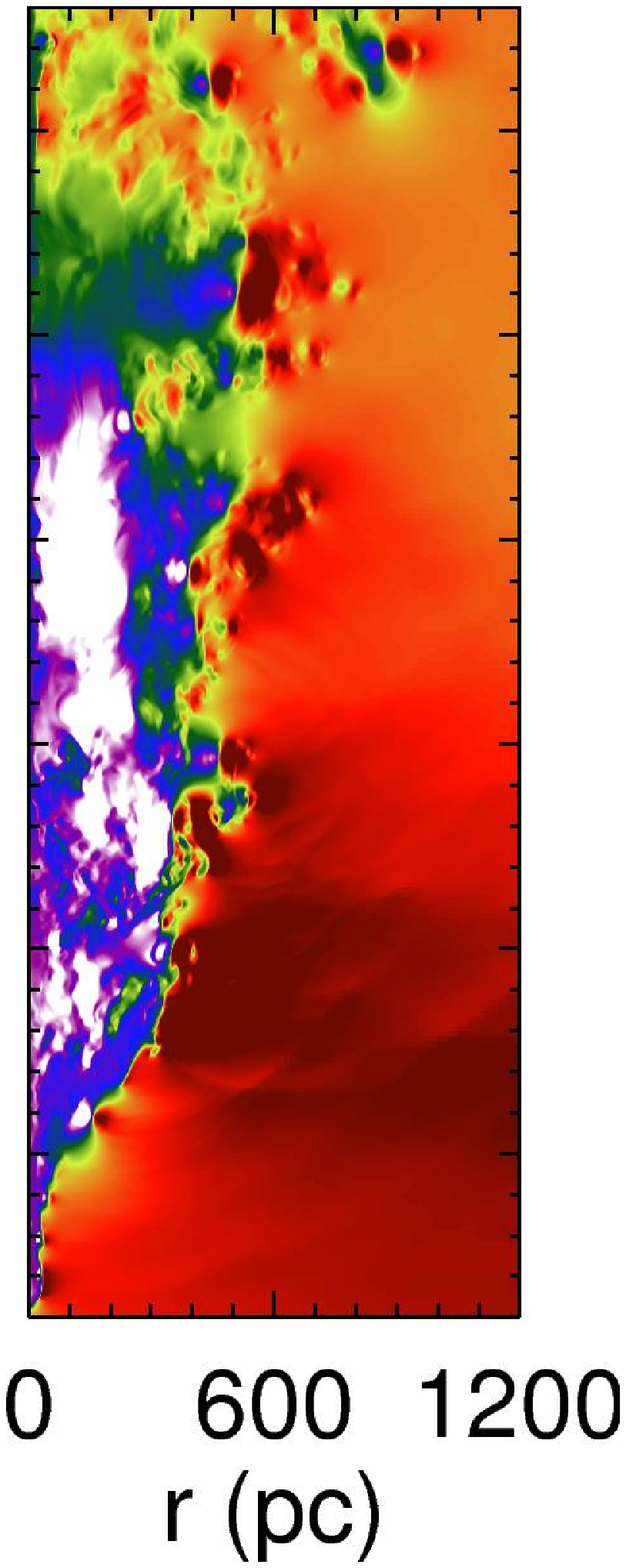}
\includegraphics[scale=0.18]{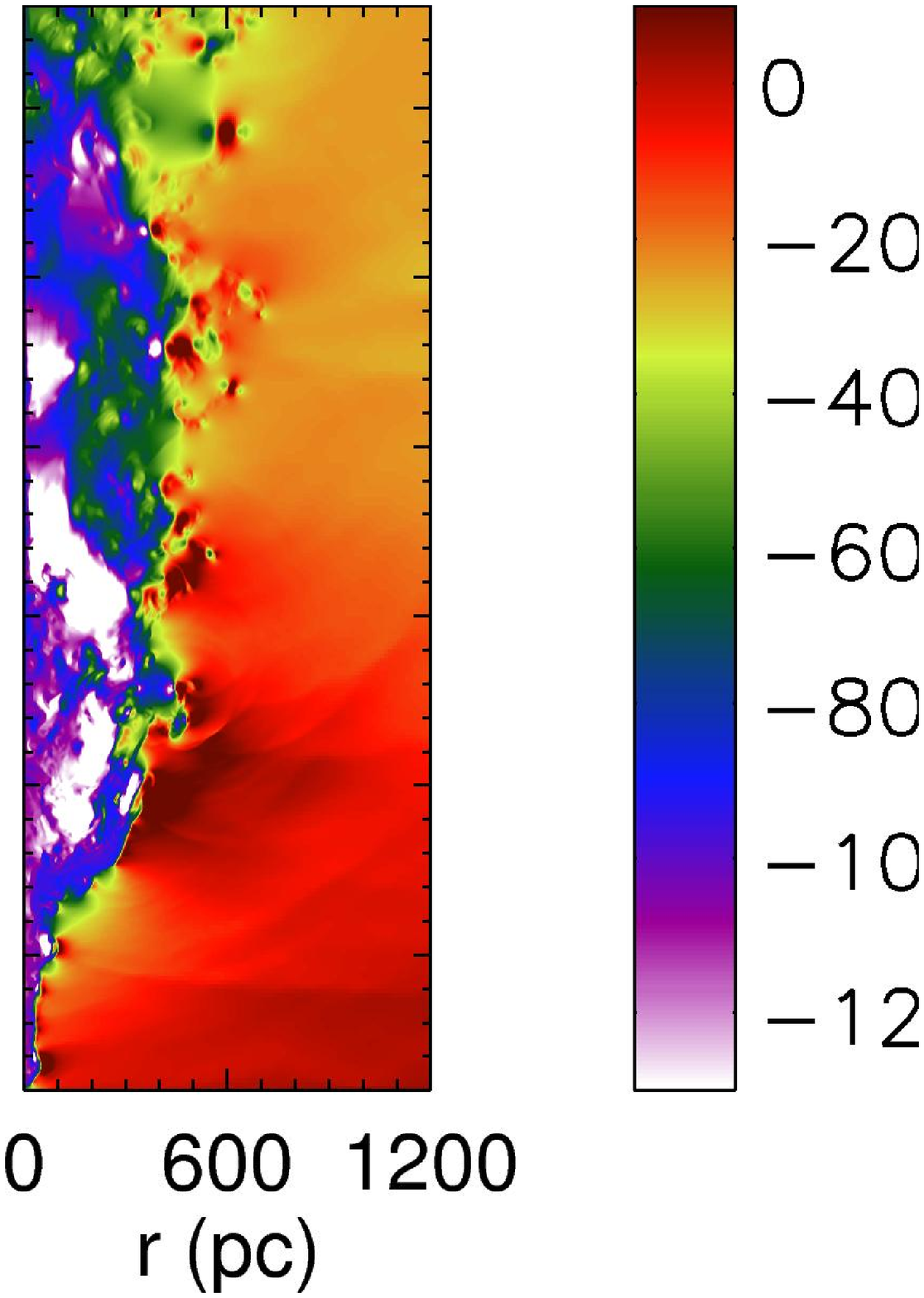} \\
\vspace{0.15in}
\caption{Cross-sections through the Model~B cloud, showing the time evolution of (from top to
  bottom) the density (expressed in terms of the hydrogen number density), temperature, \vz\ (in the
  observer's frame), \CIV\ ion fraction, \NV\ ion fraction, and \OVI\ ion fraction. The hydrogen
  number density includes both neutral and ionized hydrogen. Each variable is plotted at
  15~\Myr\ intervals from $t = 0$ to $t = 120~\Myr$. Note that all variables apart from \vz\ are
  plotted with logarithmic color scales. The arrow in the far right density plot indicates the
  protuberance mentioned in the text.
  \label{fig:CloudEvolution}}
\end{figure*}

\begin{figure*}
\figurenum{\ref{fig:CloudEvolution}}
\centering
\hspace*{0.25in}
\includegraphics[scale=0.25]{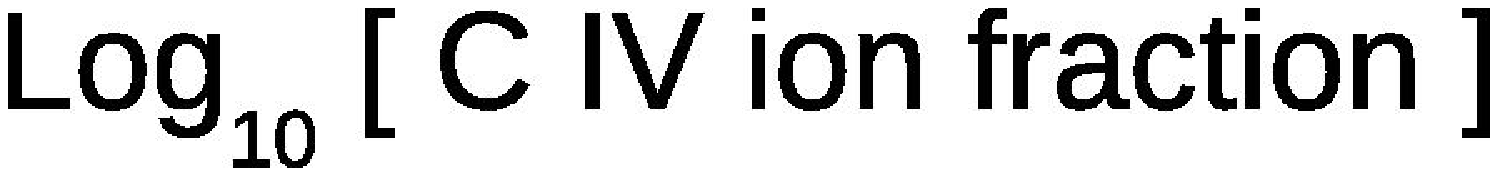} \\
\includegraphics[scale=0.18]{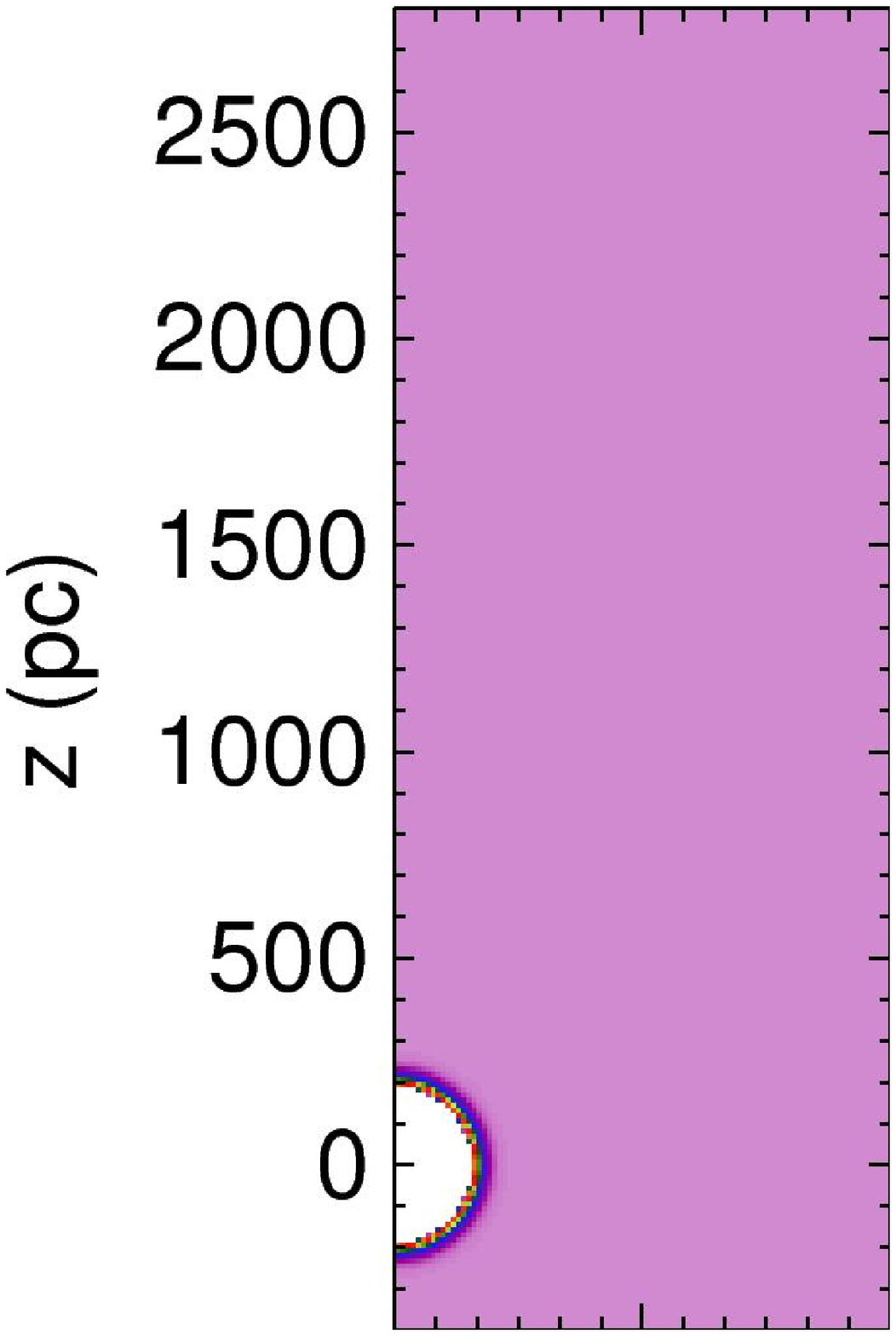}
\hspace{0.070in}
\includegraphics[scale=0.18]{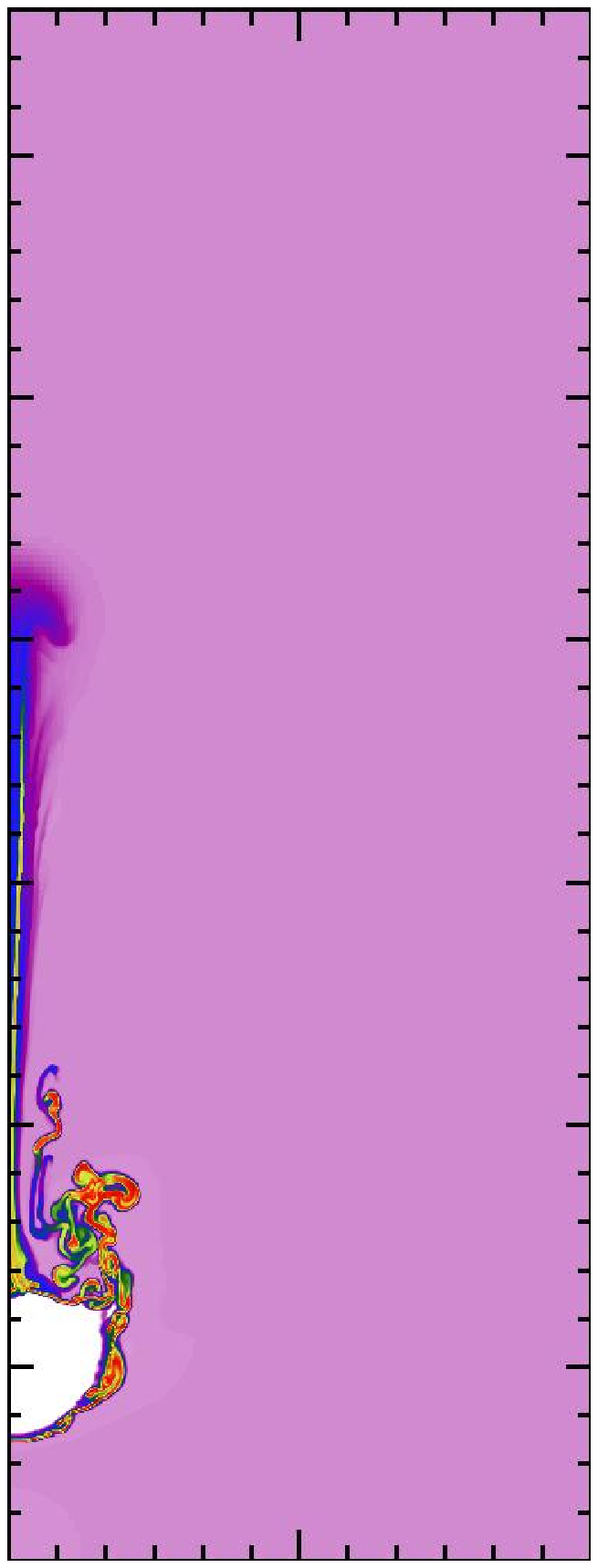}
\hspace{0.070in}
\includegraphics[scale=0.18]{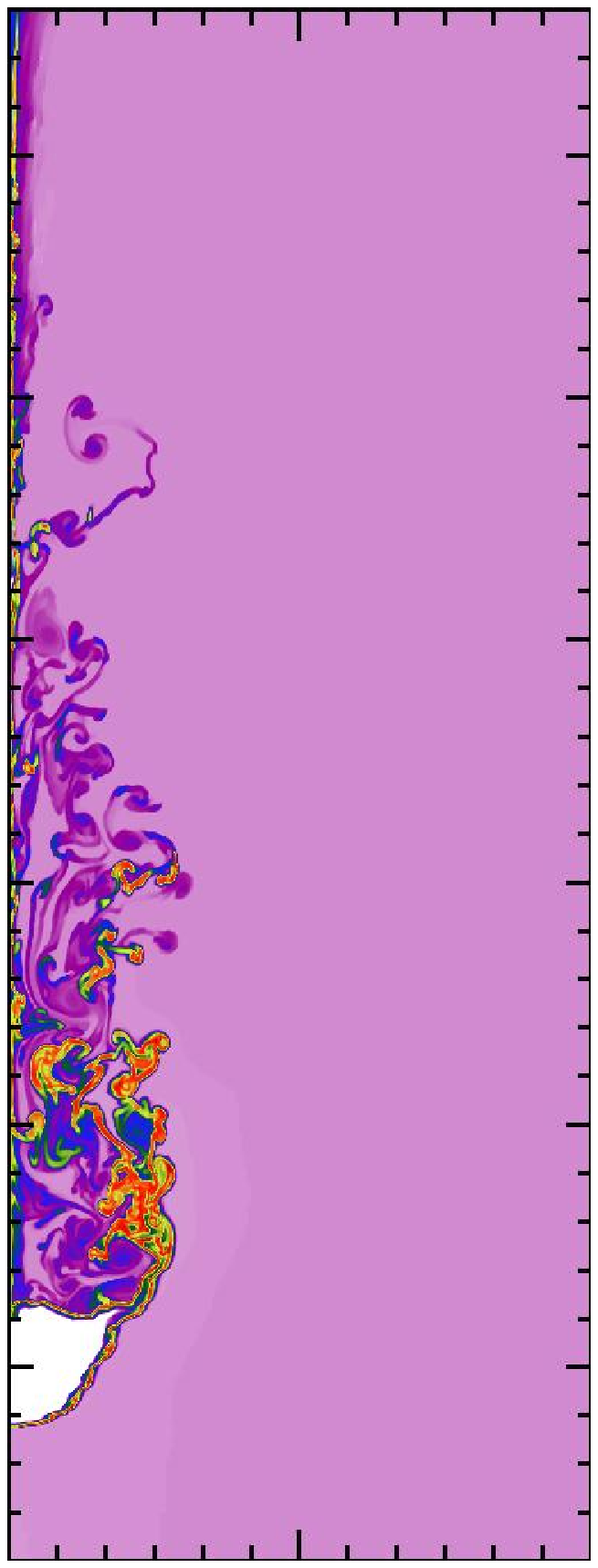}
\hspace{0.070in}
\includegraphics[scale=0.18]{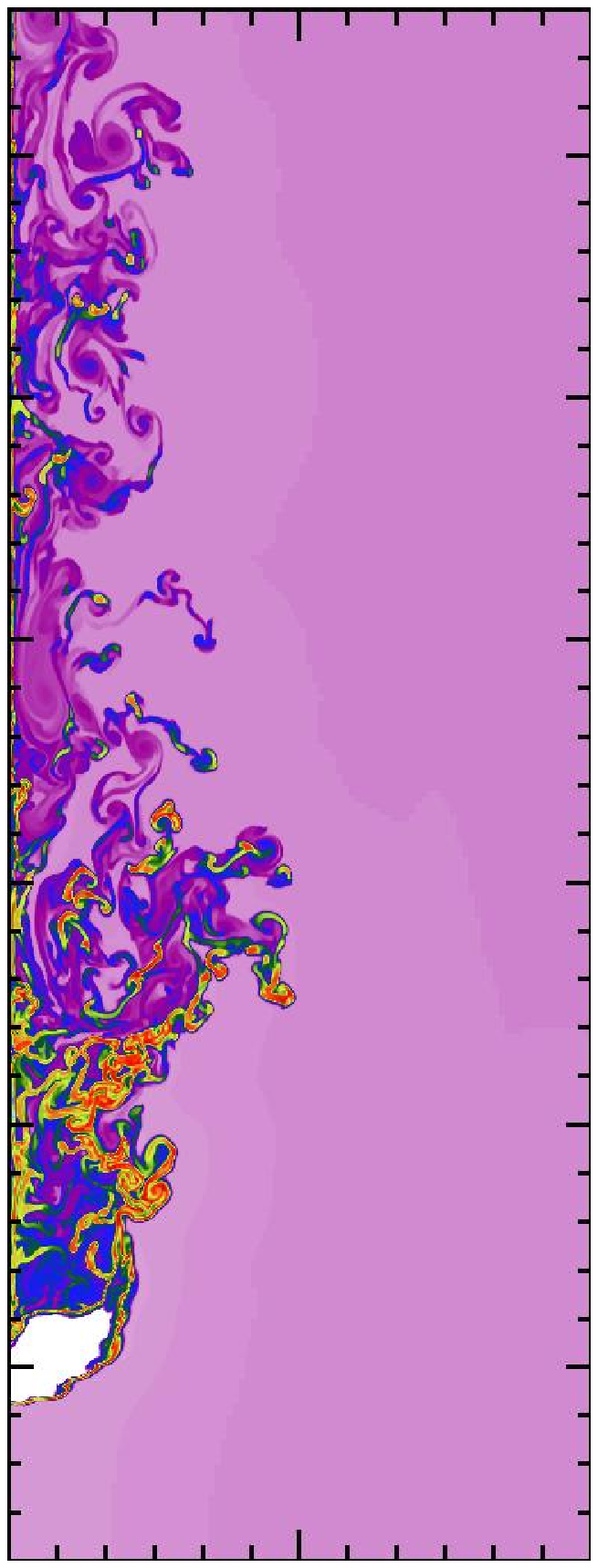}
\hspace{0.070in}
\includegraphics[scale=0.18]{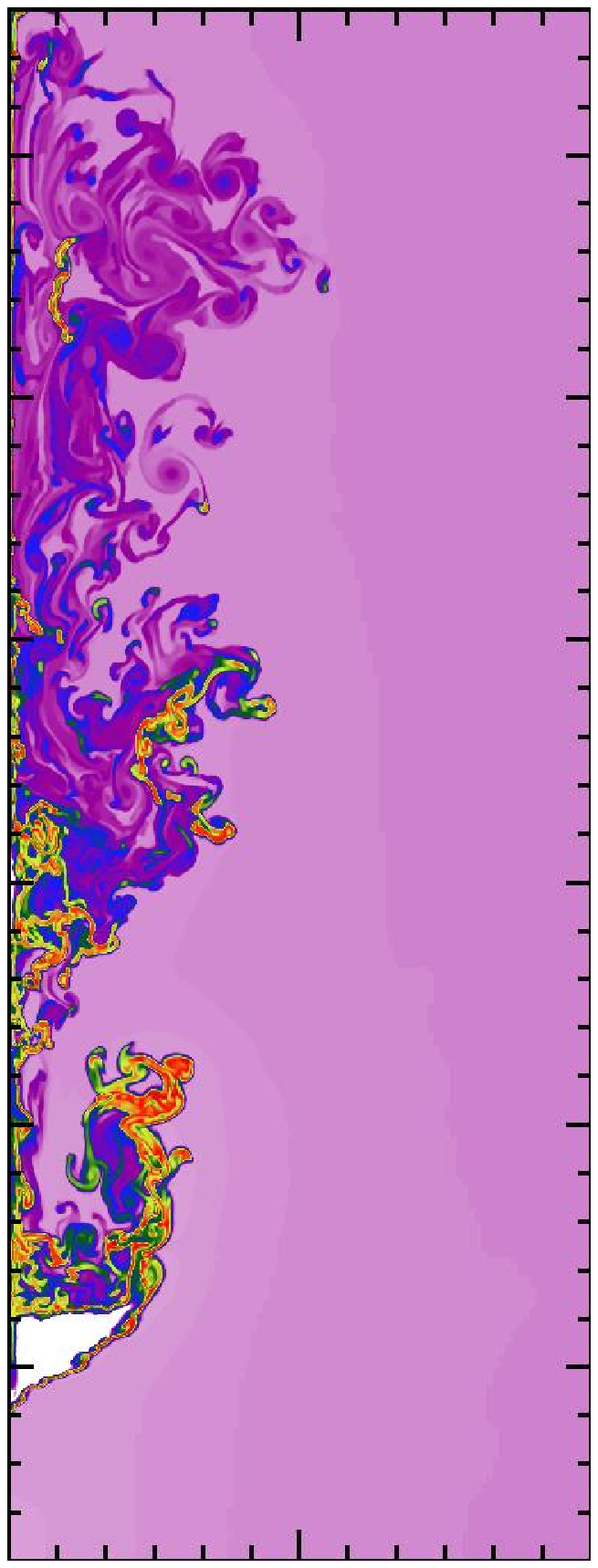}
\hspace{0.070in}
\includegraphics[scale=0.18]{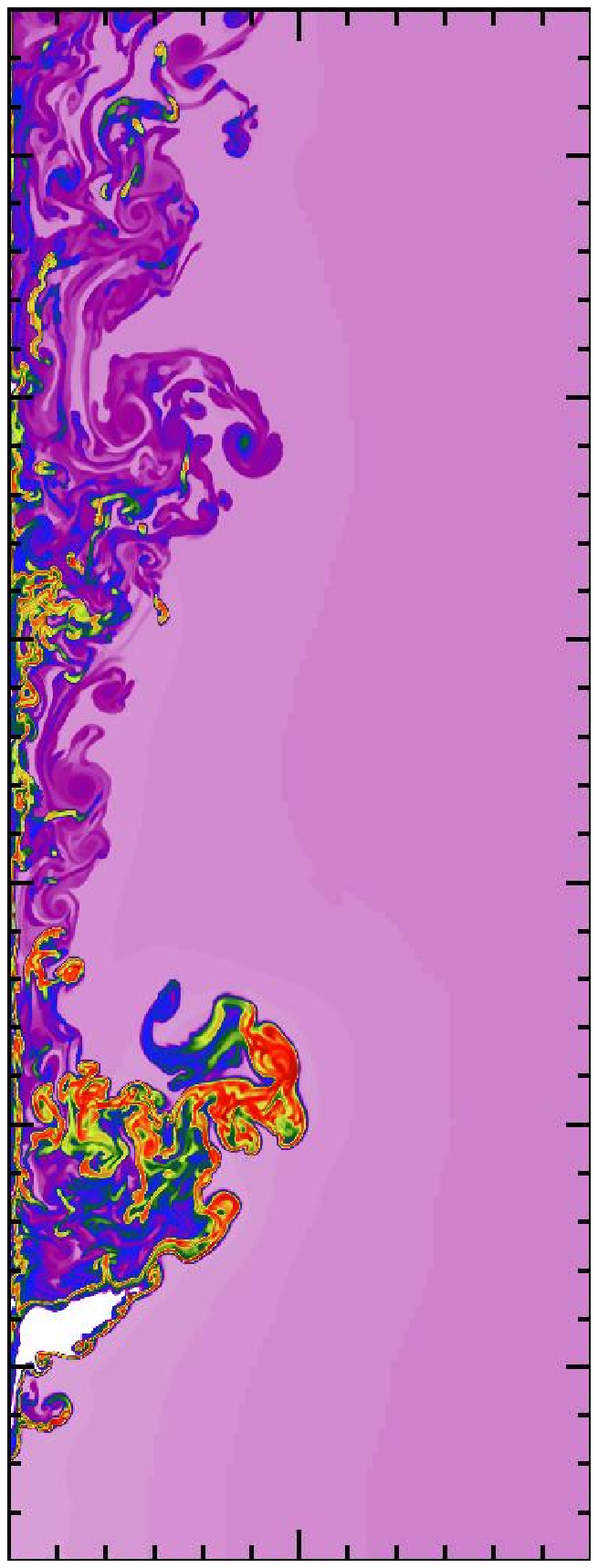}
\hspace{0.070in}
\includegraphics[scale=0.18]{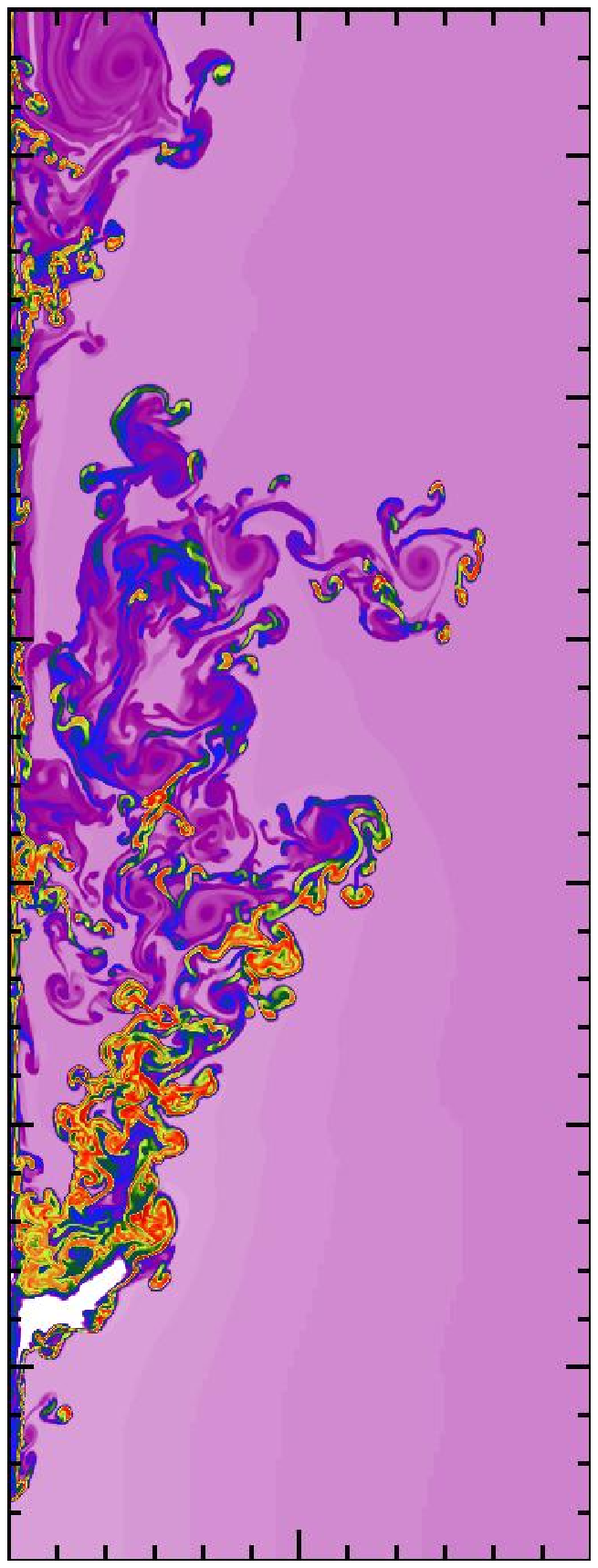}
\hspace{0.070in}
\includegraphics[scale=0.18]{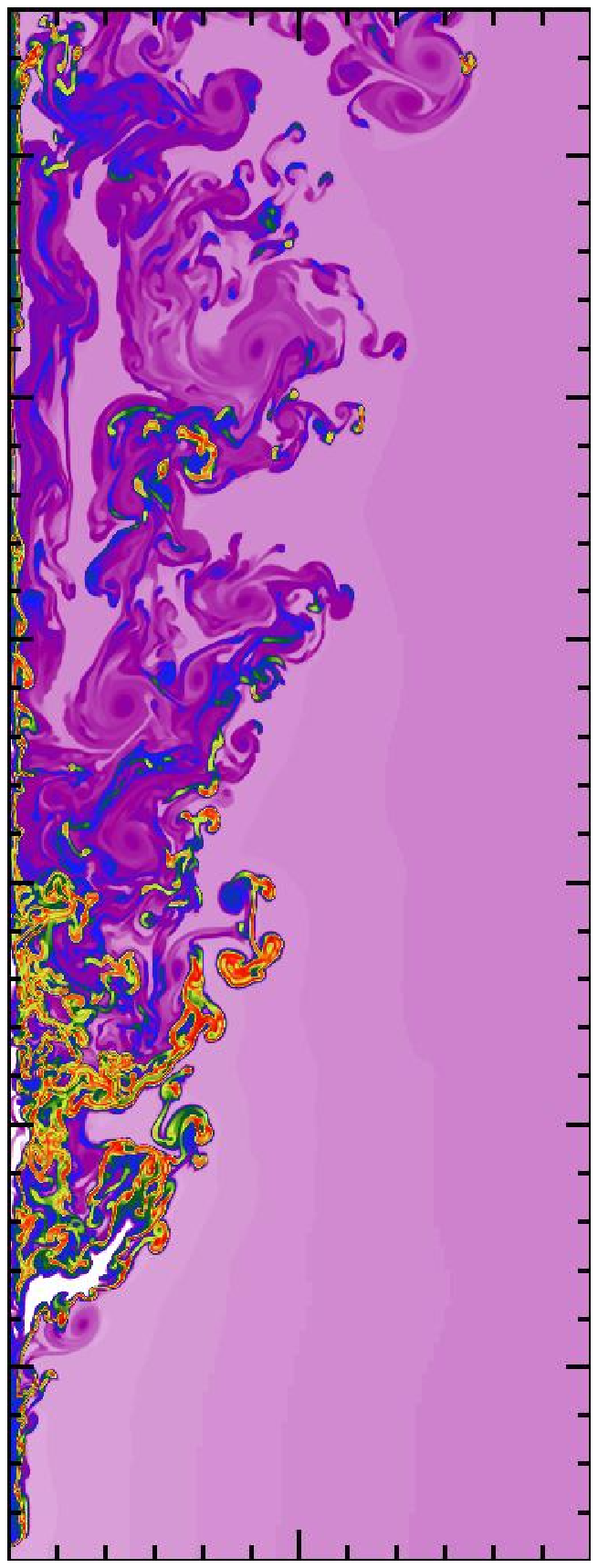}
\hspace{0.070in}
\includegraphics[scale=0.18]{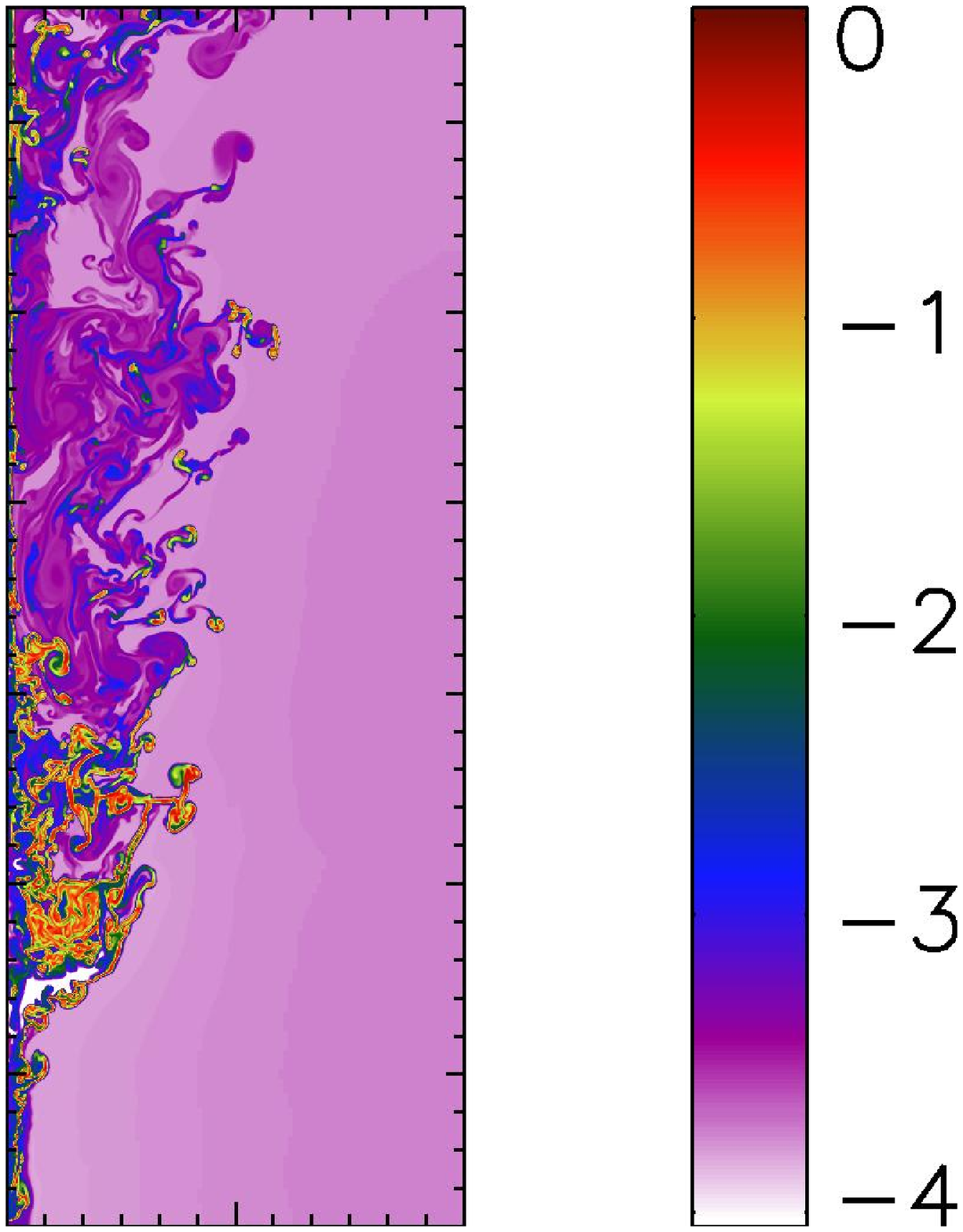} \\
\vspace{0.15in}
\hspace*{0.25in}
\includegraphics[scale=0.25]{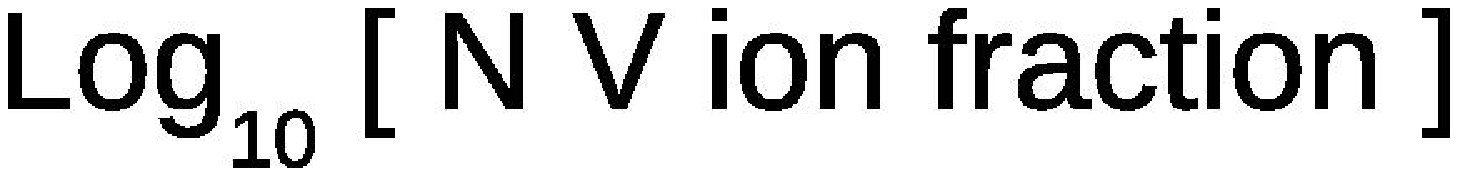} \\
\includegraphics[scale=0.18]{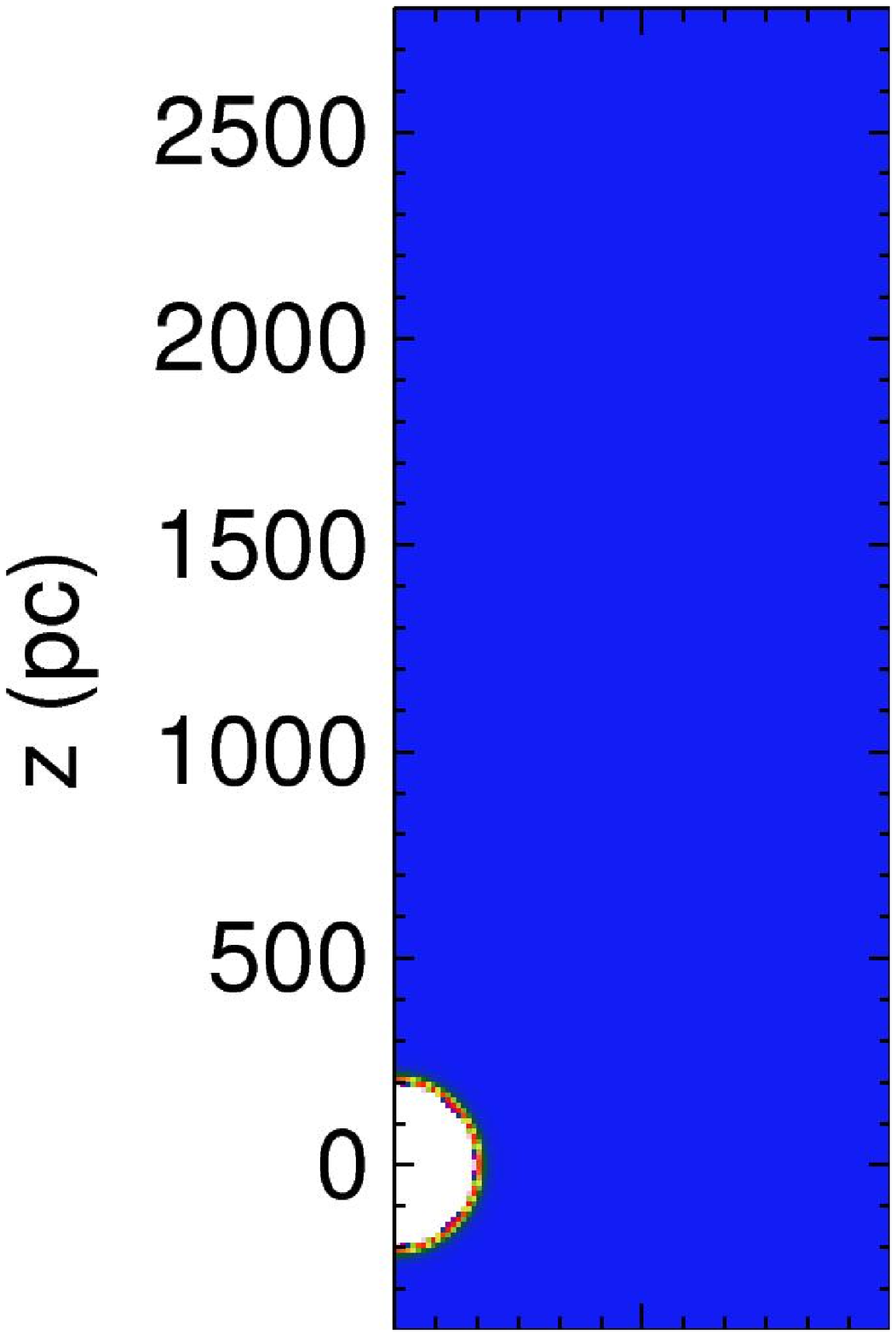}
\hspace{0.070in}
\includegraphics[scale=0.18]{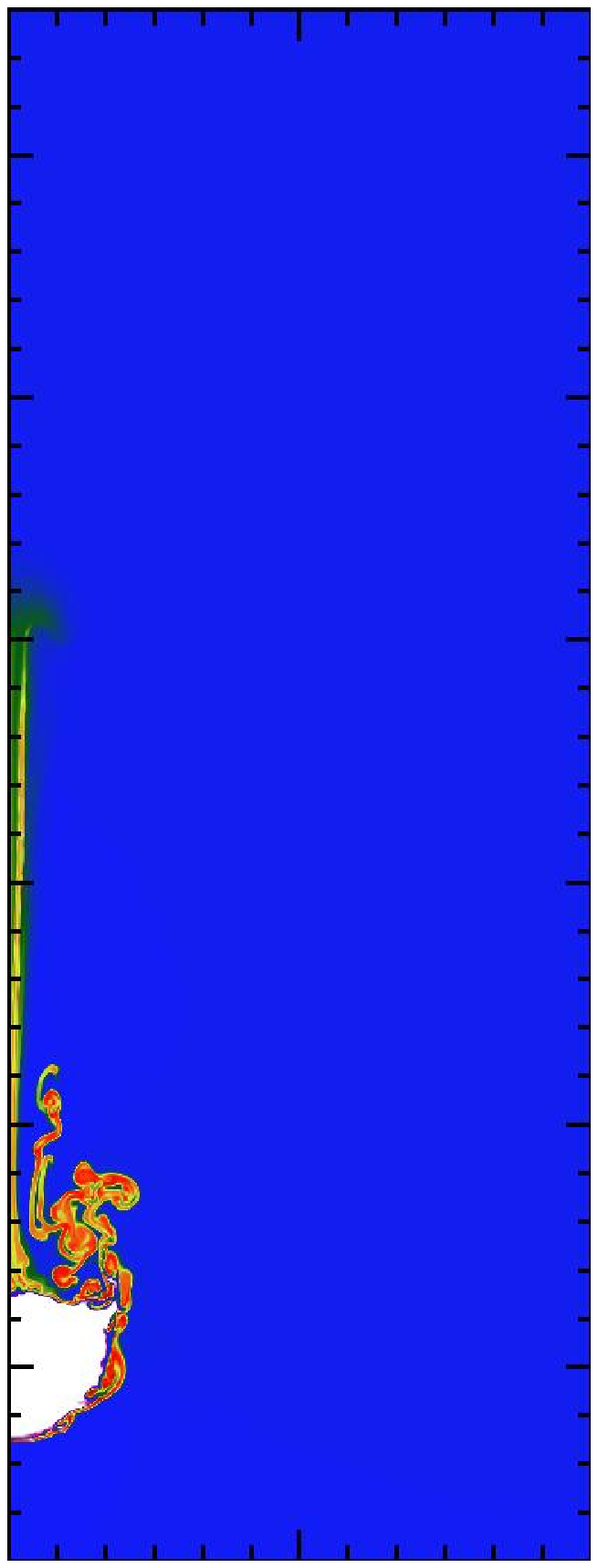}
\hspace{0.070in}
\includegraphics[scale=0.18]{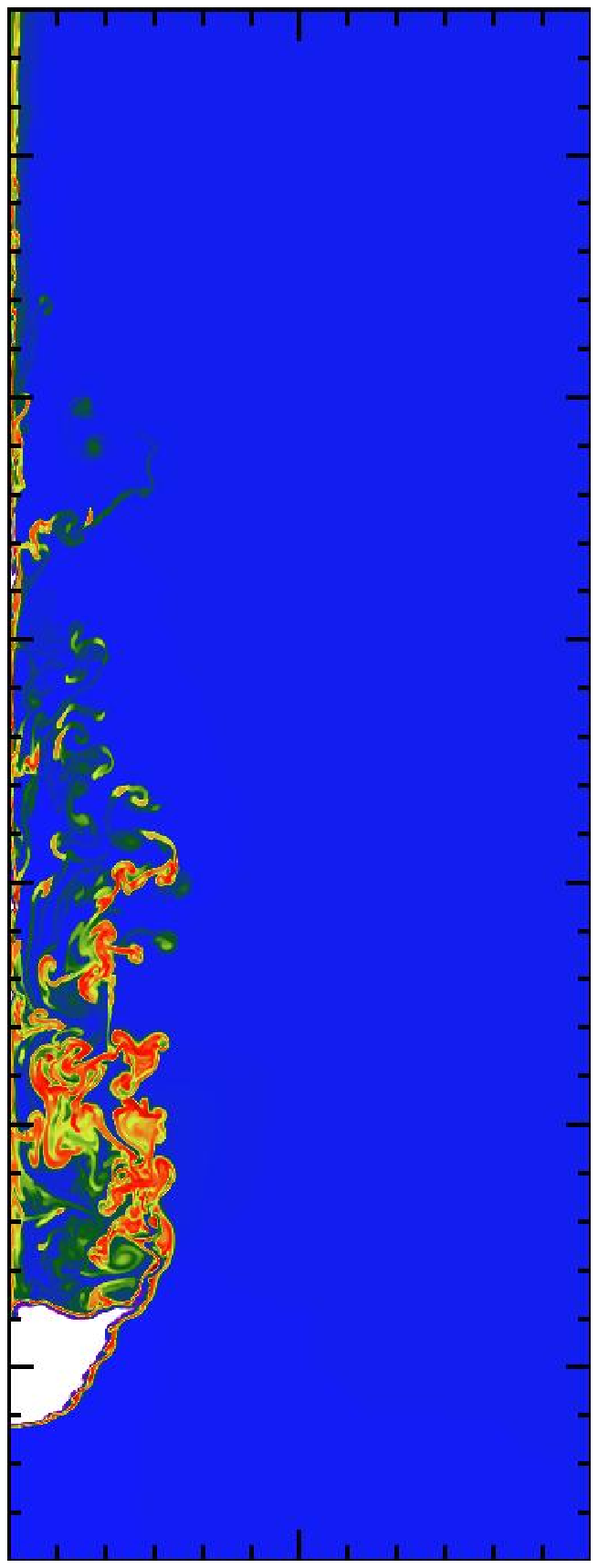}
\hspace{0.070in}
\includegraphics[scale=0.18]{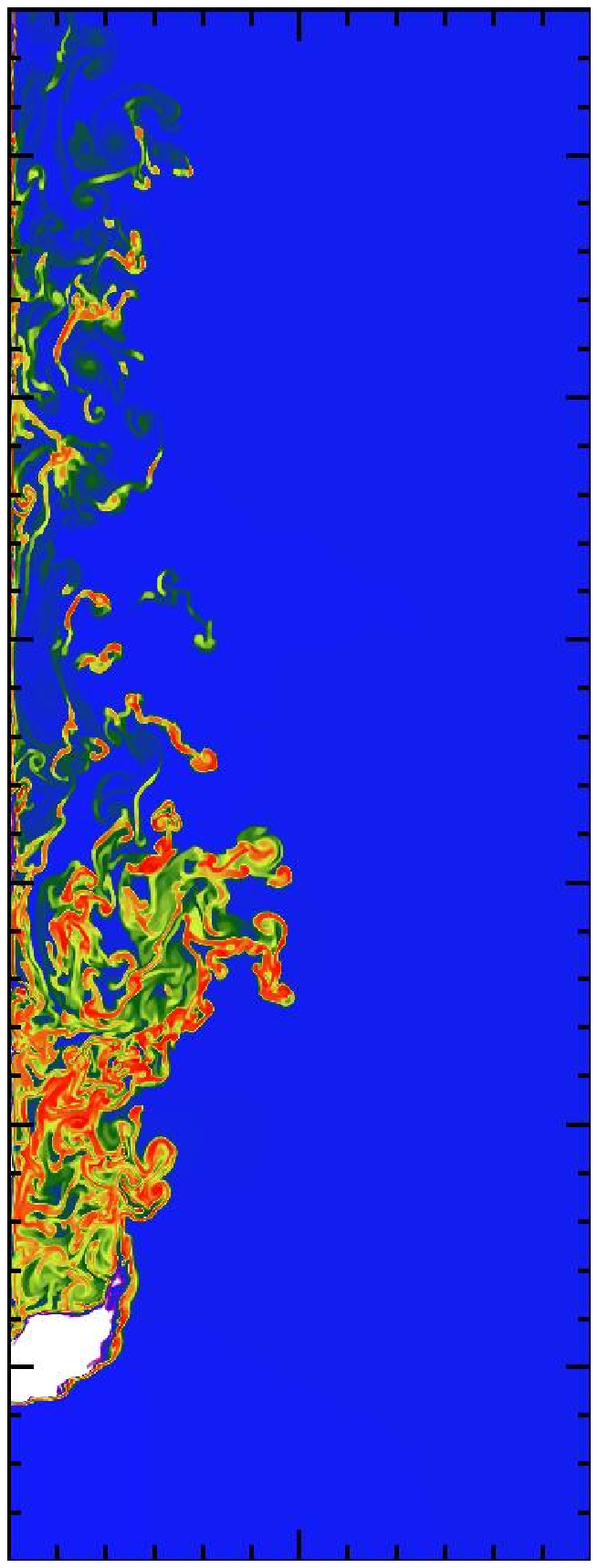}
\hspace{0.070in}
\includegraphics[scale=0.18]{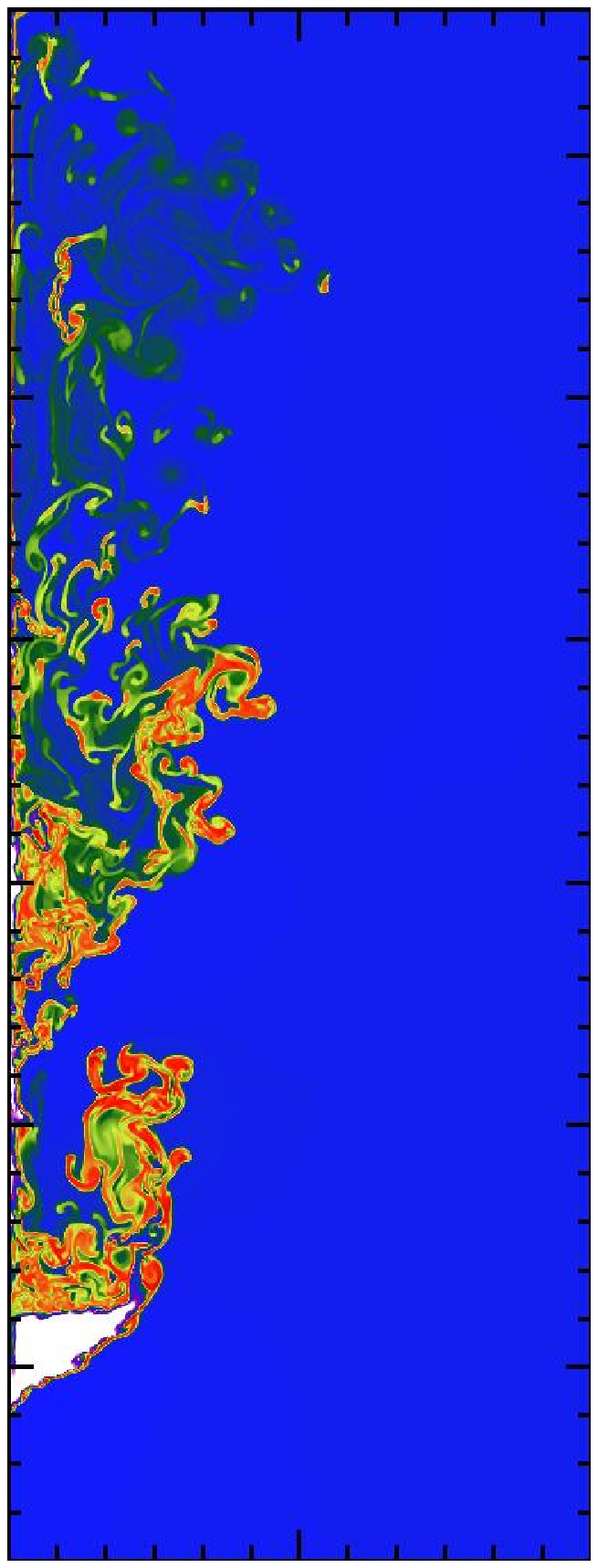}
\hspace{0.070in}
\includegraphics[scale=0.18]{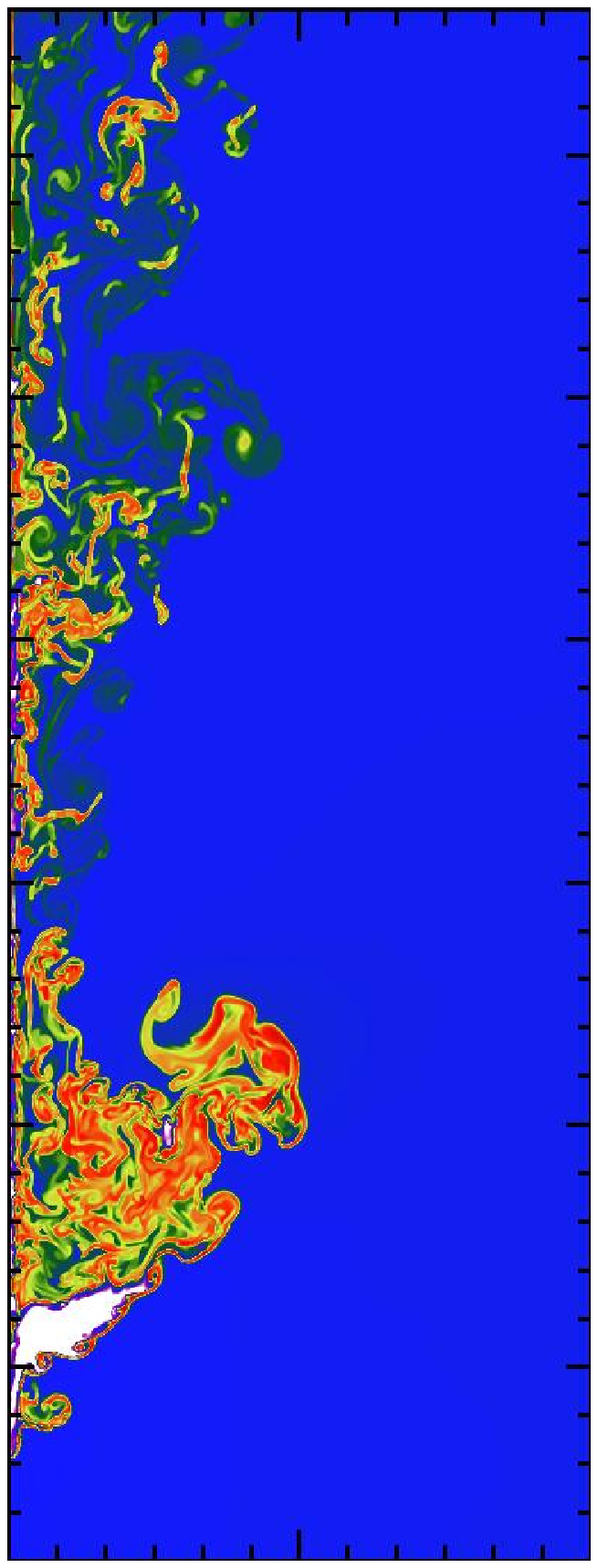}
\hspace{0.070in}
\includegraphics[scale=0.18]{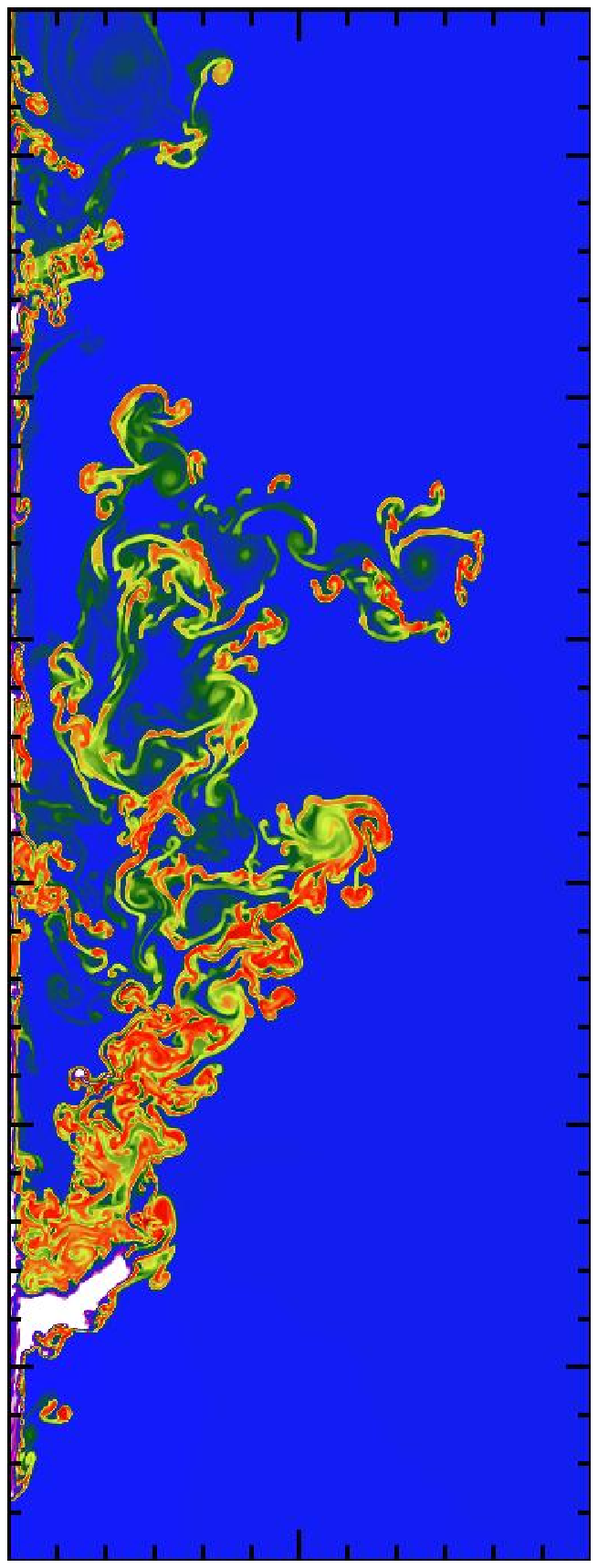}
\hspace{0.070in}
\includegraphics[scale=0.18]{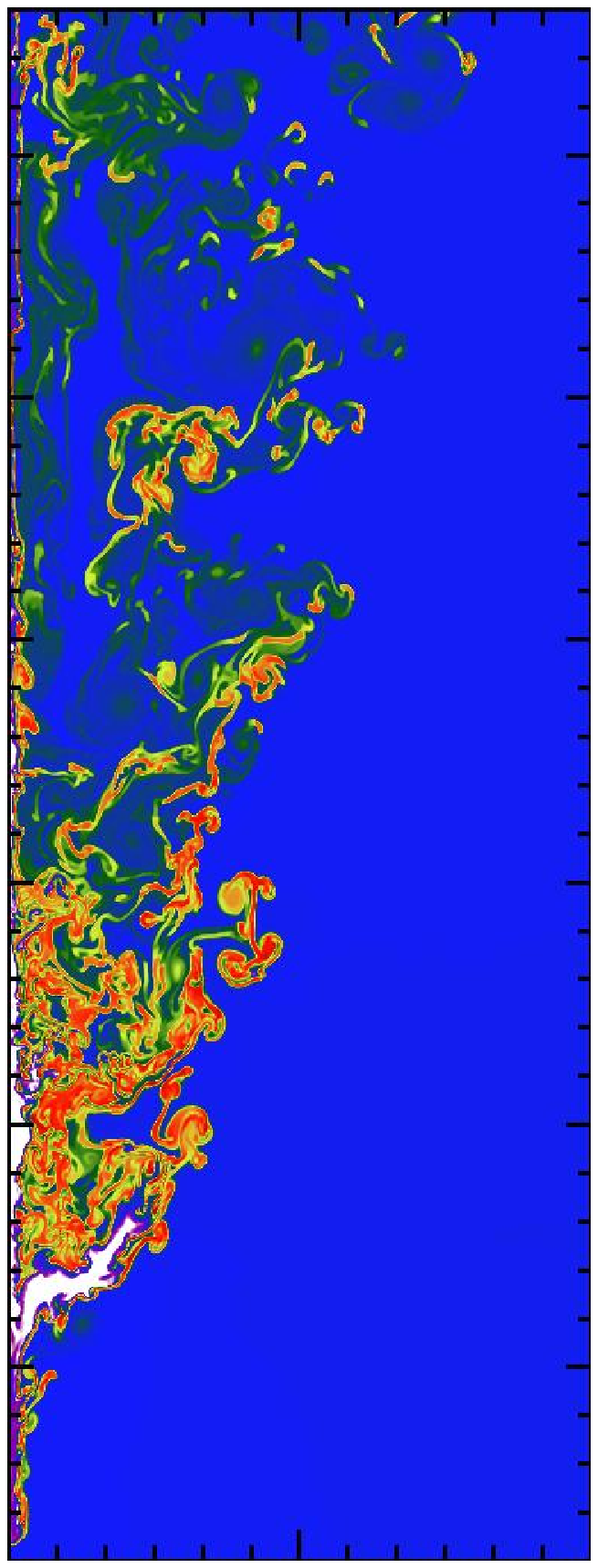}
\hspace{0.070in}
\includegraphics[scale=0.18]{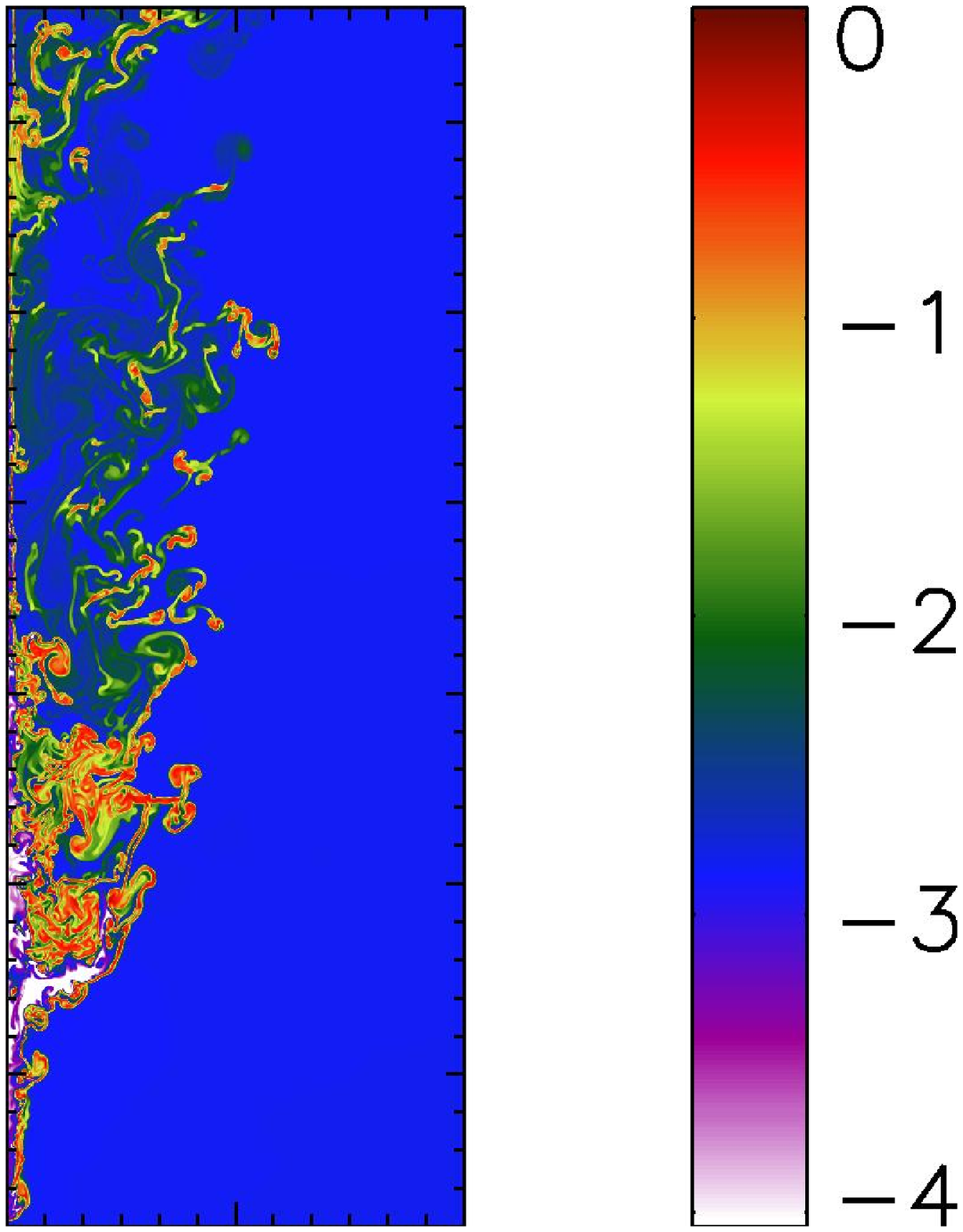} \\
\vspace{0.15in}
\hspace*{0.25in}
\includegraphics[scale=0.25]{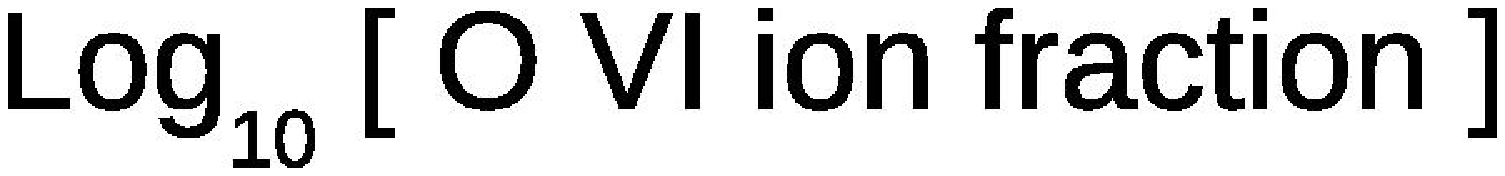} \\
\includegraphics[scale=0.18]{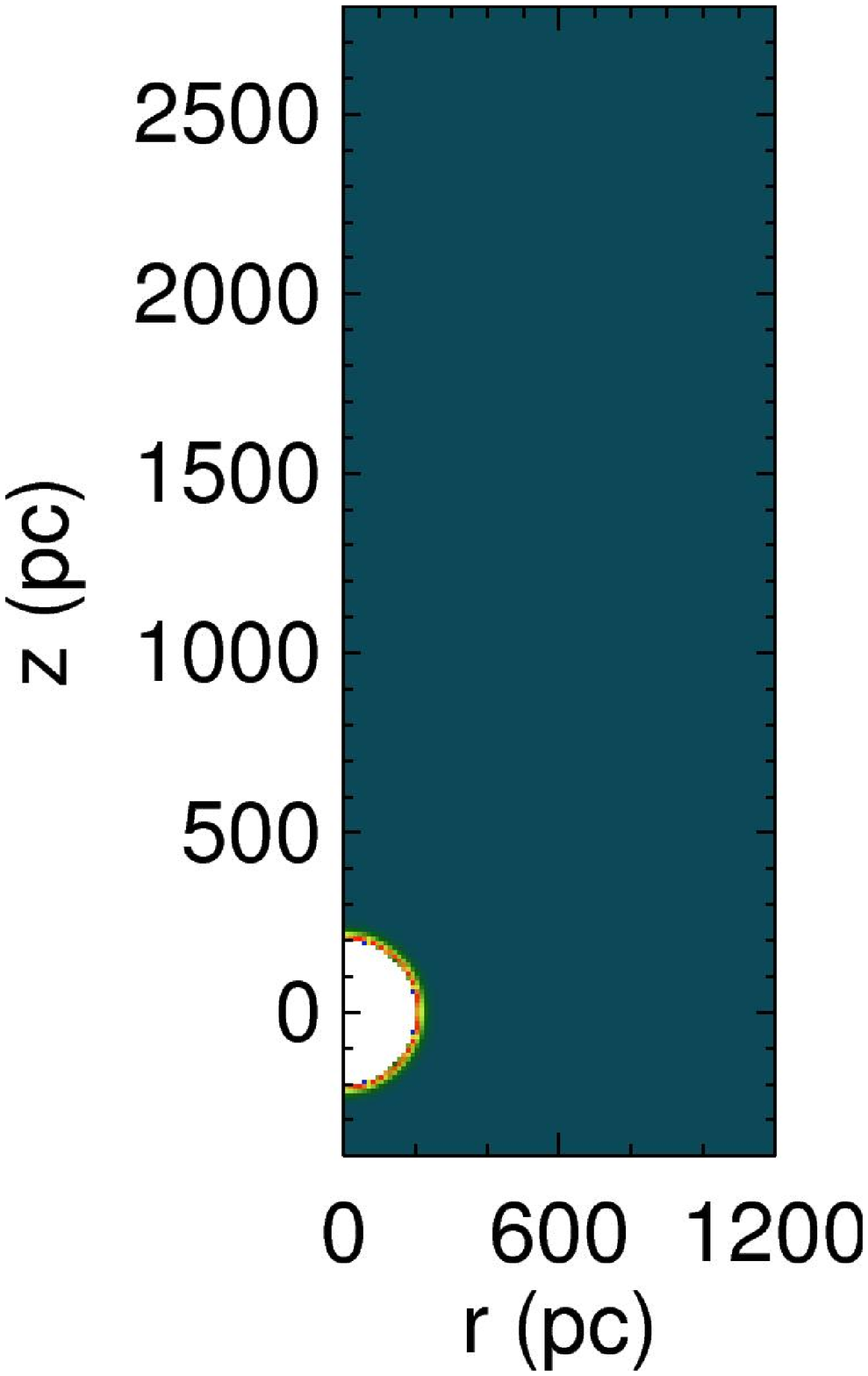}
\includegraphics[scale=0.18]{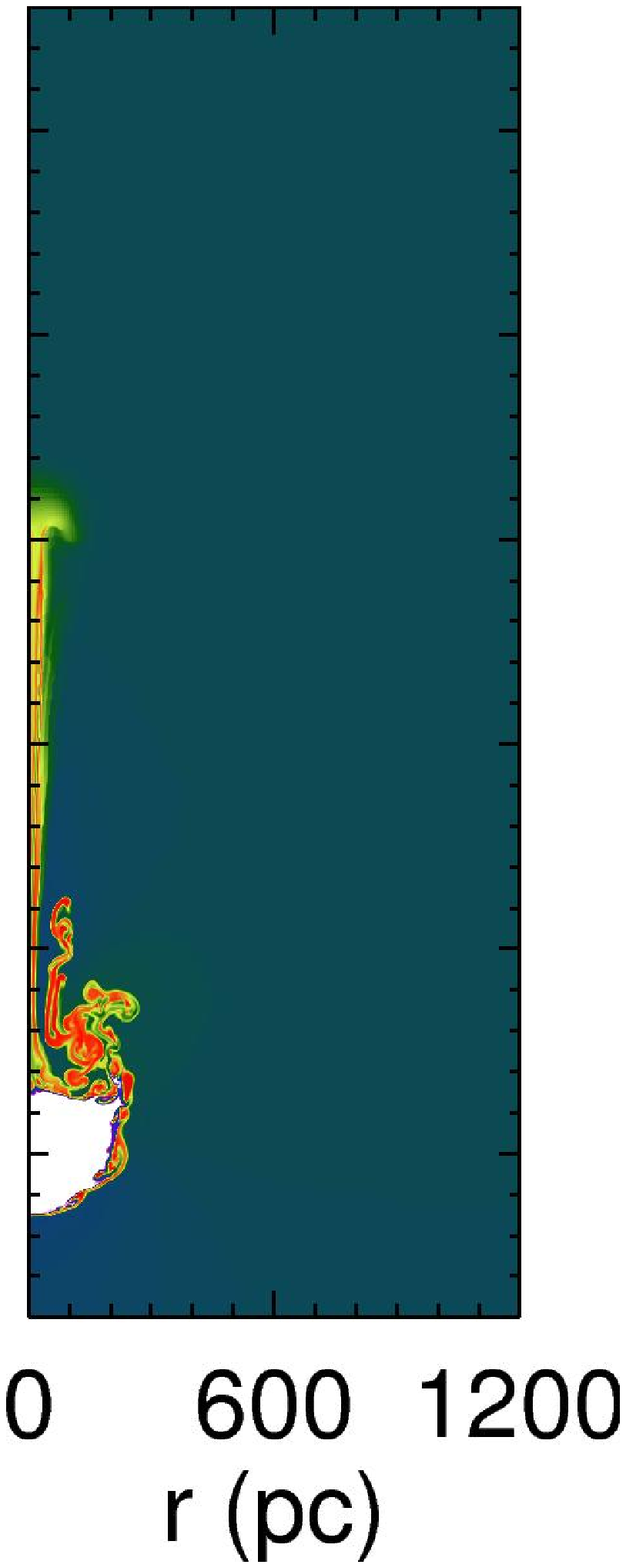}
\includegraphics[scale=0.18]{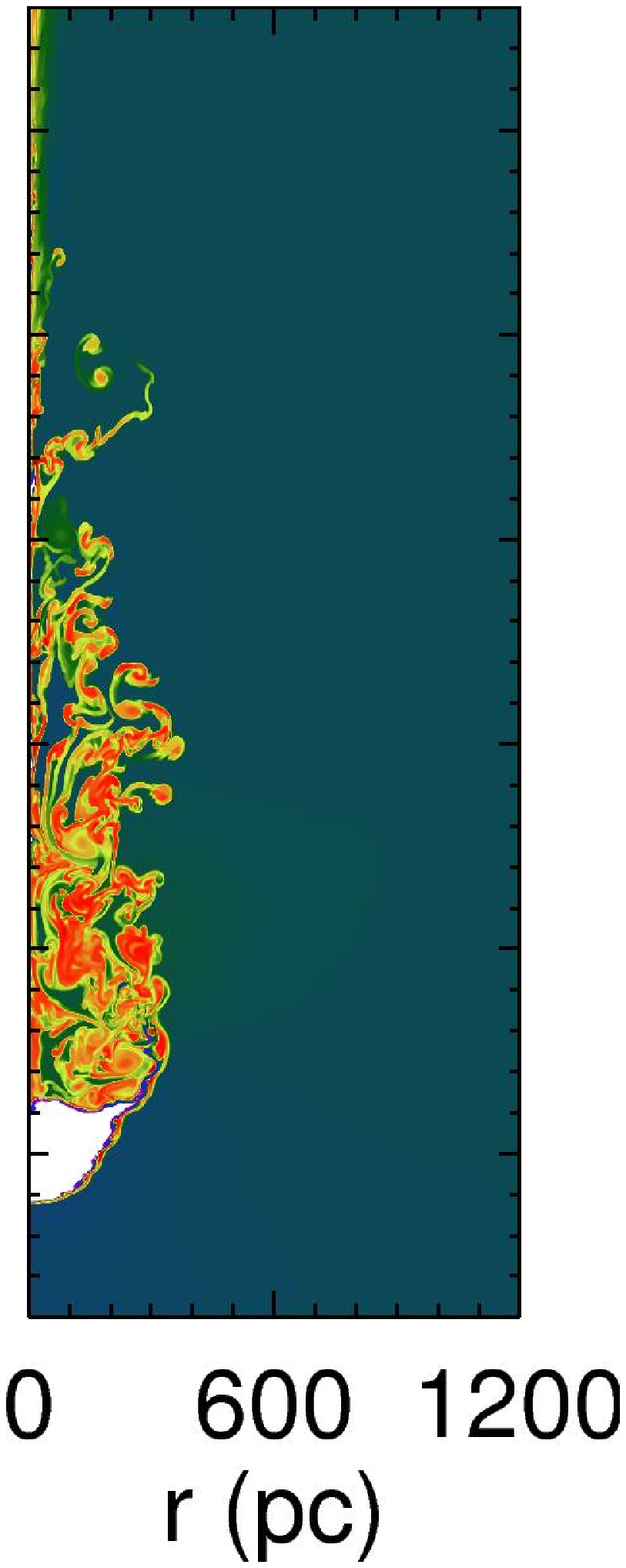}
\includegraphics[scale=0.18]{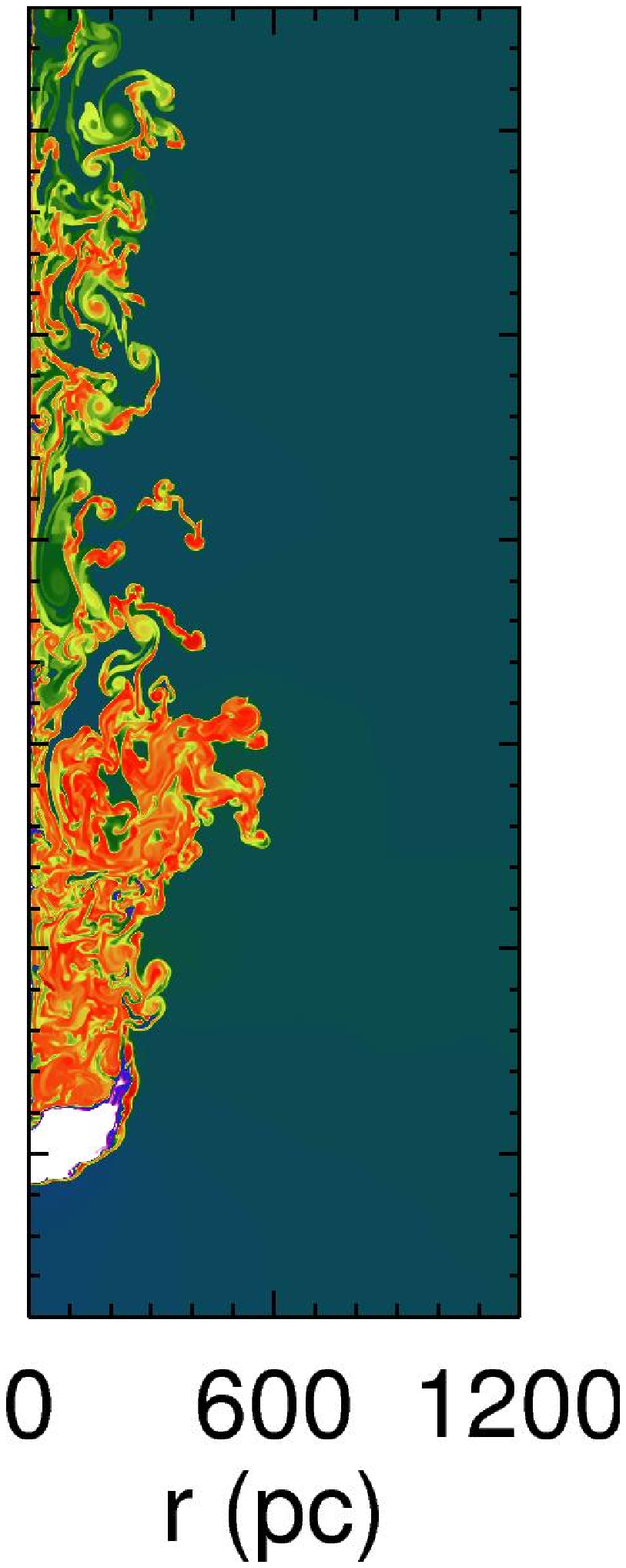}
\includegraphics[scale=0.18]{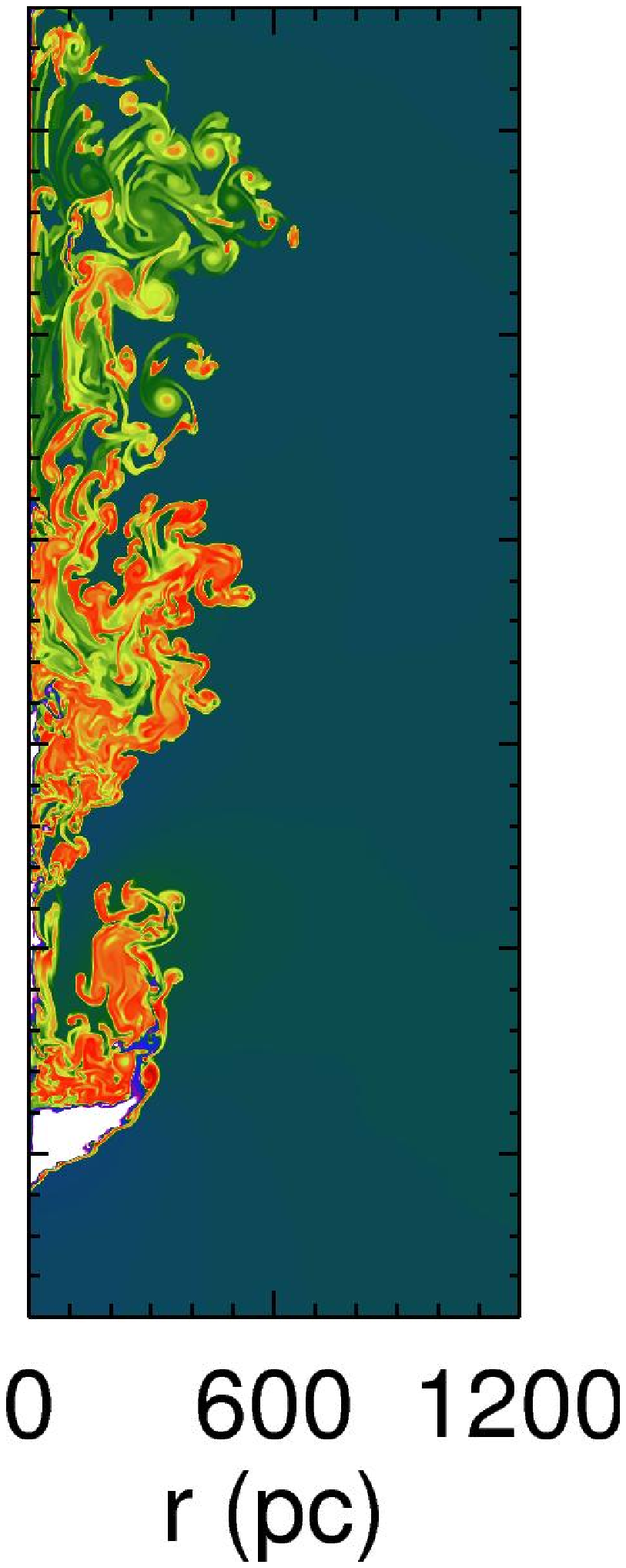}
\includegraphics[scale=0.18]{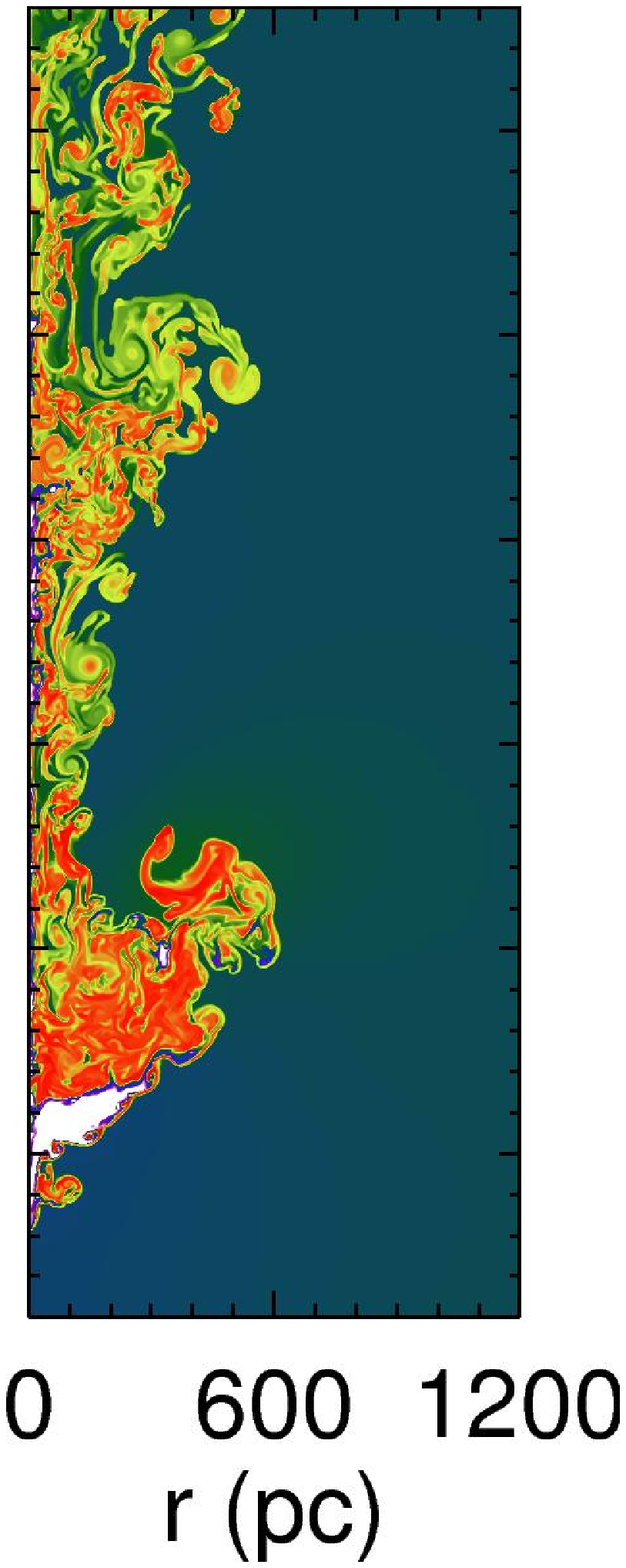}
\includegraphics[scale=0.18]{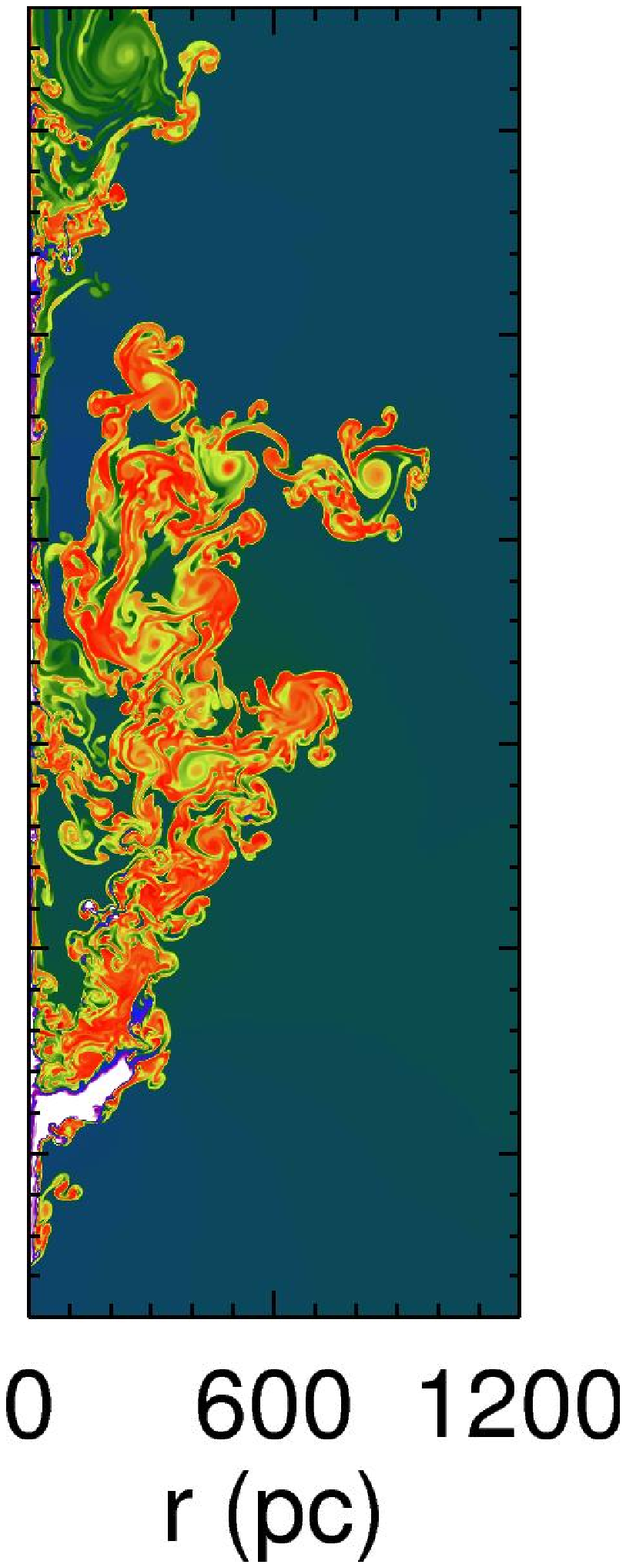}
\includegraphics[scale=0.18]{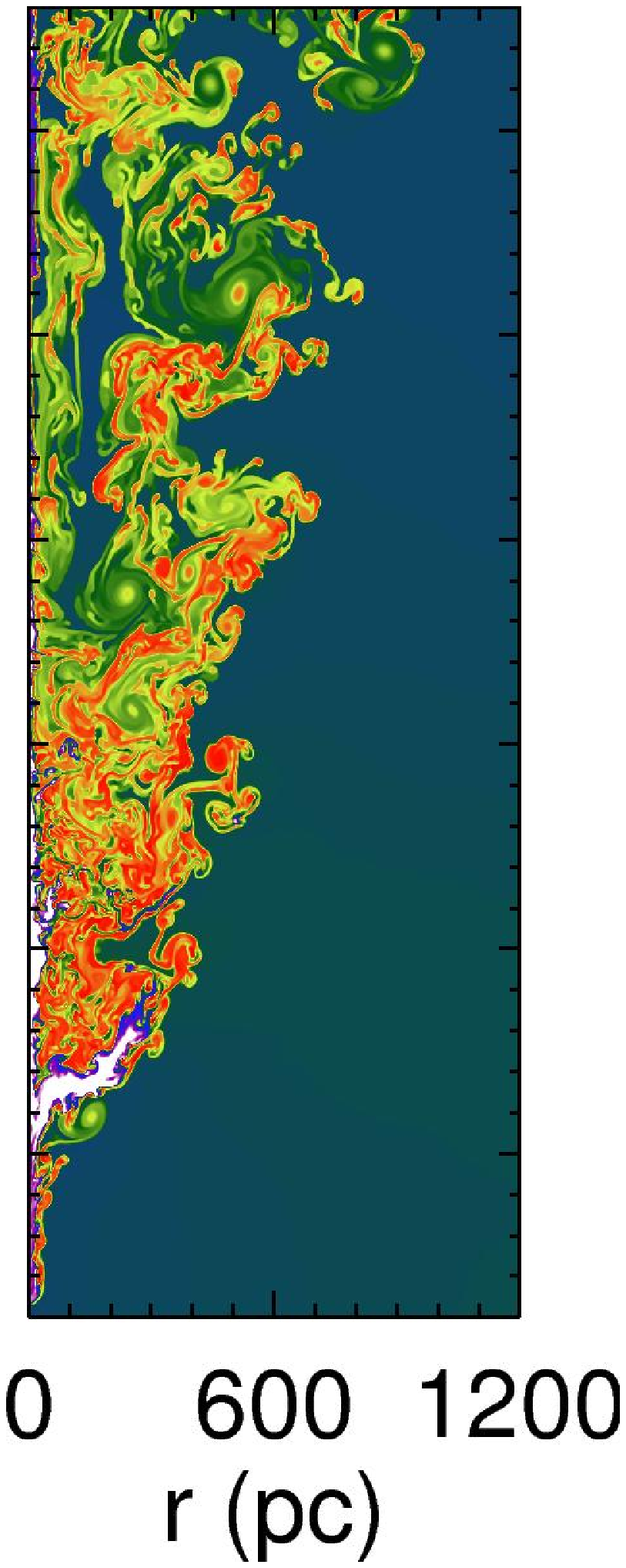}
\includegraphics[scale=0.18]{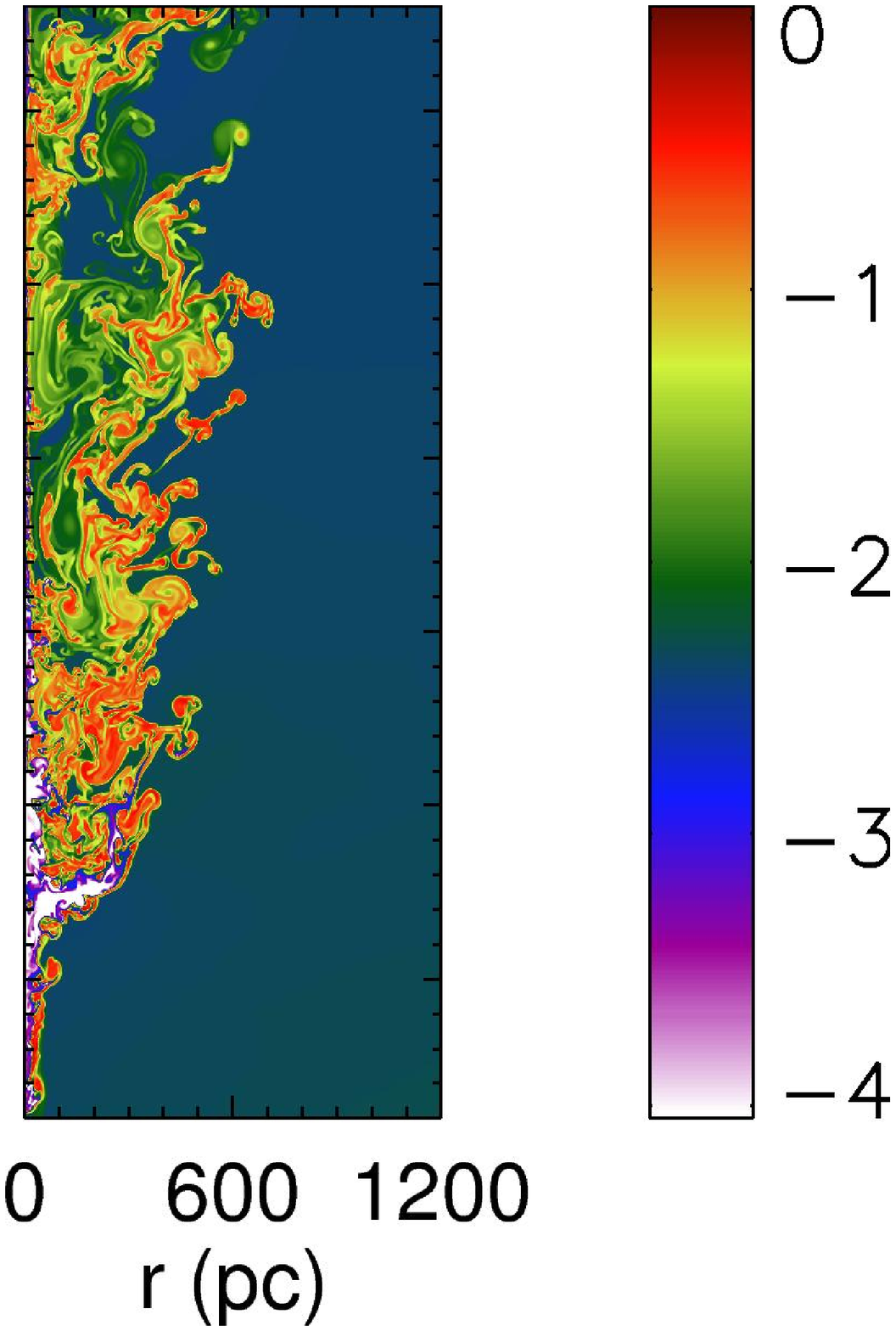} \\
\vspace{0.15in}
\caption{\textit{Continued}}
\end{figure*}

We chose Model~B as our representative model as it exhibits many of the physical processes also seen
in the other models.  In Section~\ref{subsec:CloudEvolutionDifferences}, below, we will discuss the
differences and similarities between the various models. Here, we discuss Model~B, whose evolution
is shown in Figure~\ref{fig:CloudEvolution}. Figure~\ref{fig:CloudEvolution} plots, from top to
bottom, the hydrogen number density, the temperature, \vz\ (in the observer's frame), and the ion
fractions\footnote{For example, the \CIV\ ion fraction is the fraction of all carbon atoms that are
  in the C$^{+3}$ ionization state.} of \CIV, \NV, and \OVI. Note that at $t = 0~\Myr$ the ion
fractions for all three high ions are low almost everywhere on the grid. This is because the
cloud is too cold and the ambient medium too hot for these ions.

Figure~\ref{fig:CloudEvolution} shows that the initially spherical cloud deforms during the
simulation. This is due to the Bernoulli effect. As the ISM flows around the cloud, the speed of the
ISM relative to the cloud is greater along the sides of the cloud ($r \sim 150$, $z \sim 0~\pc$)
than near the top or the bottom of the cloud ($r \sim 0$, $z \sim \pm 150~\pc$). As a result,
according to Bernoulli's equation, the pressure will be greater immediately in front of and behind
the cloud than at the edge. Examining the pressures in the output from the hydrodynamical code
confirms this expectation.

The cloud also deforms due to Kelvin-Helmholtz or shear instabilities
\citep[e.g.,][]{chandrasekhar61}, instigated by the velocity difference between the cloud and the
ISM. Because the speed of the ISM relative to the cloud is largest at the edge of the cloud, the
instabilities grow most rapidly here \citep[Section~101]{chandrasekhar61}, and this part of the
cloud is pulled outward (see the hydrogen number density and \vz\ plots in
Figure~\ref{fig:CloudEvolution}). As the ISM flows around the initially spherical cloud, a vortex
develops behind the cloud. The flow in this vortex becomes more complicated as the shear
instabilities start to ablate material from the edge of the cloud. This material impedes the flow of
the ISM into the vortex behind the cloud, while new vortices form around the ablating material. This
complicated flow causes ISM material some way behind the cloud to move in toward the $r=0$ axis;
this material then flows down (i.e., in the $-z$ direction) and then out again. As this material
flows outward along the back of the cloud, it helps to stretch the cloud out in the horizontal
direction.

The gas ablated from the cool, dense cloud ($\Tcl = 10^3$~K, $\nHcl = 0.1~\pcc$) mixes with the hot,
tenuous ISM ($\TISM = 10^6$~K, $\nHISM = 10^{-4}~\pcc$), creating mixed gas of intermediate
temperature.  Eventually, the mixed gas reaches a temperature of a few times $10^5$~K, which is
optimal for radiative cooling through line emission. As the mixed gas flows back from the edge of
the cloud, the flow splits: some of the mixed gas is drawn into the vortex behind the cloud, while
some flows further back from the cloud, continuing to mix with the hot ISM. The temperature of the
mixed gas flowing away from the cloud continues to increase above a few times $10^5$~K, as the
continued mixing raises the temperature more rapidly than the gas can radiatively cool. In contrast,
the mixed gas flowing into the vortex behind the cloud cools, due to both mixing with cooler gas and
radiative cooling. In particular, this gas cools more efficiently after it reaches the region near
the $r=0$ axis, resulting in the accumulation of gas with $T \sim 10^4~\K$ along this axis in the
temperature plots of Figure~\ref{fig:CloudEvolution}.  We traced the fractions of original cloud
material and ISM material contained in the mixed gas and found that a significant fraction of the
cool gas along the $r=0$ axis at later times was initially hot ISM, indicating that this gas is
mixed gas that has undergone radiative cooling.

The plots of the high ion fractions (the last three rows in Figure~\ref{fig:CloudEvolution}) show
that the fractions of these high ions are higher in the mixed gas than in the initially cool
cloud gas or in the initially hot ISM gas. As the cool and hot gas mix, the ions that we are
interested in are produced both by ionization during the heating of the initially cool gas, and by
recombination during the cooling of the initially hot gas. The fraction of Li-like ions for each
element depends somewhat on the temperature of the gas; for example, in the hotter gas the
\OVI\ fraction is higher than the \CIV\ fraction. The NEI ionization levels in
Figure~\ref{fig:CloudEvolution} are different from those expected from CIE, because changes in the
ionization levels lag behind the changes in temperature that are due to mixing or radiative
cooling. This is similar to what we found in our previous study of TMLs (KS10).

It is possible that our use of 2D cylindrical geometry resulted in an overestimate of the amount of
cool gas that accumulates along the $r=0$ axis. In such a geometry, material that flows toward the
$r=0$ axis tends to stick to the axis, because of the reflecting boundary condition there. The 2D
cylindrical geometry is also responsible for the protuberance at the front of the cloud seen at
later times, for example, the feature indicated by the arrow in the far right density plot in our
Figure~\ref{fig:CloudEvolution} \citep{vieser07a}. However, although the high ions contained in the
cooled gas that accumulates along the symmetry axis give rise to large ion column densities along
this sightline (see Section~\ref{subsec:HighStageIons}), this column of material does not contribute
much to the total number of high ions on the grid.

\subsection{Evolution of the Number of \HI\ Atoms}
\label{subsec:LossOfHI}

The default FLASH NEI module assumes that hydrogen is fully ionized, and so we were unable to trace
the ionization evolution of hydrogen. We therefore assumed that the hydrogen in gas with $T < 10^4$~K
is entirely neutral (\HI), while the hydrogen in hotter gas is fully ionized. Our initial clouds
``lose'' their \HI\ content via the physical processes discussed above, namely ablation of material
from the cloud and its subsequent temperature increase due to mixing with the hot ISM. (Note,
however, that \HI\ can also be replenished when hot gas cools below $10^4$~\K, and such gains can
count against the losses.) Here, we consider the loss of \HI\ from the clouds due only to heating,
and the loss of \HI\ due to heating and/or ablation.  In the former case, we consider all
\HI\ atoms, regardless of their velocity. In the latter case, we consider only \HI\ atoms moving
with HVC-like velocities: $\vz \le -80~\kmps$ for Models~A, B, F, and G (in which the initial
velocity of the cloud was $\vzcl = -100~\kmps$), and $\vz \le -100~\kmps$ for Models~C, D, and E (in
which \vzcl\ was $-150$ or $-300~\kmps$). These definitions were chosen to correspond with
observational analyses, in which one can distinguish between high- and low-velocity \HI.

To investigate the loss of \HI\ from the clouds, for each model we calculate the ratio, $\beta(t)$,
of the number of neutral hydrogen atoms lost since the beginning of the simulations to the initial
number of neutral hydrogen atoms, i.e.,
\begin{equation}
  \beta (t) \equiv \frac{\nHIinit - \nHI (t)}{\nHIinit},
  \label{eq:beta}
\end{equation}
where \nHIinit\ and $\nHI (t)$ are the initial and current numbers of \HI\ atoms, respectively.  As
noted above, we have two different ways of counting $\nHI (t)$, leading to two different values of
$\beta(t)$: $\betaall(t)$ for \HI\ at all velocities, and $\betaHVC(t)$ for HVC-like \HI. Note that
the initial number of \HI\ atoms, \nHIinit, is the same in both cases, as all the \HI\ is
high-velocity at the start of the simulation. The cool gas that accumulates along the $r=0$ axis
after ablation, mixing, and cooling has low velocities; it is thus included in $\betaall(t)$ but not
$\betaHVC(t)$. We did not include material that escapes from the top of the computational
domain. This means that the true value of $\nHI(t)$ is larger than the value obtained from the
computational domain, and so our estimates of $\beta(t)$ are upper limits. However, the material
that escapes from the top of the domain has low densities and low velocities, and so the amounts of
HVC-like material are not significantly affected.

\begin{figure}
\plotone{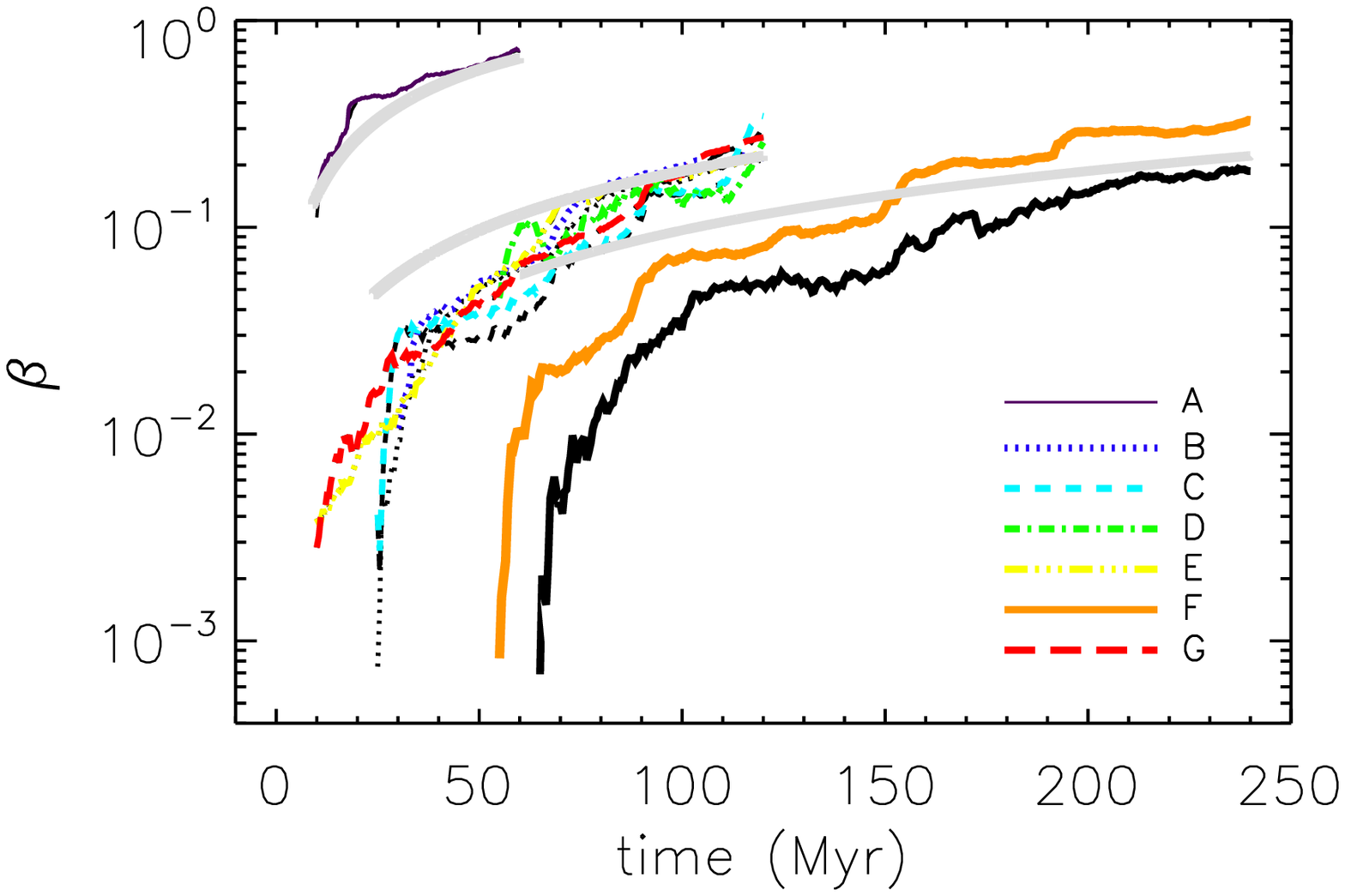}
\caption{Ratio $\beta$ of the number of \HI\ atoms lost to the initial number of \HI\ atoms, as a
  function of time (see Equation~\ref{eq:beta}). The different line styles correspond to the
  different cloud models, as indicated in the key (thin solid line: Model~A; dotted line: Model~B;
  short dashed line: Model~C; dot-dashed line: Model~D; triple dot-dashed line: Model~E; thick solid
  line: Model~F; long dashed line: Model~G). For each model, the black line shows $\beta$ for
  \HI\ with all velocities (\betaall), while the colored line shows $\beta$ for \HI\ with HVC-like
  velocities (\betaHVC), respectively.  The smooth thick light gray curves show $\beta(t)$ for
  spherical clouds with initial radii of 20, 150, and 300~\pc\ (top to bottom) that lose mass at
  rates proportional to their surface areas (see Section~\ref{subsubsec:EffectOfSize} for details).
  \label{fig:HIloss}}
\end{figure}

Figure~\ref{fig:HIloss} shows $\beta(t)$ for each of our 7 models. The black lines show
$\betaall(t)$, and the colored lines show $\betaHVC(t)$. In all cases, $\beta$ generally increases
with time, indicating that \HI\ is lost throughout the simulation. For Models~A through E plus G,
the black and colored lines are similar to each other (they are almost identical for Model~A),
indicating that most of the ablated material is ``hot'' (i.e., above $10^4~\K$). However, for
Model~F the colored line is clearly above the black line, indicating that some of the ablated
material is not yet ionized, or that some of the ablated gas has been heated and subsequently
cooled. We find that the latter explanation is the more important: the amount of radiatively cooled
ablated gas that accumulates along the $r=0$ axis is larger in Model~F than in the other models.

Here we note a number of features from Figure~\ref{fig:HIloss}: (1) $\beta$ is relatively
insensitive to the cloud's initial velocity, over a wide range of velocities (100--300~\kmps;
compare Models~B, C, and D). (2) $\beta$ is also relatively insensitive to the cloud's initial
density profile (compare Models~C and E) and the cloud and ISM's initial densities (compare Models~B
and G). (3) A smaller cloud loses its \HI\ content, as a fraction of its initial mass, faster than a
larger cloud (compare Models~A, B, and F). With these various trends in mind, we will discuss the
effects of the different model parameters in more detail in the following section.

\subsection{Differences between the Models -- The Effects of Different Model Parameters}
\label{subsec:CloudEvolutionDifferences}

\begin{figure*}
\hspace*{0.70in}
\includegraphics[scale=0.2]{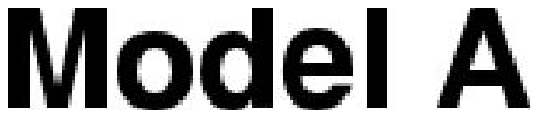}
\hspace{2.80in}
\includegraphics[scale=0.2]{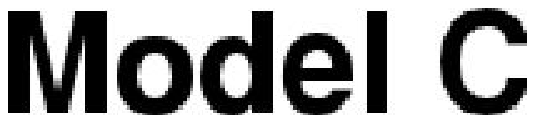} \\
\hspace*{0.3in}
\includegraphics[scale=0.2]{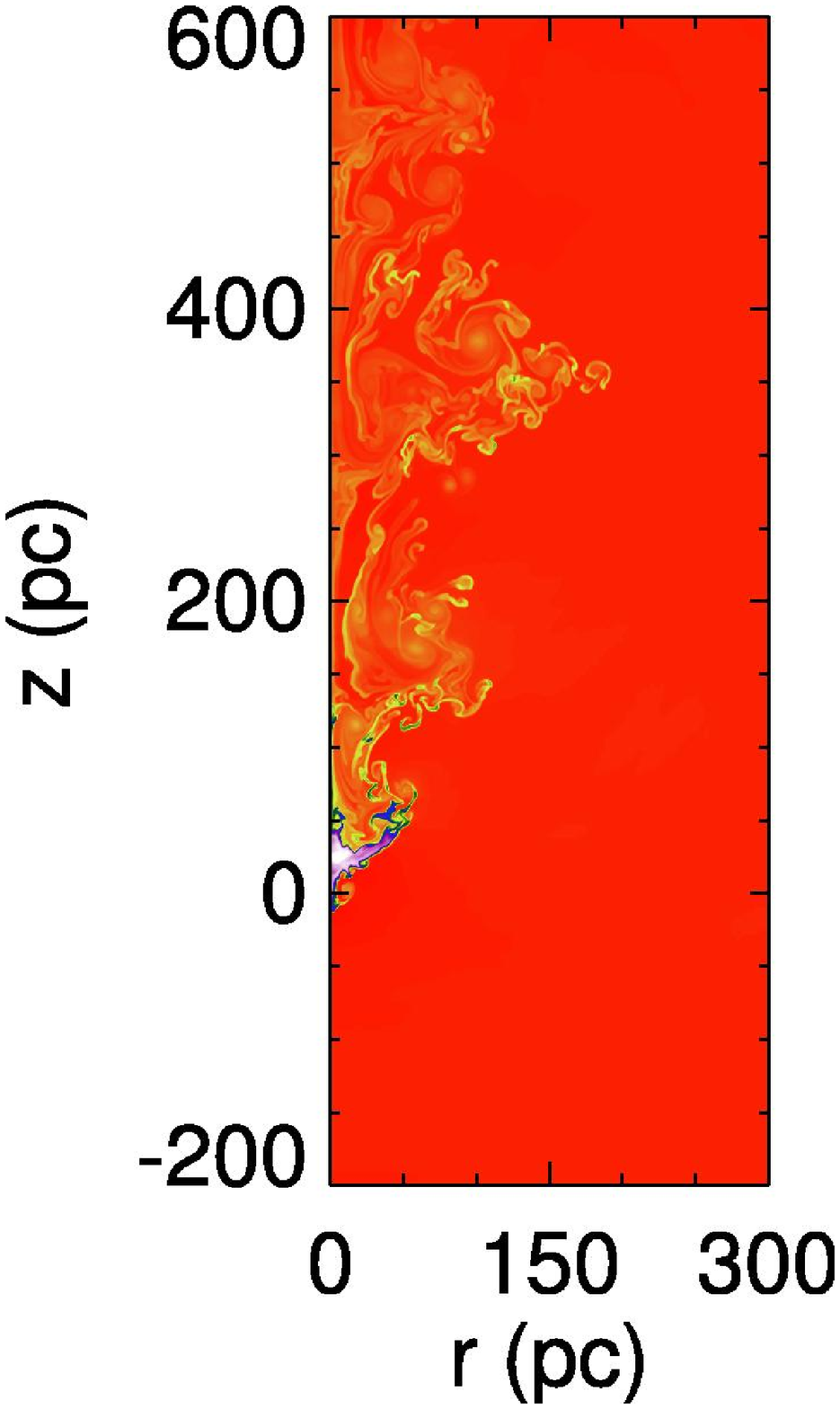}
\hspace{0.01in}
\includegraphics[scale=0.2]{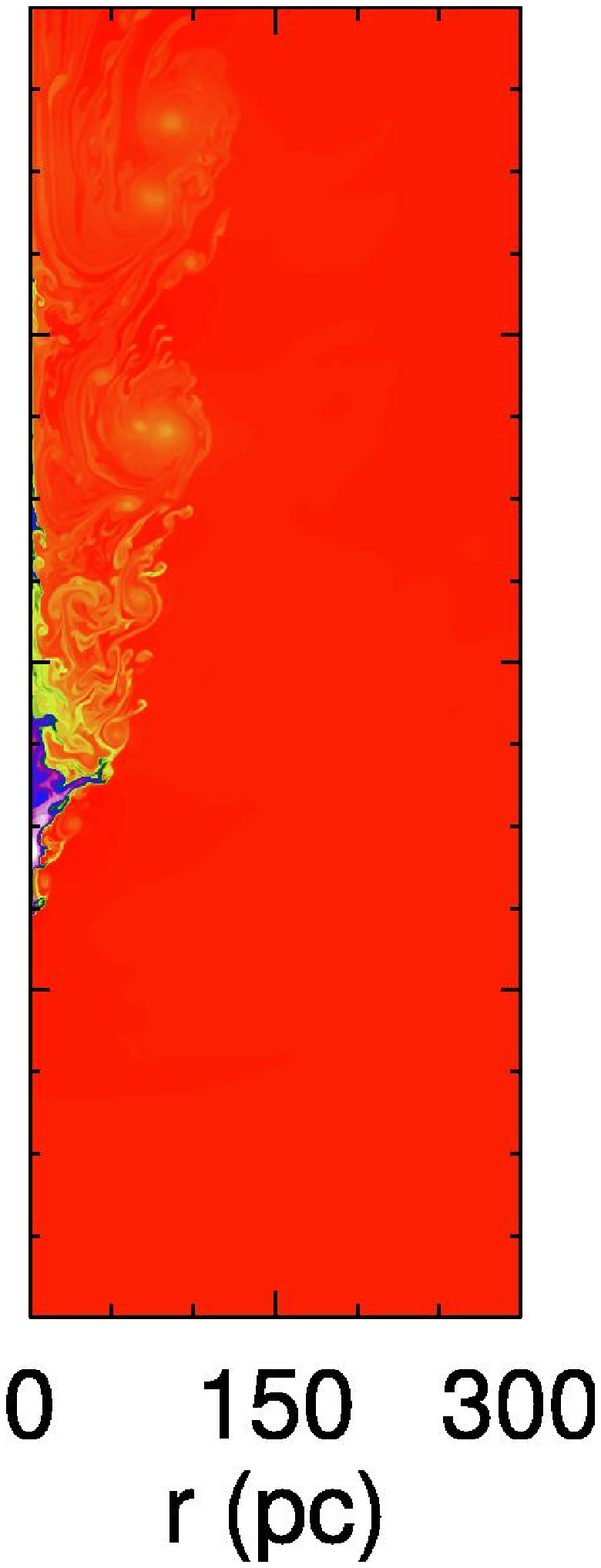}
\hspace{0.01in}
\includegraphics[scale=0.2]{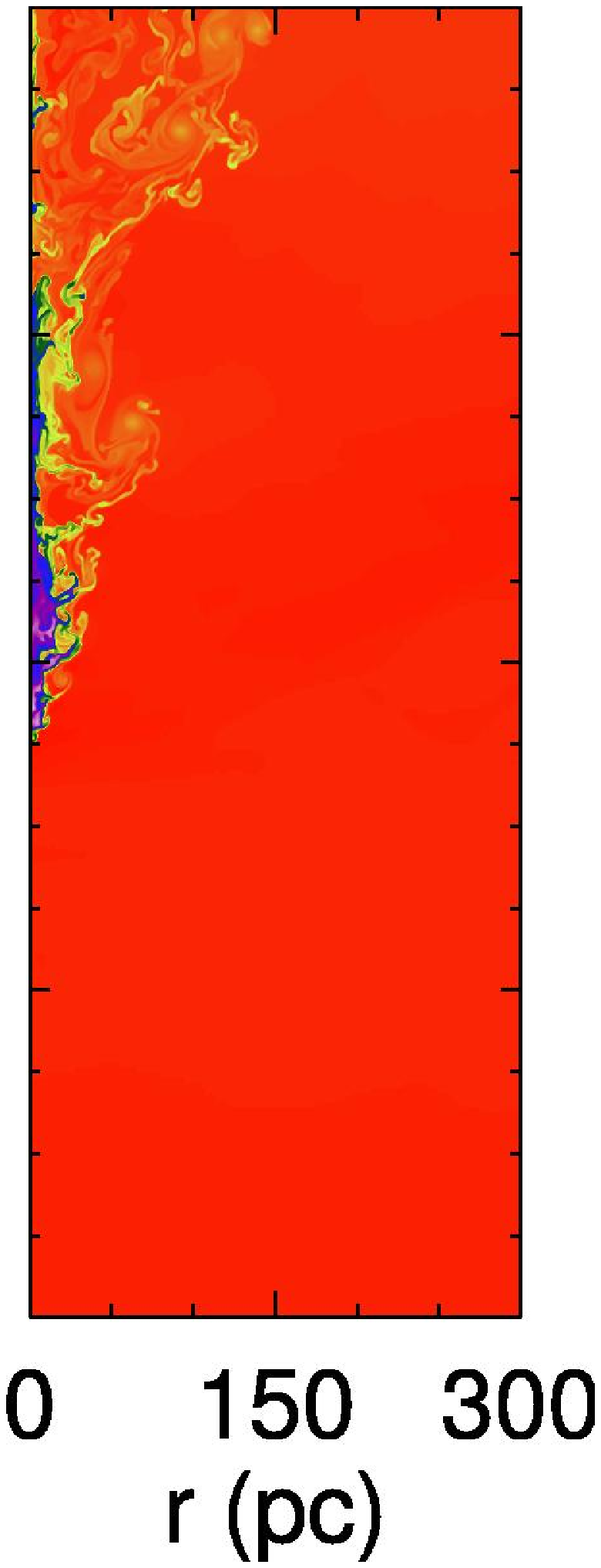}
\hspace{0.01in}
\includegraphics[scale=0.2]{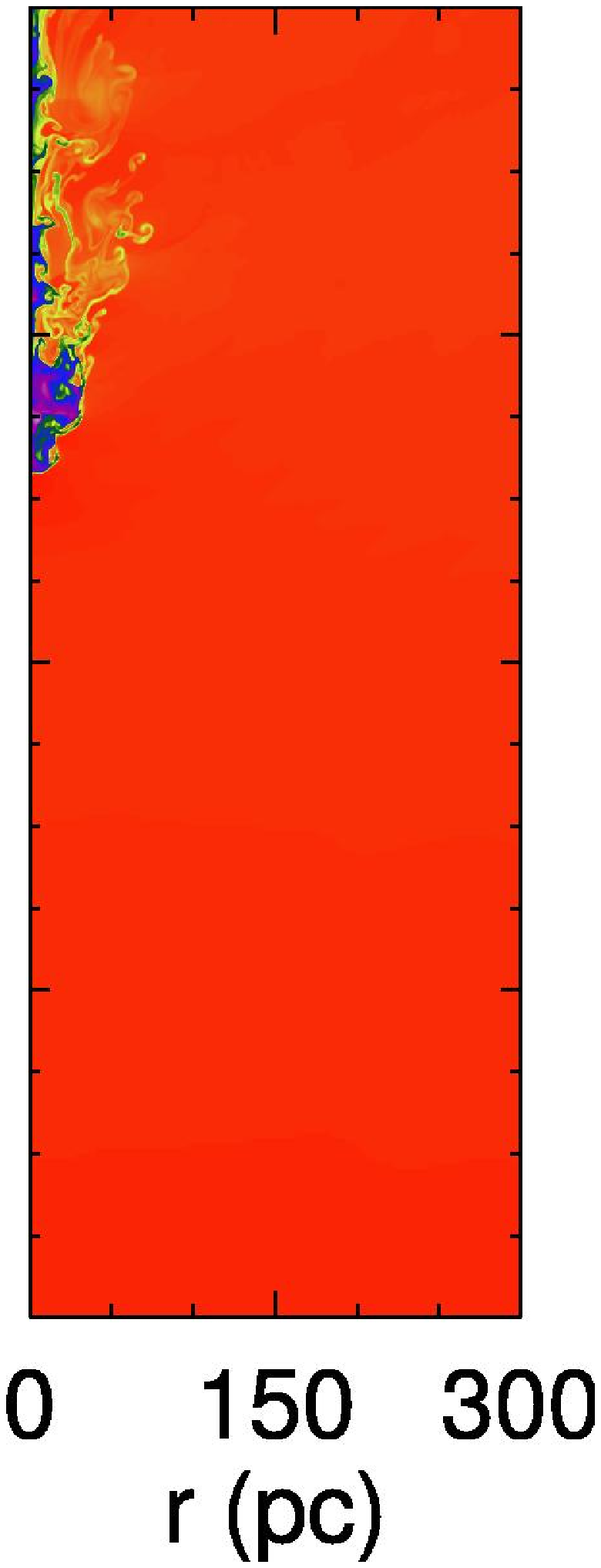}
\includegraphics[scale=0.2]{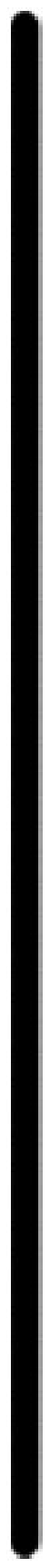}
\includegraphics[scale=0.2]{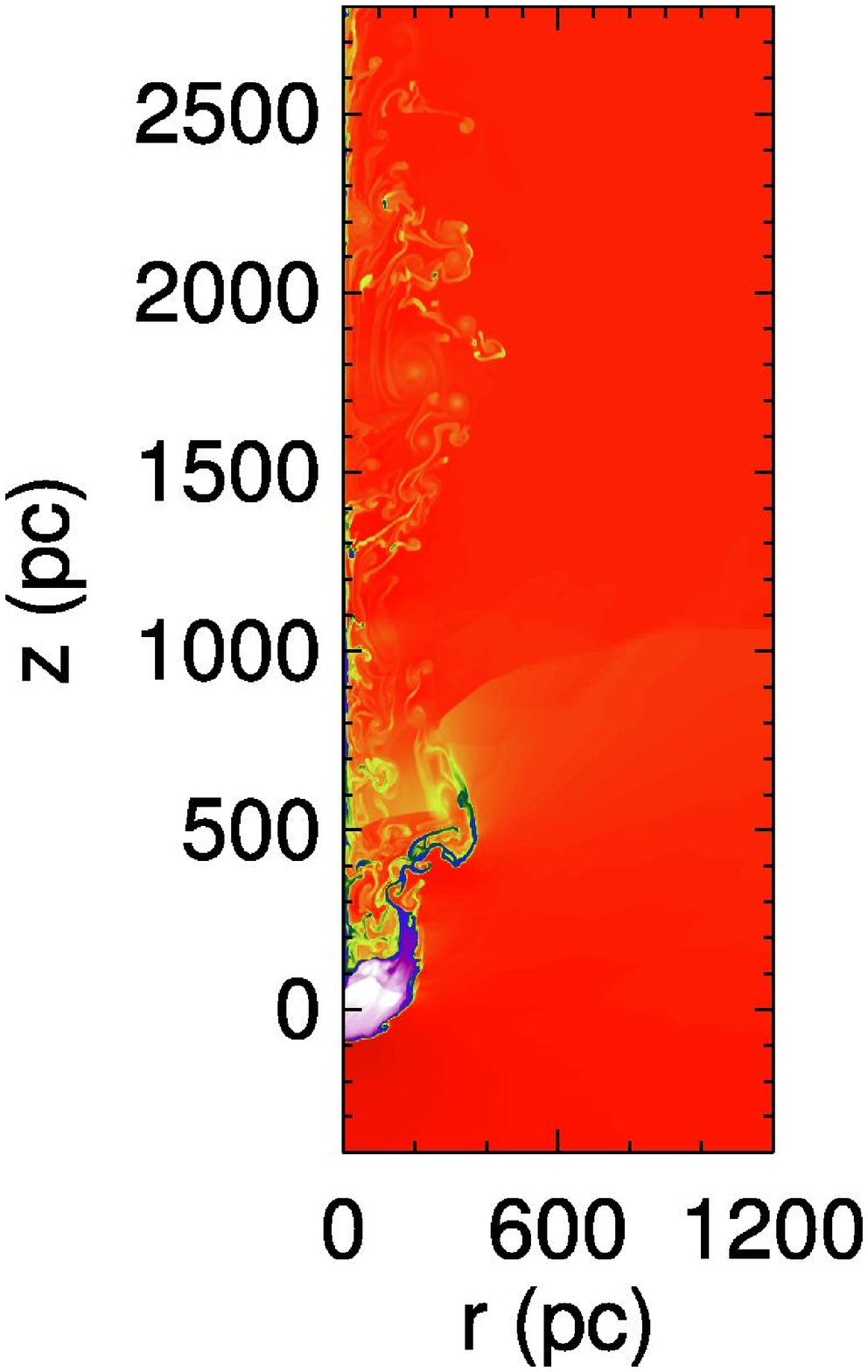}
\includegraphics[scale=0.2]{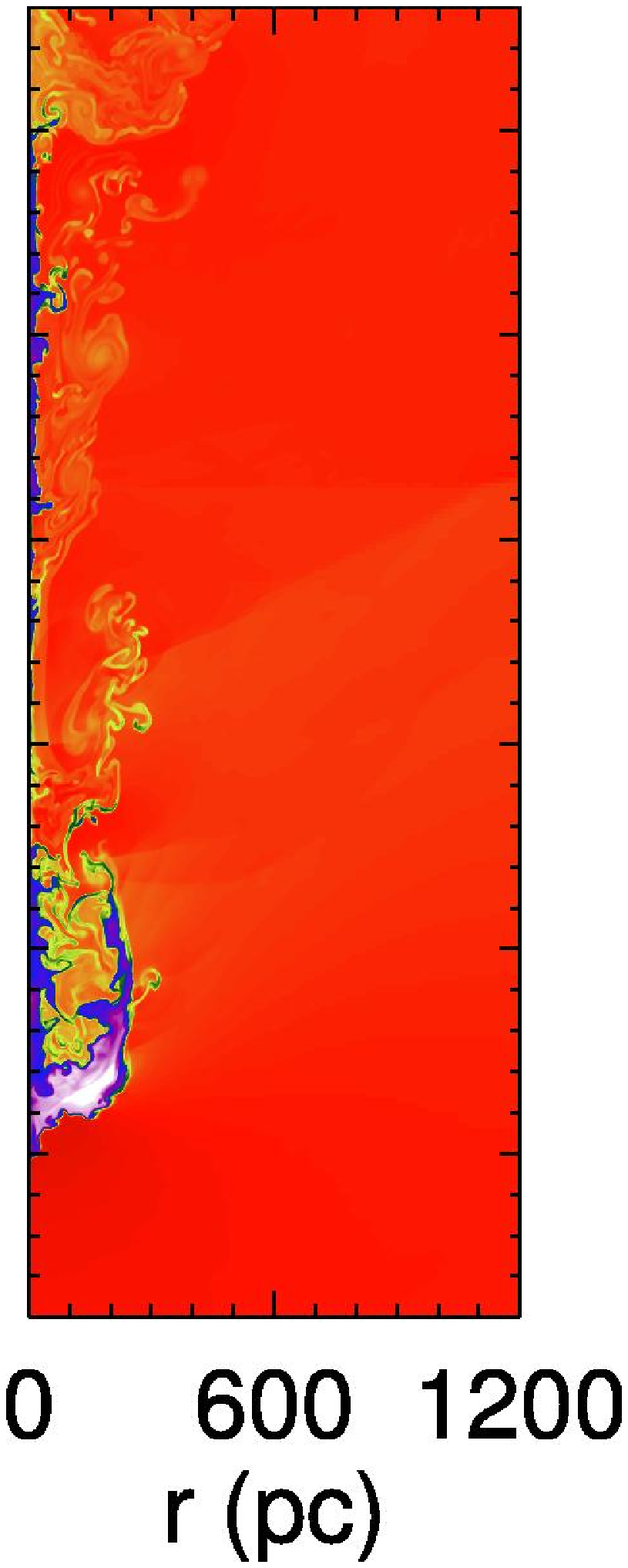}
\includegraphics[scale=0.2]{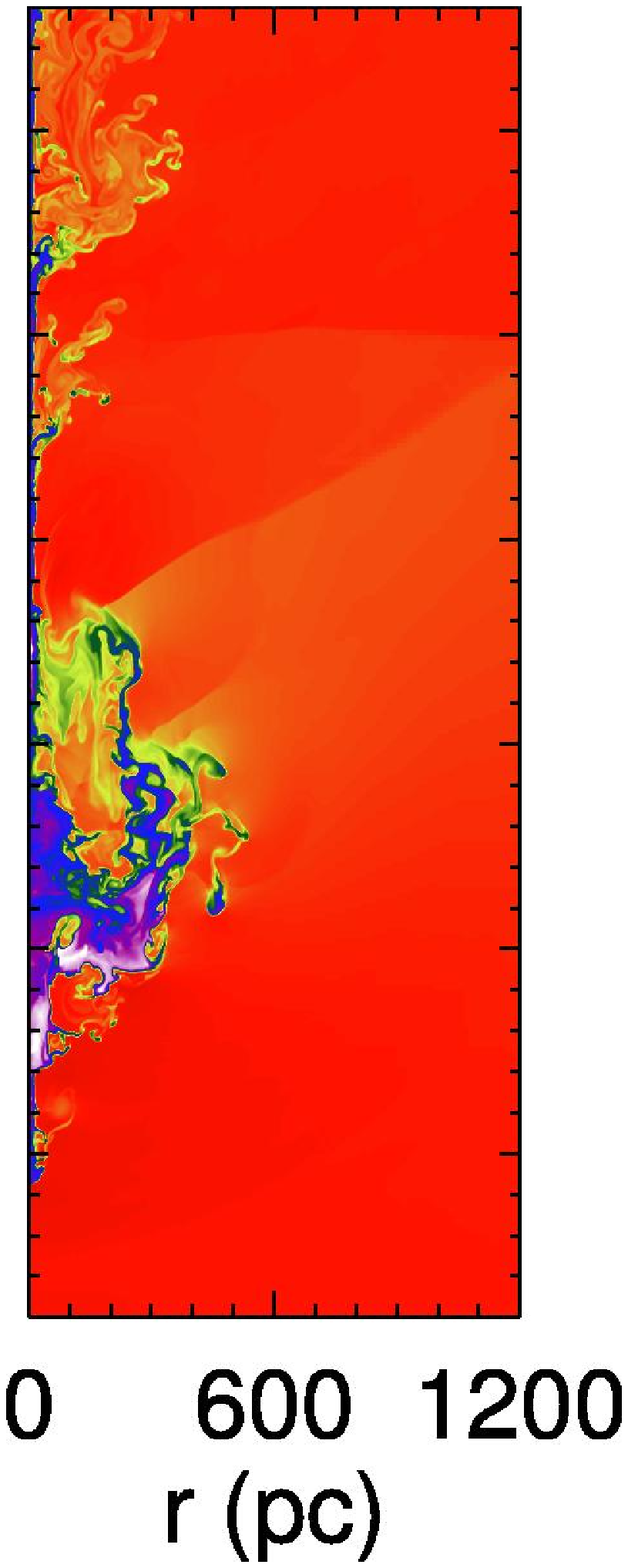}
\includegraphics[scale=0.2]{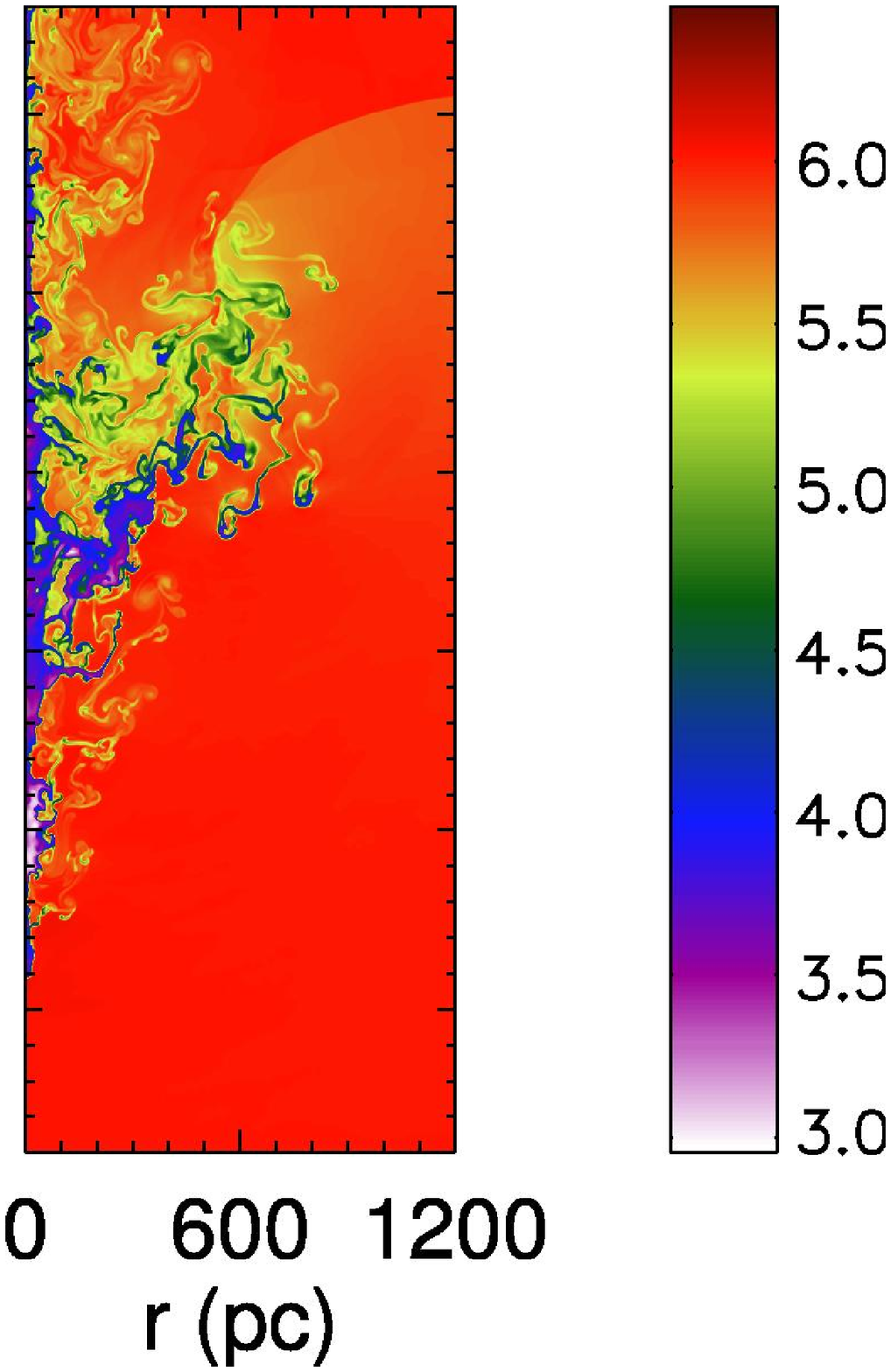} \\

\vspace{0.1in}
\hspace*{0.70in}
\includegraphics[scale=0.2]{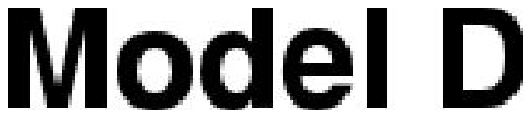}
\hspace{2.80in}
\includegraphics[scale=0.2]{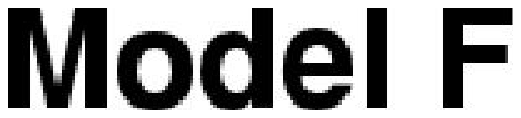} \\
\hspace*{0.3in}
\includegraphics[scale=0.3]{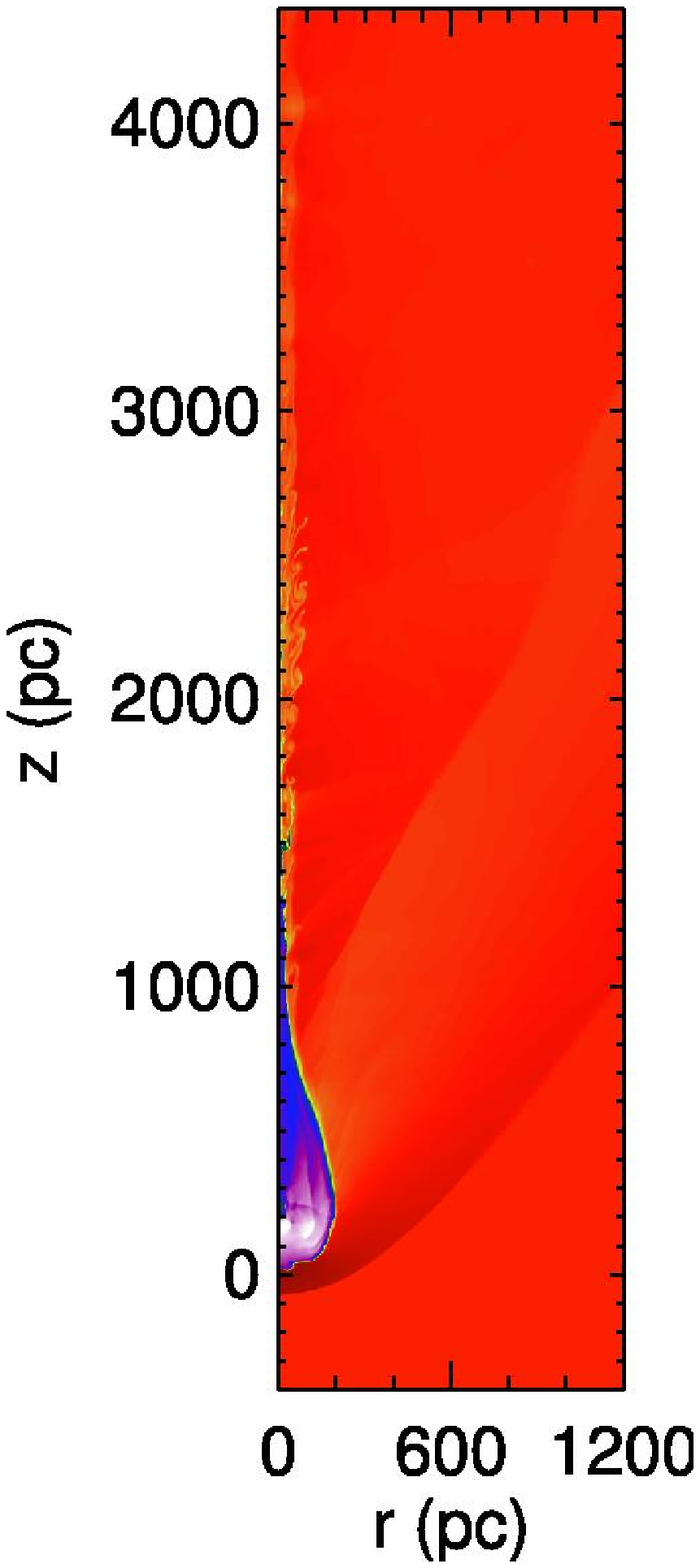}
\includegraphics[scale=0.3]{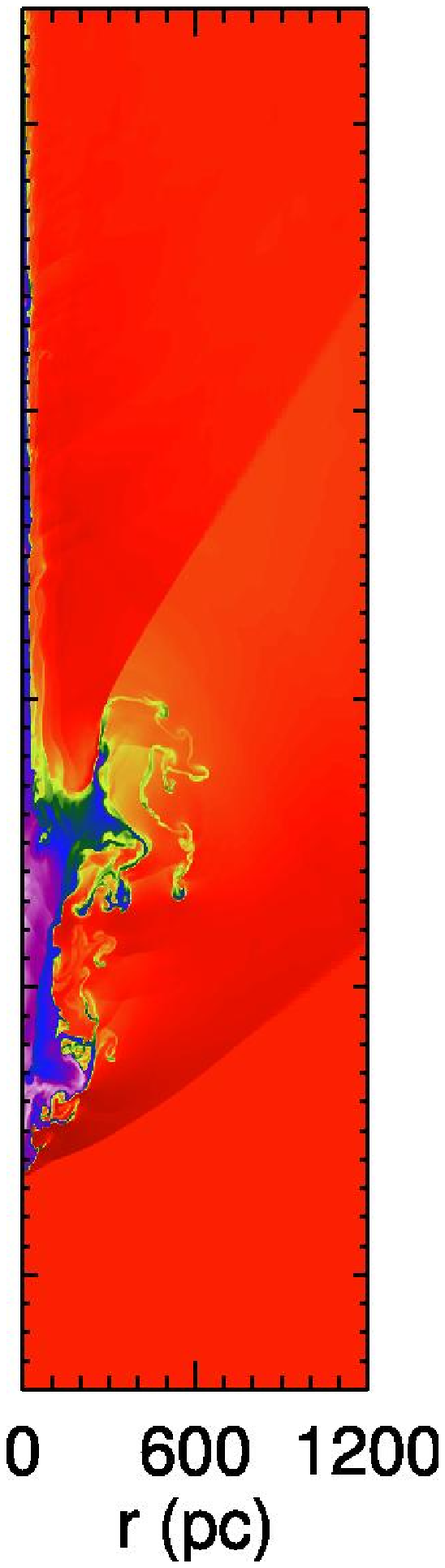}
\includegraphics[scale=0.3]{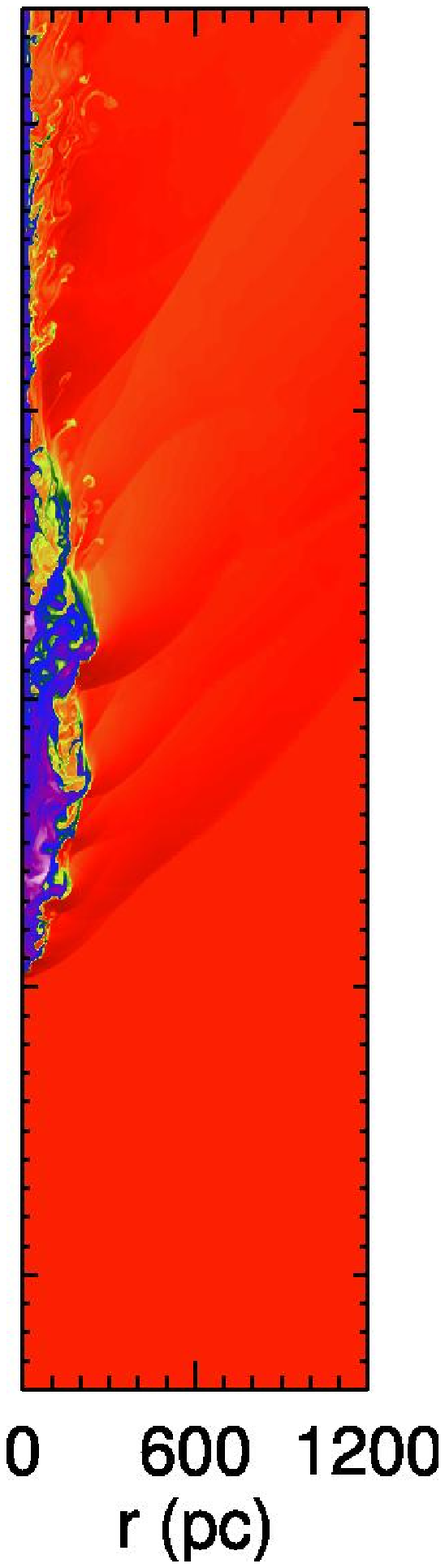}
\includegraphics[scale=0.3]{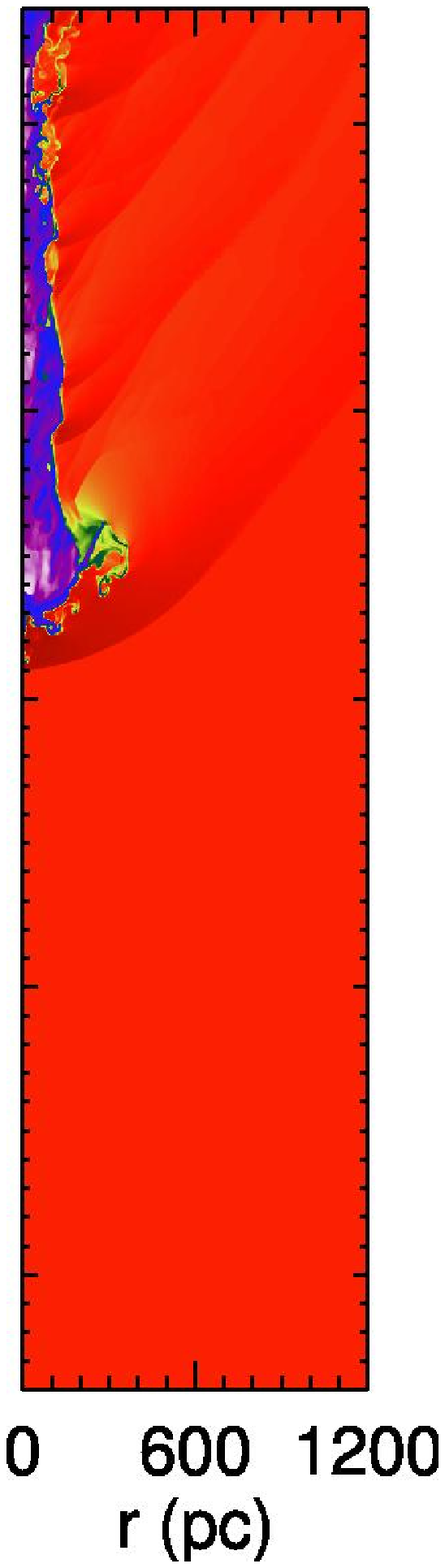}
\includegraphics[scale=0.2]{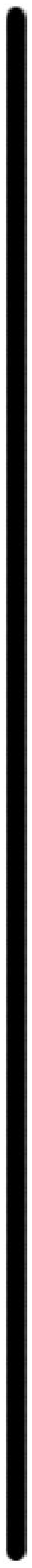}
\includegraphics[scale=0.3]{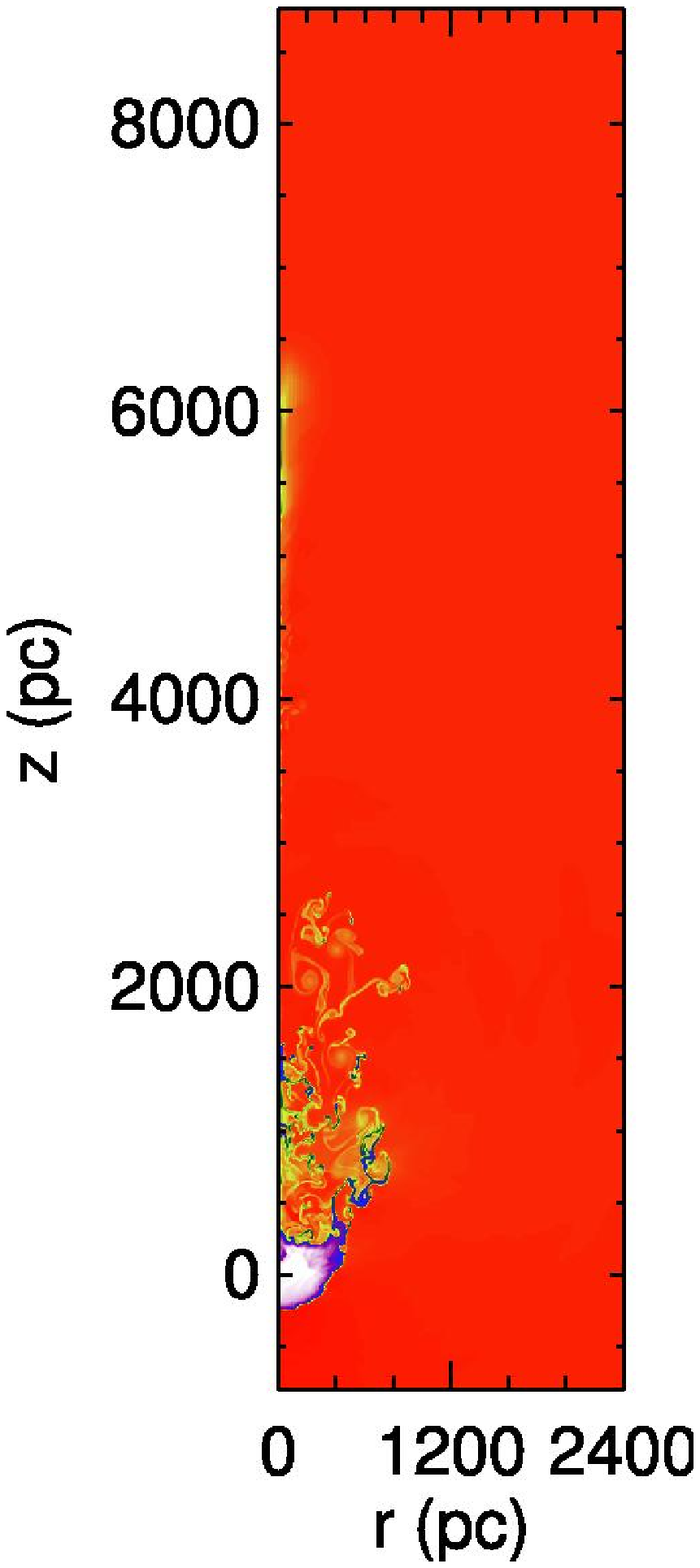}
\includegraphics[scale=0.3]{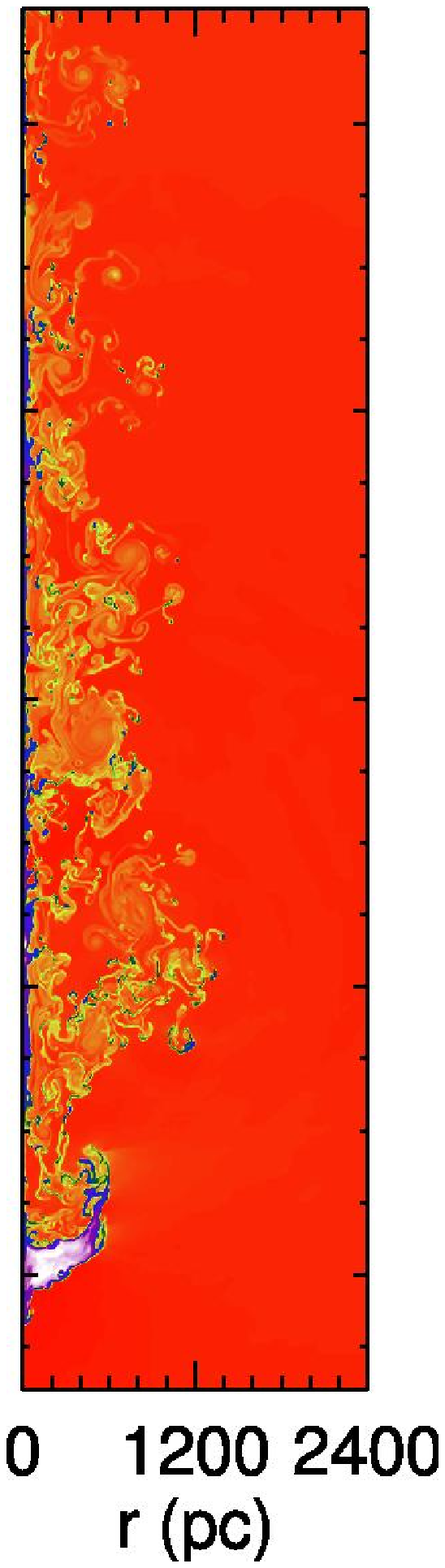}
\includegraphics[scale=0.3]{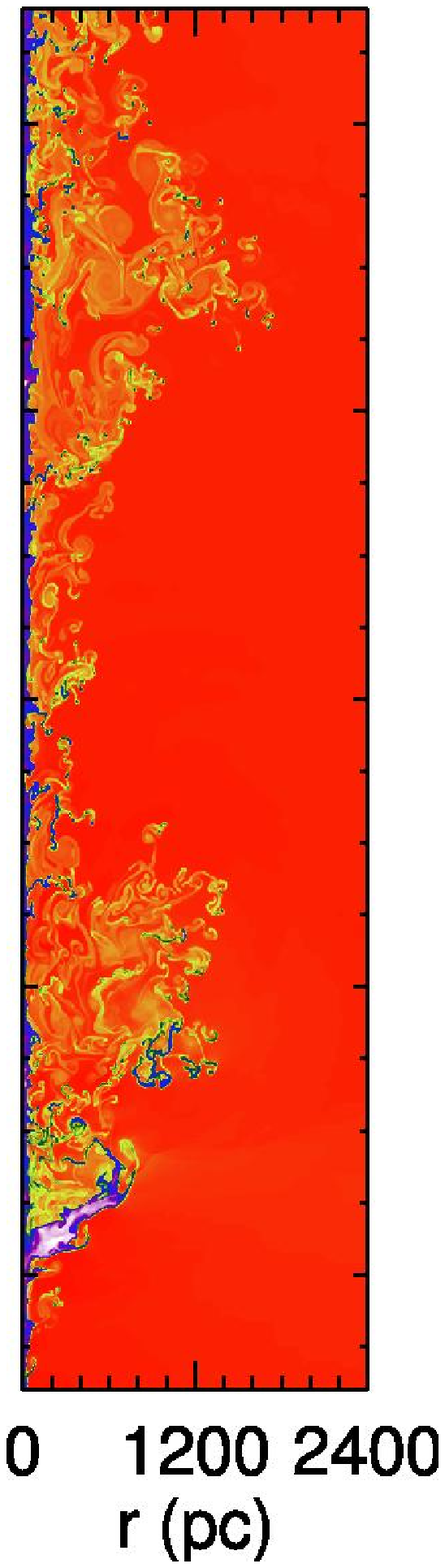}
\includegraphics[scale=0.3]{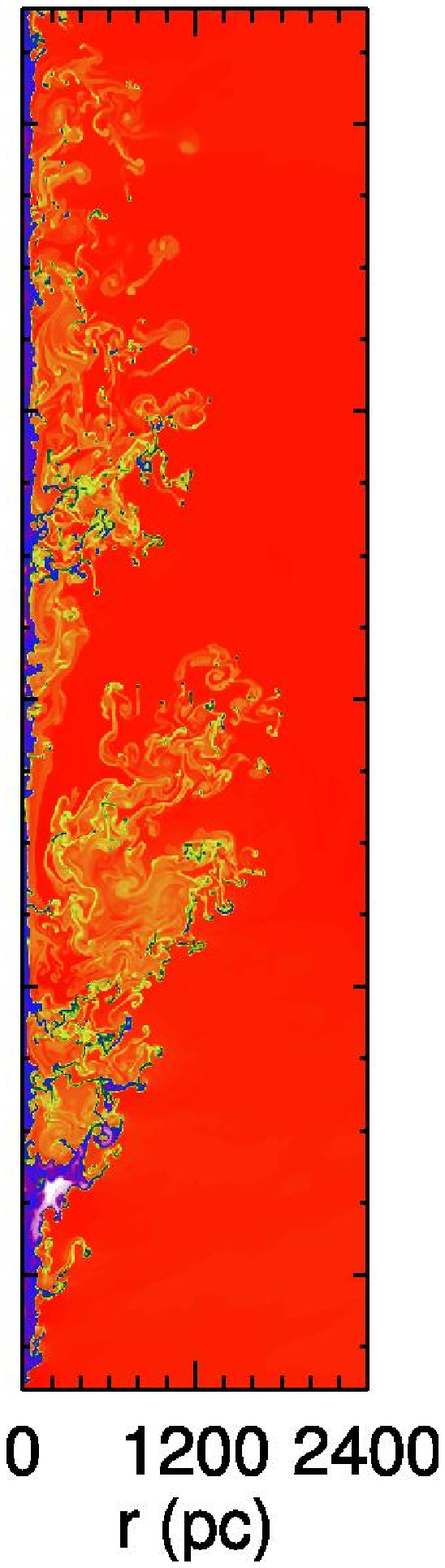} \\

\vspace{0.1in}
\hspace*{0.70in}
\includegraphics[scale=0.2]{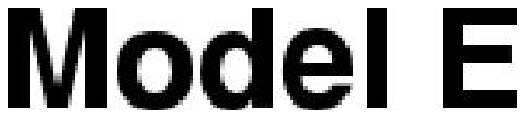}
\hspace{2.80in}
\includegraphics[scale=0.2]{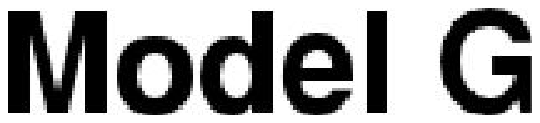} \\
\hspace*{0.3in}
\includegraphics[scale=0.2]{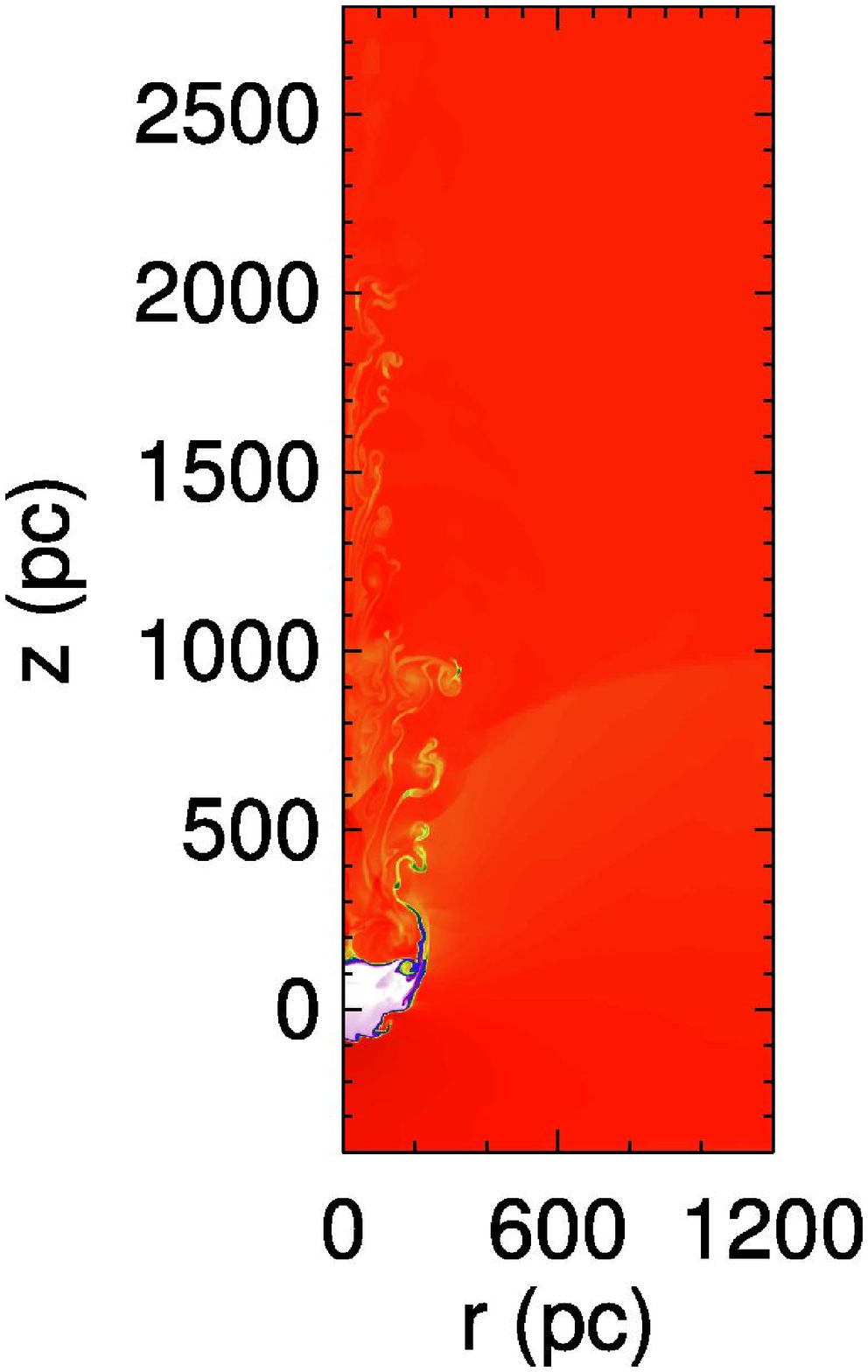}
\includegraphics[scale=0.2]{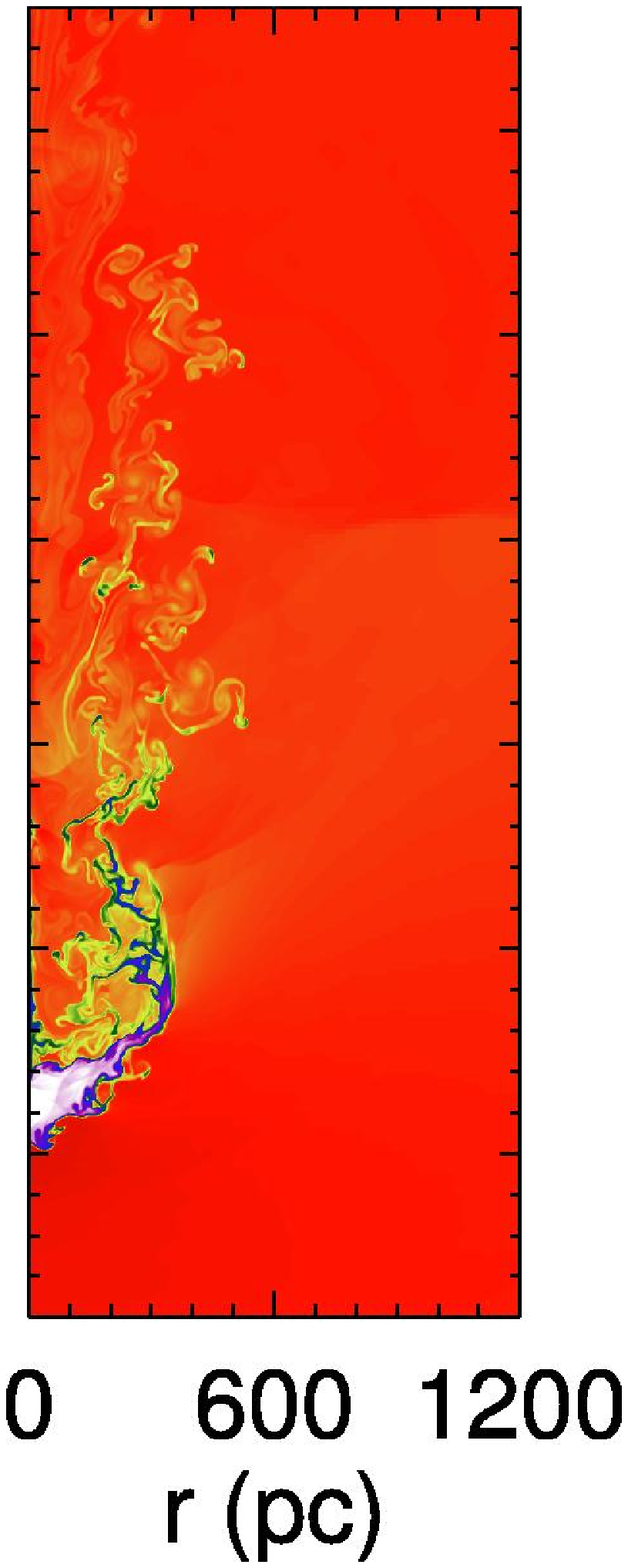}
\includegraphics[scale=0.2]{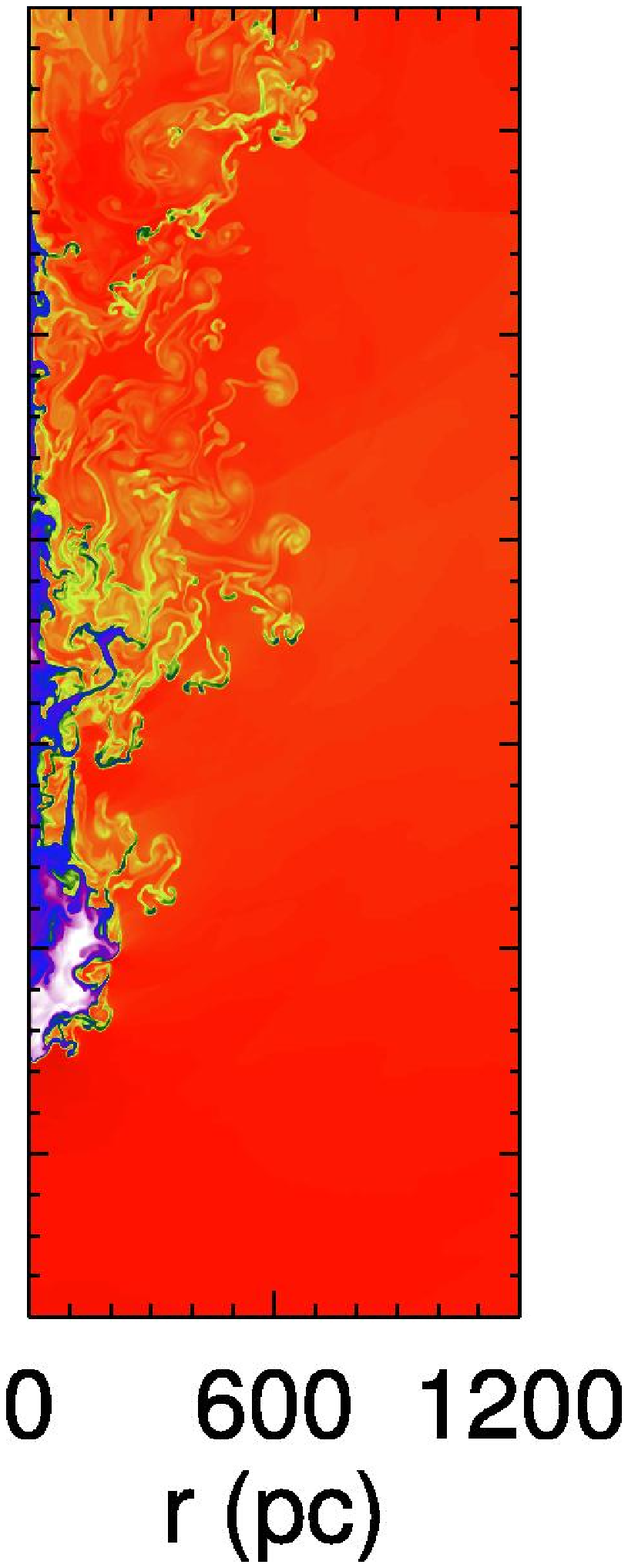}
\includegraphics[scale=0.2]{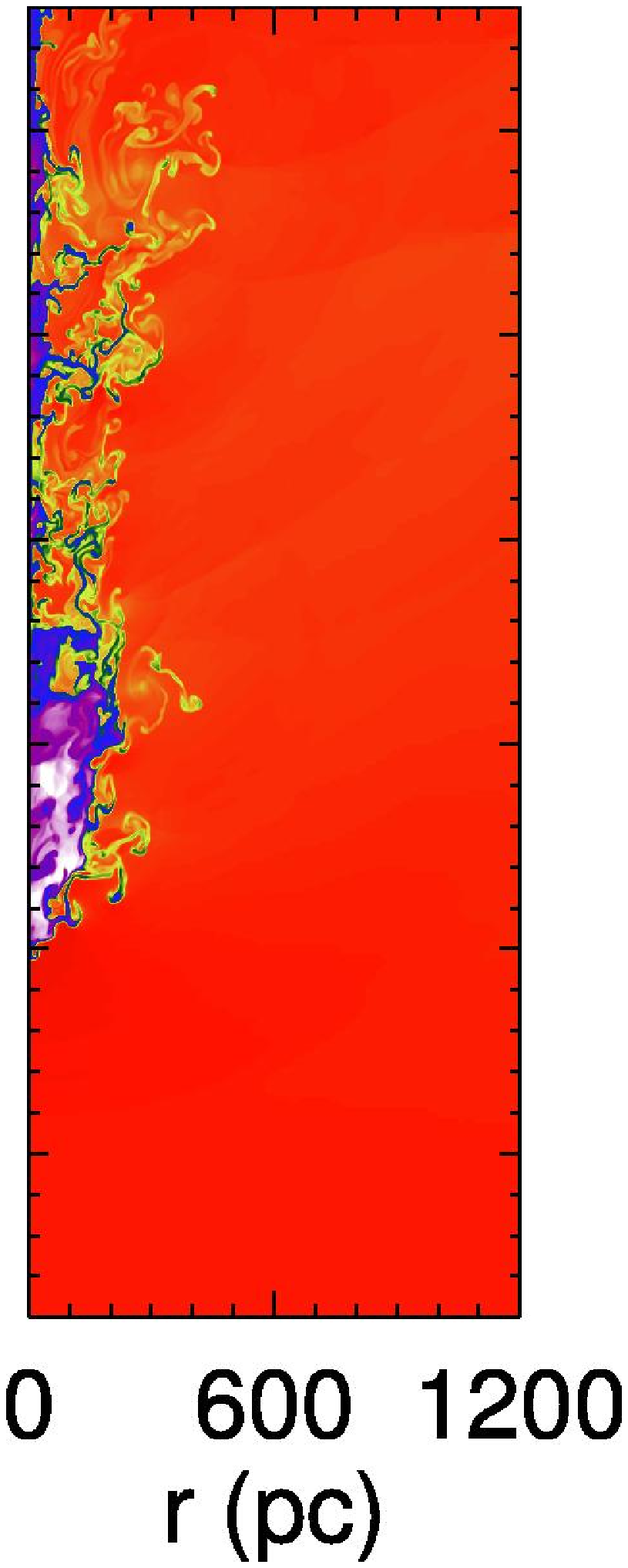}
\includegraphics[scale=0.2]{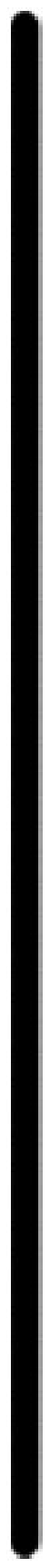}
\includegraphics[scale=0.2]{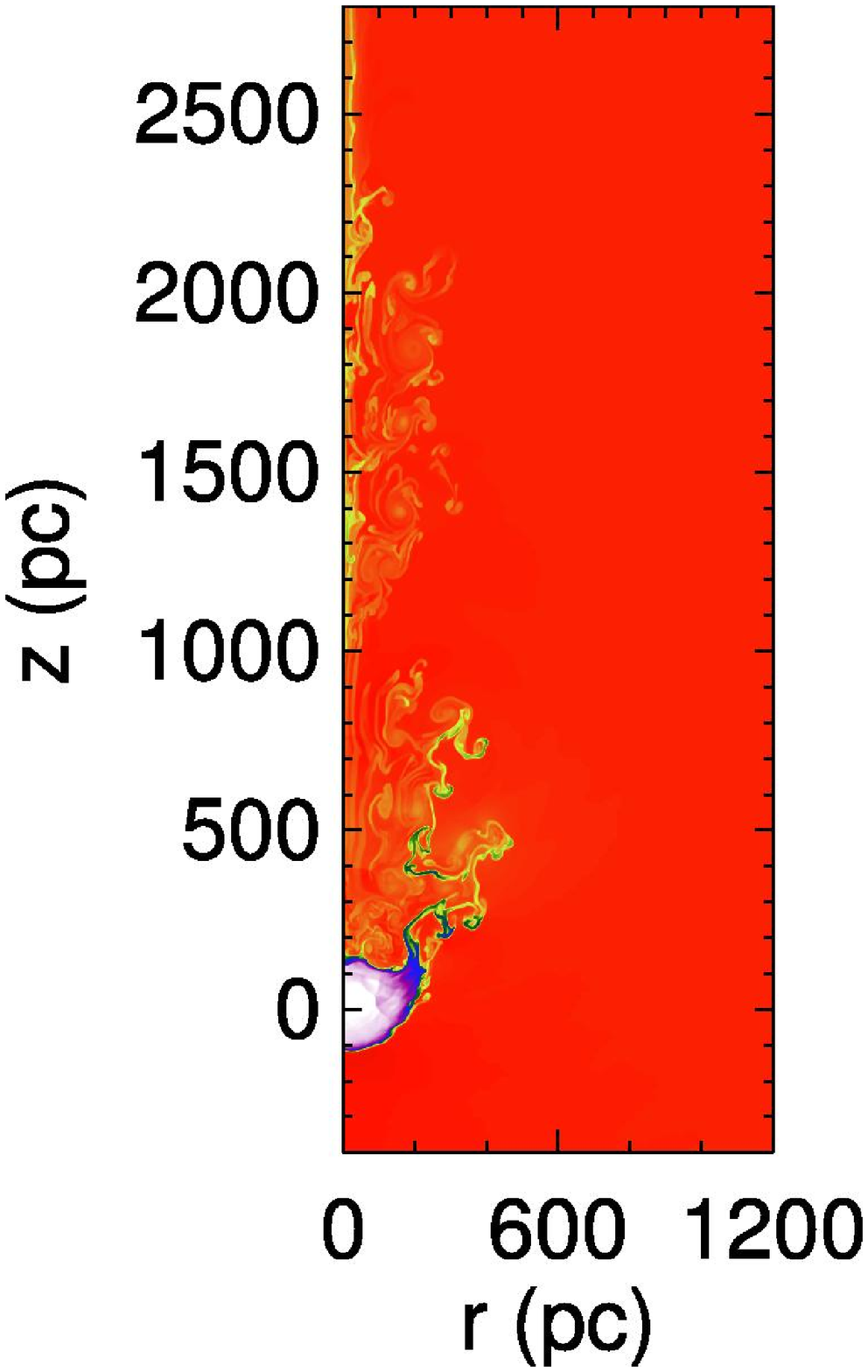}
\includegraphics[scale=0.2]{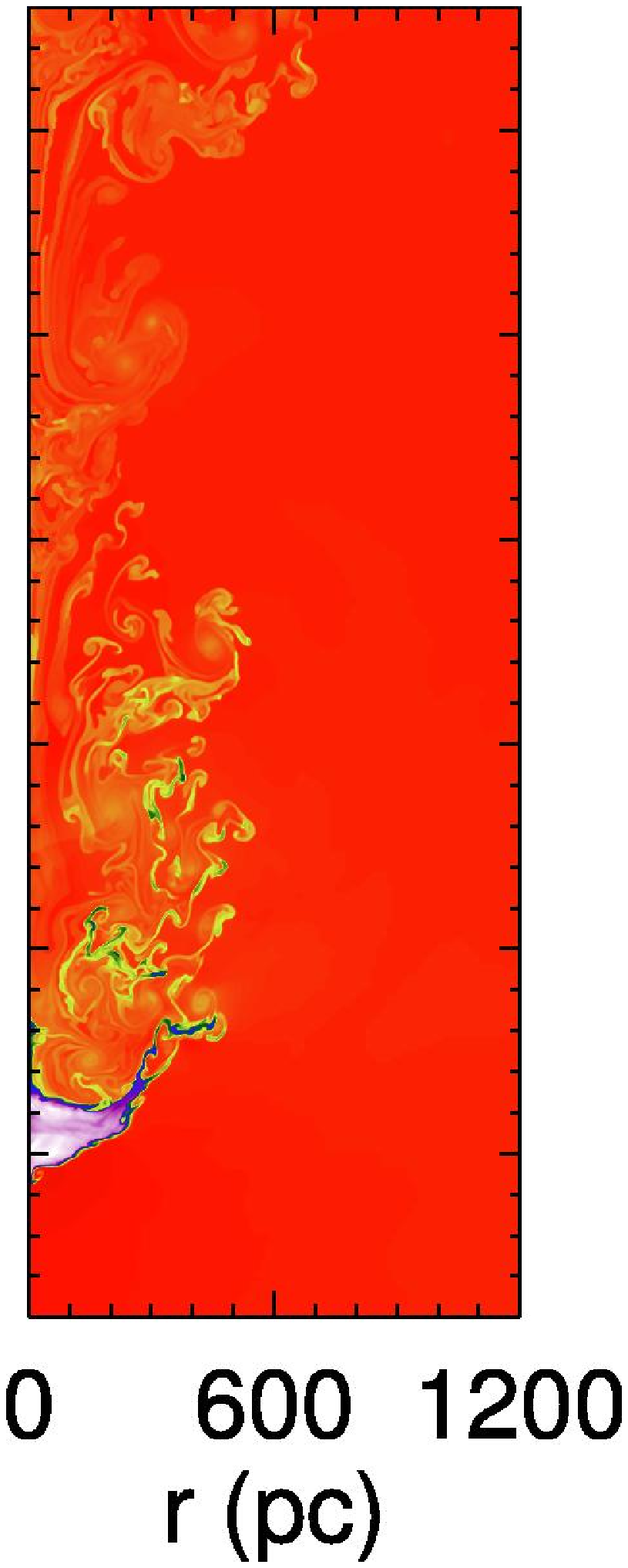}
\includegraphics[scale=0.2]{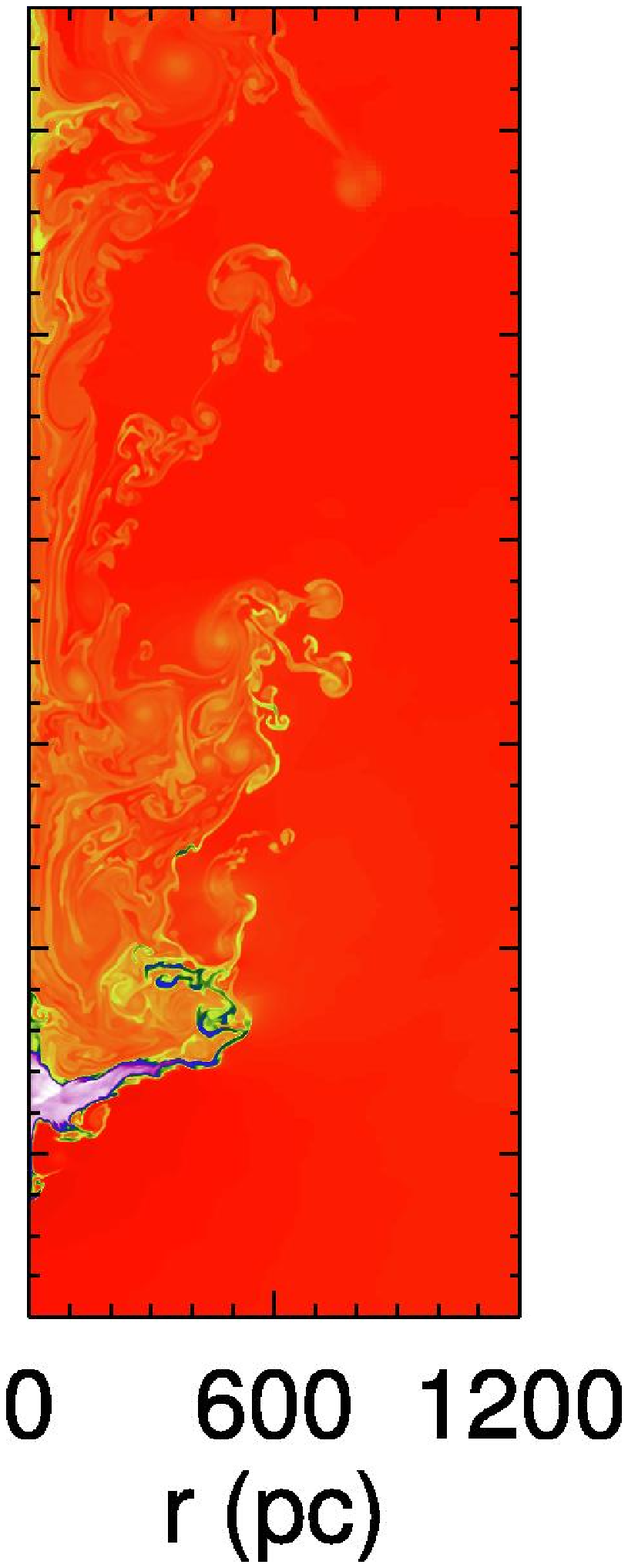}
\includegraphics[scale=0.2]{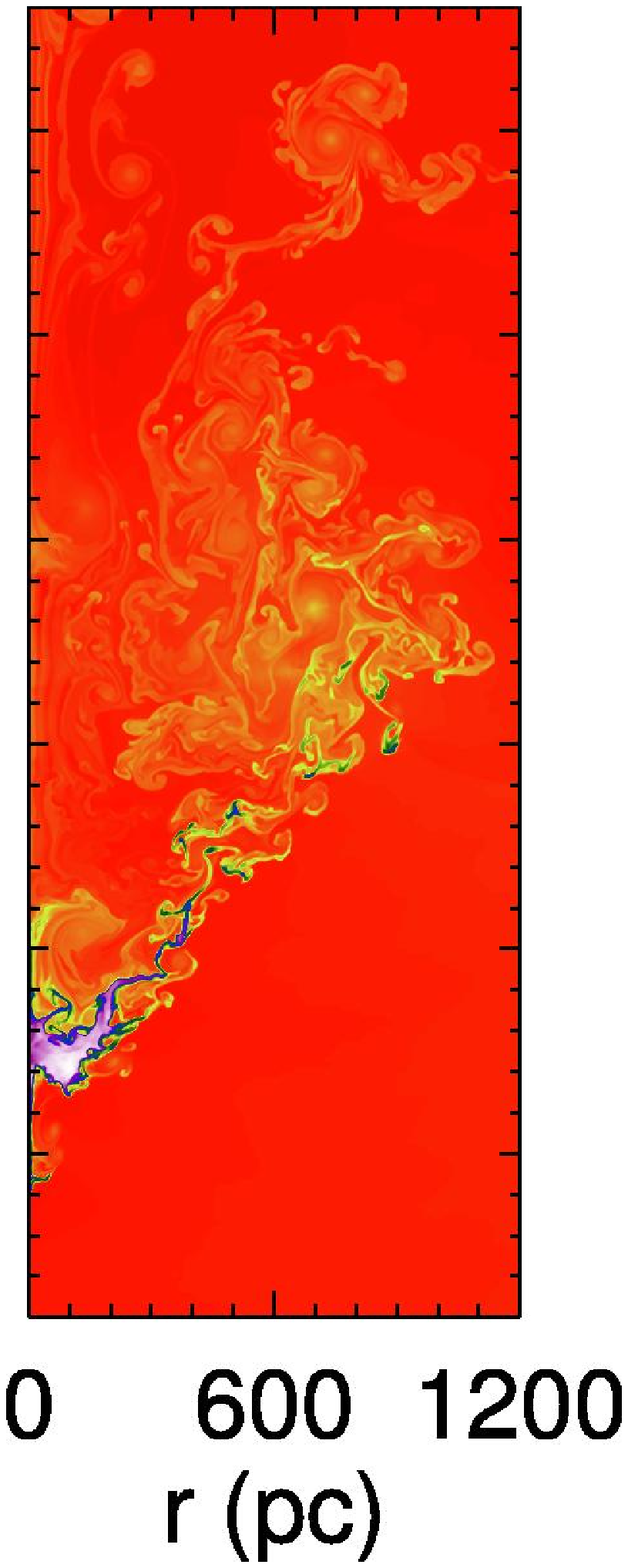} \\
\vspace{0.1in}
\caption{Comparison of the evolution of the cloud in Models~A, C, D, E, F, and G
  (cf.\ Figure~\ref{fig:CloudEvolution}). Cross-sections through the clouds showing the logarithm of
  the temperature are plotted for Model~A (top left), Model~C (top right), Model~D (middle left),
  Model~F (middle right), Model~E (bottom left), and Model~G (bottom right). For Model~A, the panels
  correspond to $t=15$, 30, 45, and 60~\Myr, respectively, from left to right.  For Models~C, D, E,
  and G, the panels correspond to $t=30$, 60, 90, and 120~\Myr, and for Model~F they correspond to
  $t=60$, 120, 180, and 240~\Myr.  The same color scale is used for all plots (see the color bar in
  the extreme top right).  Models~C, E, and G have the same size domain as Model~B
  (Figure~\ref{fig:CloudEvolution}).  The Model~A domain has a smaller height and width than the
  Model~B domain, while the Model~F domain has a larger height and width. The Model~D domain has the
  same width as but a larger height than the Model~B domain.
  \label{fig:CloudEvolutionDifferences}}
\end{figure*}

Figure~\ref{fig:CloudEvolutionDifferences} compares the evolution of Models~A, C, D, E, F, and G by
plotting the temperature on a logarithmic scale at various times (the evolution of Model~B is shown
in Figure~\ref{fig:CloudEvolution}). Note that for Models~C through G, the four panels correspond to
similar stages in these clouds' evolution, while the first Model~A panel corresponds to a similar
stage of evolution as the final panels for the other models (see
Section~\ref{subsubsec:EffectOfSize}). The computational domain for Model~D has a larger height
than those for Models~B, C, E, and G, because of the higher initial velocity of the cloud. In
simulations with higher initial cloud velocities (Models~C, D, and E), the location of the cloud
shifts upward further than in the simulations with lower cloud velocities (Models~B, F, and G). This
upward shift is because, in our simulations, the cloud is initially stationary while the ISM flows
upward, pushing the cloud upward (see Section~\ref{sec:Method}). However, despite the low velocity
in Model~A, the cloud in this simulation is still shifted upward significantly, due to its low
inertia relative to the ISM.

\subsubsection{The Effect of the Cloud Velocity (Models~B, C, and D)}
\label{subsubsec:EffectOfVelocity}

The initial conditions for the cloud and the ISM are the same in Models~B, C, and D, apart from the
initial velocity of the cloud: $-100$, $-150$, and $-300~\kmps$ in the observer's frame,
respectively. Because the sound speed of the ISM ($T = 10^6~\K$) is $\sim$150~\kmps, these
velocities correspond to the subsonic, transonic, and supersonic regimes, respectively. We would
therefore expect a bow shock to develop in Model~D (supersonic case);
Figure~\ref{fig:CloudEvolutionDifferences} shows that a bow shock does indeed develop at early times
in this model, and persists until the end of the simulation.

Apart from the formation of a bow shock in Model~D, Models~C and D both evolve with similar
hydrodynamical processes that were seen in Model B: Bernoulli's effect, ablation of the cloud's
material due to shear instabilities, mixing of the ablated gas with the hot ISM, and cooling of the
mixed gas. However, the shear instabilities grow more rapidly the larger the velocity difference
between the cloud and the ISM \citep[Section~101]{chandrasekhar61}. The faster-growing instabilities
clearly affect the evolution of the cloud: a faster cloud disrupts more violently than a slower
cloud. The temperature plots show that the amount of cool cloud material (shown in white) is smaller
for a faster cloud at a given time: compare the 3rd, 5th, 7th, and 9th panels in the
Figure~\ref{fig:CloudEvolution} temperature plot (Model~B) with the Model~C and D panels in
Figure~\ref{fig:CloudEvolutionDifferences} (these sets of panels correspond to the same times: $t =
30$, 60, 90, and 120~\Myr). These differences in the severity of the cloud disruption due to shear
instabilities lead to different morphologies for the clouds with different velocities.

\begin{figure}
\centering
\includegraphics[width=0.5\linewidth]{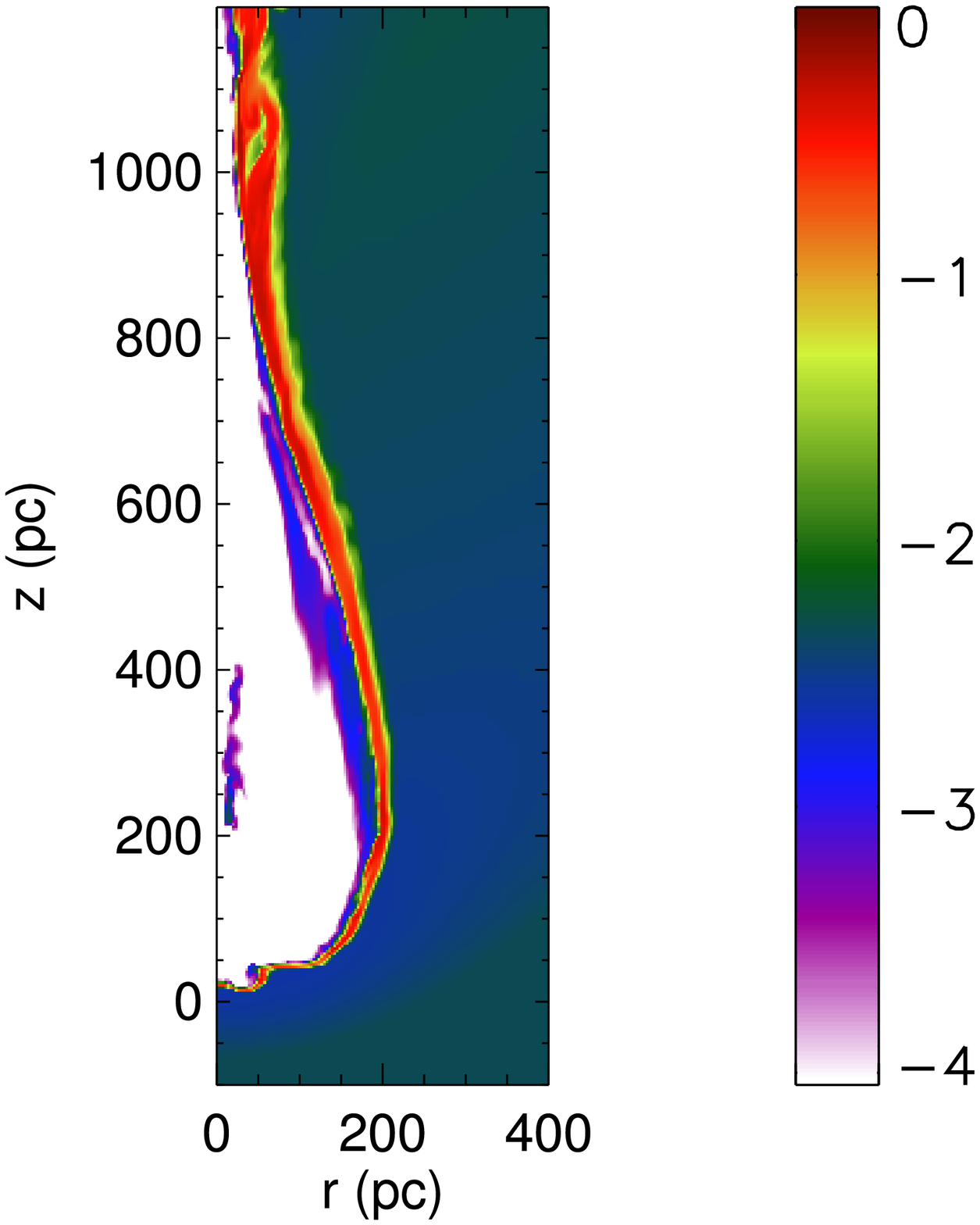}
\vspace{0.35in}
\caption{\OVI\ ion fraction for Model~D at $t = 30~\Myr$ (the same time as the first Model~D panel
  in Figure~\ref{fig:CloudEvolutionDifferences}).
  \label{fig:ModelDOVI}}
\end{figure}

Although different cloud velocities lead to different cloud morphologies, we find that the
faster-growing instabilities do not necessarily lead to material being ablated and/or ionized at
larger rates for the faster clouds. Figure~\ref{fig:HIloss} shows that the rates at which the
Model~B, C, and D clouds lose their \HI\ to ablation and/or ionization, offset by material that has
cooled, are rather similar. Note that in Model~D, the bow-shock that forms in front of the cloud
helps protect the cloud from ablation, because the shocked ISM has a lower velocity than the
unshocked ISM.  Note also that in Model~D, the mixing of the ablated gas with the hot ISM is also
different from the mixing in Models~B and C: the fast-moving ISM constrains the ablated material so
that it remains close to the cloud. This gas mixes and cools as it moves along the cloud
periphery. The mixed gas is so closely constrained to the edge of the cloud that it cannot be
clearly seen in the temperature plots (Figure~\ref{fig:CloudEvolutionDifferences}), but high
ions are abundant along the periphery of the cloud where the mixed gas exists (see
Figure~\ref{fig:ModelDOVI}).

\subsubsection{The Effect of the Cloud Density Profile (Models~C and E) and the Cloud Density (Models~B and G)}
\label{subsubsec:EffectOfDensity}

Models~C and E have the same initial conditions, apart from the initial density profile of the
cloud: in Model~C, the cloud density decreases smoothly at the edge of the cloud until it equals the
ambient density, whereas in Model~E the cloud has a uniform density and a sharp edge (see
Figure~\ref{fig:CloudProfile}). Models~E and C allow us to compare the results for a uniform,
sharp-edged density profile with those for a more realistic density profile. Note that similar
comparisons have been made in previous works \citep{vieser07a,heitsch09}.

The first two panels for Model~C in Figure~\ref{fig:CloudEvolutionDifferences} show that a tail of
low-temperature gas propagates upward immediately behind the cloud, whereas Model~E has not
developed such a tail at early times. These tails are composed of the low-density gas that was
initially at the edge of cloud in Model~C, due to the smooth transition between the cloud density
and the ISM density at the cloud edge. Similar tails are also seen in the other models at early
times, except in Model~E. The low-density gas at the edge of the Model~C cloud is ablated via shear
instabilities more quickly than the high-density gas at the edge of the Model~E cloud. This ablated
material accumulates near the $r=0$ axis at early times after mixing with the ISM and radiatively
cooling, and moves upward along this axis, forming the tail. Cool gas does eventually start
accumulating along the $r=0$ axis in Model~E, but at later times (see the third Model~E panel in
Figure~\ref{fig:CloudEvolutionDifferences}, corresponding to $t = 90~\Myr$).

Models~B and G have the same initial conditions, except for the cloud and ISM densities being an
order of magnitude lower in Model~G. The Model~G cloud generally evolves in a similar fashion to the
Model~B cloud; this is because the ratio of densities across the cloud-ISM interface are the same in
both models, and so the growth of perturbations due to the shear instability should occur at the
same rate \citep[Section~101]{chandrasekhar61}. However, the lower density in Model~G reduces the
cooling rates, so the mixed gas cools more slowly.  As a result, very little of the ablated gas that
accumulates along the $r=0$ axis in Model~G is cool.

As mentioned at the end of Section~\ref{subsec:LossOfHI}, the detailed differences in the
hydrodynamical evolution of the cloud between Models~C and E as well as between Models~B and G do
not affect $\beta$ significantly (Figure~\ref{fig:HIloss}), indicating that the rate at which the
cloud loses its \HI\ does not strongly depend on either the initial density profile or the initial
density of the cloud, provided that the density contrast between the cloud and the ISM is the same.

\subsubsection{The Effect of the Cloud Size (Models~A, B, and F)}
\label{subsubsec:EffectOfSize}

Models~A, B, and F have the same initial conditions, apart from the size of the cloud; in these
three models, the radii of the initial clouds are $\approx$20, $\approx$150 and $\approx$300~\pc,
respectively (see Figure~\ref{fig:CloudProfile}). As these clouds all have the same initial
velocity ($-100~\kmps$), they should all be subject to the same physical processes.

As noted at the end of Section~\ref{subsec:LossOfHI}, a smaller cloud loses its \HI\ content more
rapidly than a larger cloud. In order to better understand this trend, we consider the simple case
of a uniform spherical cloud losing mass at a rate proportional to its surface area, i.e., $dM(t) /
dt \propto -4 \pi r^2(t)$. In this case, the radius decreases linearly with time, $r(t) = r_0 - kt$,
where $r_0$ is the initial radius of the cloud and $k$ is the rate at which the radius
decreases. Because the cloud in this simple model is uniform and spherical, $\nHI(t) \propto
r^3(t)$, and so $\beta(t) = 1 - (1 - kt/r_0)^3$ (from Equation~\ref{eq:beta}).

The smooth thick light gray curves in Figure~\ref{fig:HIloss} show $\beta(t)$ calculated according
to this simple model with $k = 0.1~\pc~\Myr^{-1}$ and $r_0 = 20$, 150, and 300~\pc\ (top to
bottom). This simple model represents the loss of \HI\ from the cloud due to all processes, and so
should be compared with the colored (HVC-like \HI) curves; as previously noted, these curves
represent \HI\ lost to ablation and to ionization. Note that this model is not an accurate physical
model -- Figures~\ref{fig:CloudEvolution} and \ref{fig:CloudEvolutionDifferences} show that the
cloud does not remain spherical during its evolution. As a result, we did not fit the thick light
gray curves in Figure~\ref{fig:HIloss} to the corresponding curves derived from the hydrodynamical
simulations.  Nevertheless, this simple model can provide some insight into the relative behavior of
Models~A, B, and F.

The curve derived from this simple model for $r_0 = 20~\pc$ is in reasonably good agreement with the
Model~A curve. For Models~B and F, the simple model overestimates the mass-loss at earlier times,
and for Model~F it slightly underestimates the mass-loss at later times. Despite these shortcomings,
this simple model indicates that a major reason that a smaller cloud loses its mass more rapidly is
because it has a larger surface area relative to its mass.

If the clouds evolve according to this simple model, with the same value of $k$ for all clouds, then
the Model~A cloud will be at a similar phase in its evolution (i.e., at the same value of $r(t) /
r_0$) at $t = 16~\Myr$ as the Model~B cloud at $t = 120~\Myr$ and as the Model~F cloud at $t =
240~\Myr$. We find that the morphology of the cloud in all three models varies in a similar fashion
according to this timescale; i.e., the Model~A, B, and F clouds have similar shapes at $t=16$,
$t=120$, and $t=240~\Myr$, respectively. To see this, compare the first Model~A panel in
Figure~\ref{fig:CloudEvolutionDifferences} ($t = 15~\Myr$), the final temperature panel in
Figure~\ref{fig:CloudEvolution} (Model~B at $t = 120~\Myr$), and the final Model~F panel in
Figure~\ref{fig:CloudEvolutionDifferences} ($t = 240~\Myr$). Therefore, the Model~A cloud after
$t=16~\Myr$ (second through fourth Model~A panels in Figure~\ref{fig:CloudEvolutionDifferences}) is
at a later phase in its evolution than the Model~B cloud at $t = 120~\Myr$ and the Model~F cloud at
$t = 240~\Myr$.

At a given time $t$, although a smaller cloud will have lost more of its \HI\ mass relative to its
initial mass than a larger cloud, the above simple model predicts that the larger cloud will have
lost more mass. Our simulations bear out this prediction -- the total mass lost from the Model~F
cloud is larger than the mass lost from the other, smaller model clouds. Furthermore, if the ablated
material were to mix and cool at the same rate in all models, then the fraction of the ablated gas
that is cool would be the same in all models, regardless of the cloud's initial size. This would
result in the ratio of \betaall\ to \betaHVC\ being the same in all models. However, Model~F, which
has the largest cloud, yields larger ratios than the other models; i.e., the difference between the
black and colored lines for Model~F in Figure~\ref{fig:HIloss} is larger than for the other
models. This probably indicates that more radiatively cooling takes place in the ablated material in
Model~F than in the models with smaller clouds.

\subsection{The Fate of High-Velocity Clouds}
\label{subsec:FateOfHVCs}

\citet{heitsch09} modeled HVCs of various sizes (initial \HI\ mass, $\MHIinit =
10^{3.1}$--$10^{4.6}$~\Msol), and found that HVCs with $\MHIinit < 10^{4.5}~\Msol$ will lose
all their \HI\ after traveling $<$$10~\kpc$. Thus, smaller HVCs are unlikely to reach the disk as
neutral hydrogen; nor will larger HVC complexes, if they are in fact composed of small
cloudlets. However, it is possible that the HVC material could still reach the disk in the form of
warm ionized material \citep{heitsch09,shull09,blandhawthorn09}.

Our suite of models include one cloud (Model~F) with $\MHIinit = 10^{5.5}~\Msol$,\footnote{Note that
  the total cloud masses in Table~\ref{tab:ModelParameters} have to be divided by 1.4 to give the
  \HI\ masses.} which is larger than the masses of the clouds simulated by \citet{heitsch09}.  The
Model~F cloud has $\betaHVC \approx 0.3$ at the end of the simulation ($t = 240~\Myr$); i.e.,
$\approx$70\%\ of the cloud's initial high-velocity \HI\ mass remains at $T < 10^4~\K$ at this time,
although it is possible that some of this high-velocity material has broken off the main cloud. The
cloud will travel much further before completely dissipating -- the value of \betaHVC\ from the end
of the Model~A simulation and the scaling discussed in Section~\ref{subsubsec:EffectOfSize} imply
that $\sim$30\%\ of the Model~F cloud's initial high-velocity \HI\ will remain below $10^4~\K$ at $t
\sim 900~\Myr$.

As well as our simulating a more massive cloud, there are other differences between our and
\citeauthor{heitsch09}'s simulations: we used a 2D geometry and a cooling curve calculated with
solar abundances, whereas \citeauthor{heitsch09} carried out 3D simulations with 1/10-solar
abundances. Both of these differences would tend to stabilize our clouds against disruption
(Section~\ref{sec:NeglectedPhysics}), leading to longer-lived clouds in our simulations.  However,
comparing similar-sized clouds in our and \citeauthor{heitsch09}'s simulations indicates that the
cloud lifetimes agree within a factor of $\sim$3--5. Therefore, even taking this into account, our
Model~F simulations indicate that very large clouds ($\ga$$10^{5.5}~\Msol$) will live for at least a
few hundred megayears, and travel a few tens of kiloparsecs (assuming a speed of
$\sim$100~\kmps). Such large cloud masses are not implausible: e.g., masses of $\ga$$2 \times
10^6~\Msol$, $\sim$$10^7~\Msol$, and $\sim$$10^7~\Msol$ has been measured for the Smith Cloud
\citep{nichols09}, Complex~C \citep{wakker07,thom08}, and Complex~H \citep{lockman03}, respectively
(although the complexes are not single clouds). The largest HVCs may therefore survive as far
as the Galactic disk at least partially as neutral hydrogen.

\subsection{High Ions}
\label{subsec:HighStageIons}

In our simulations, the gas that ablates from the cold cloud and mixes with the hot ISM is rich in
high ions (\CIV, \NV, and \OVI).  In this section we investigate the properties of these high ions
in more detail. First, we estimate the quantities of high ions that are produced via ablation and
ionization of the cloud material, and then we discuss the variation of column density with radius
for each high ion; these column densities can be directly compared with observations.

\begin{figure}
\centering
\plotone{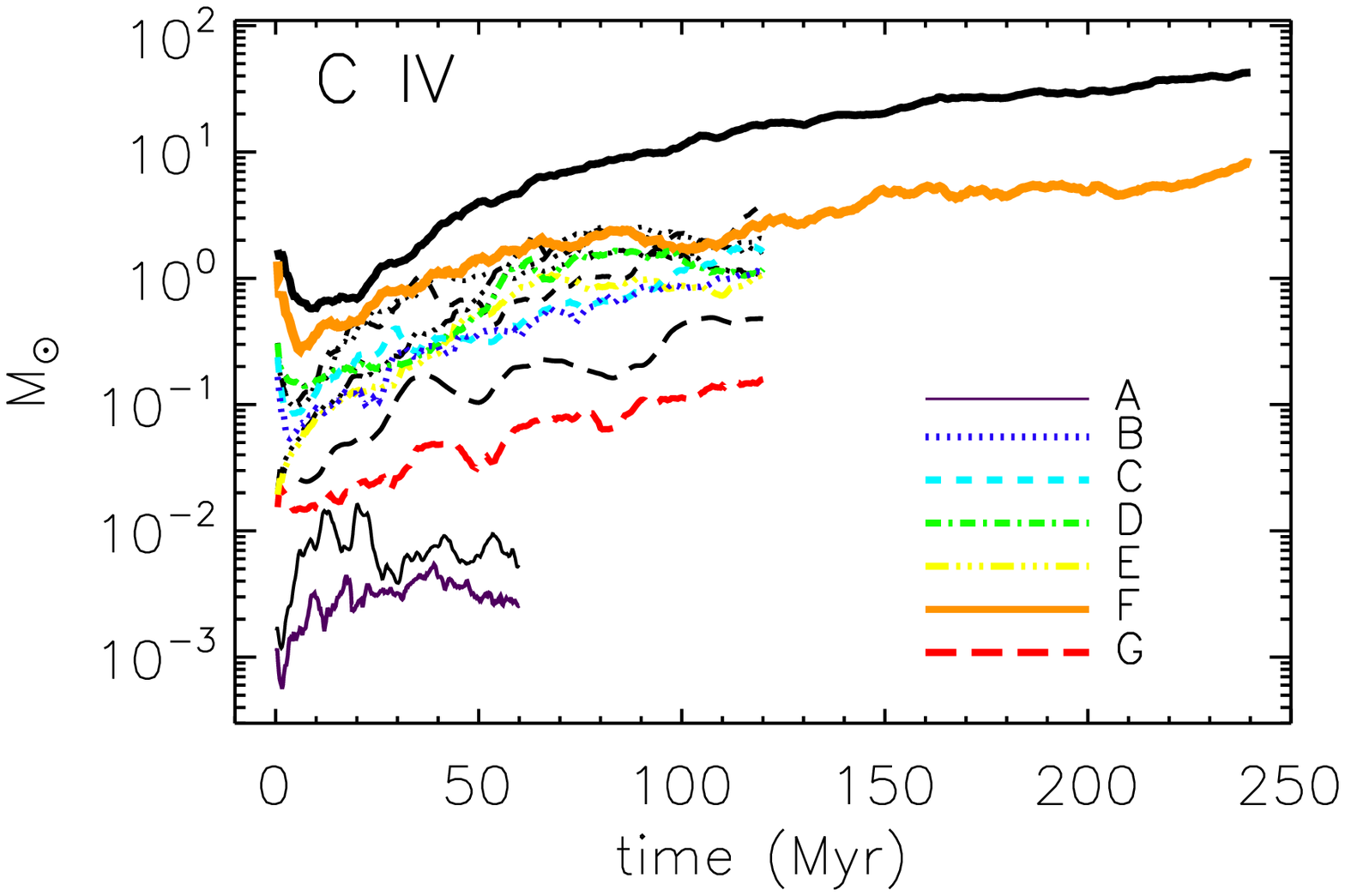}
\plotone{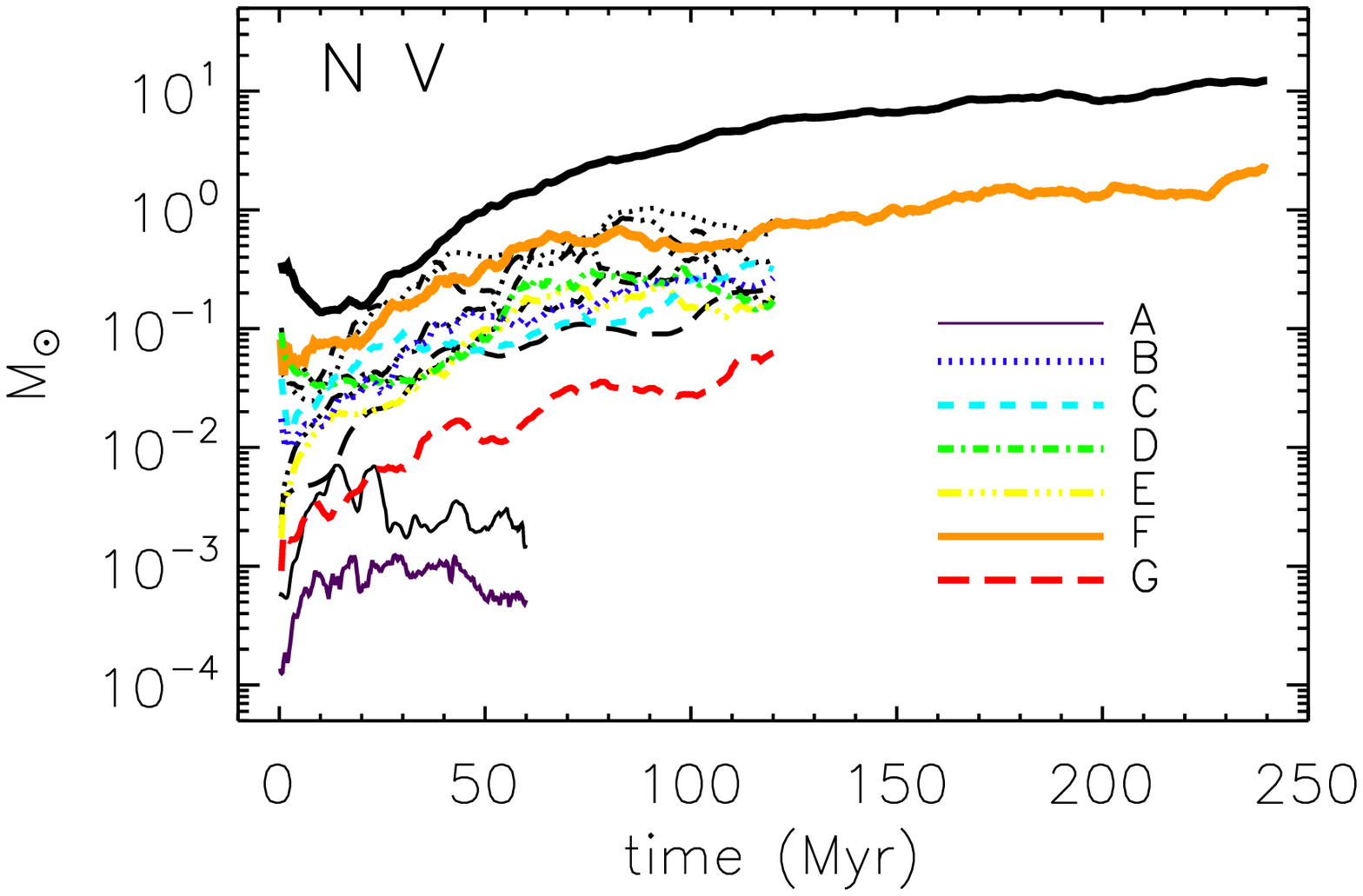}
\plotone{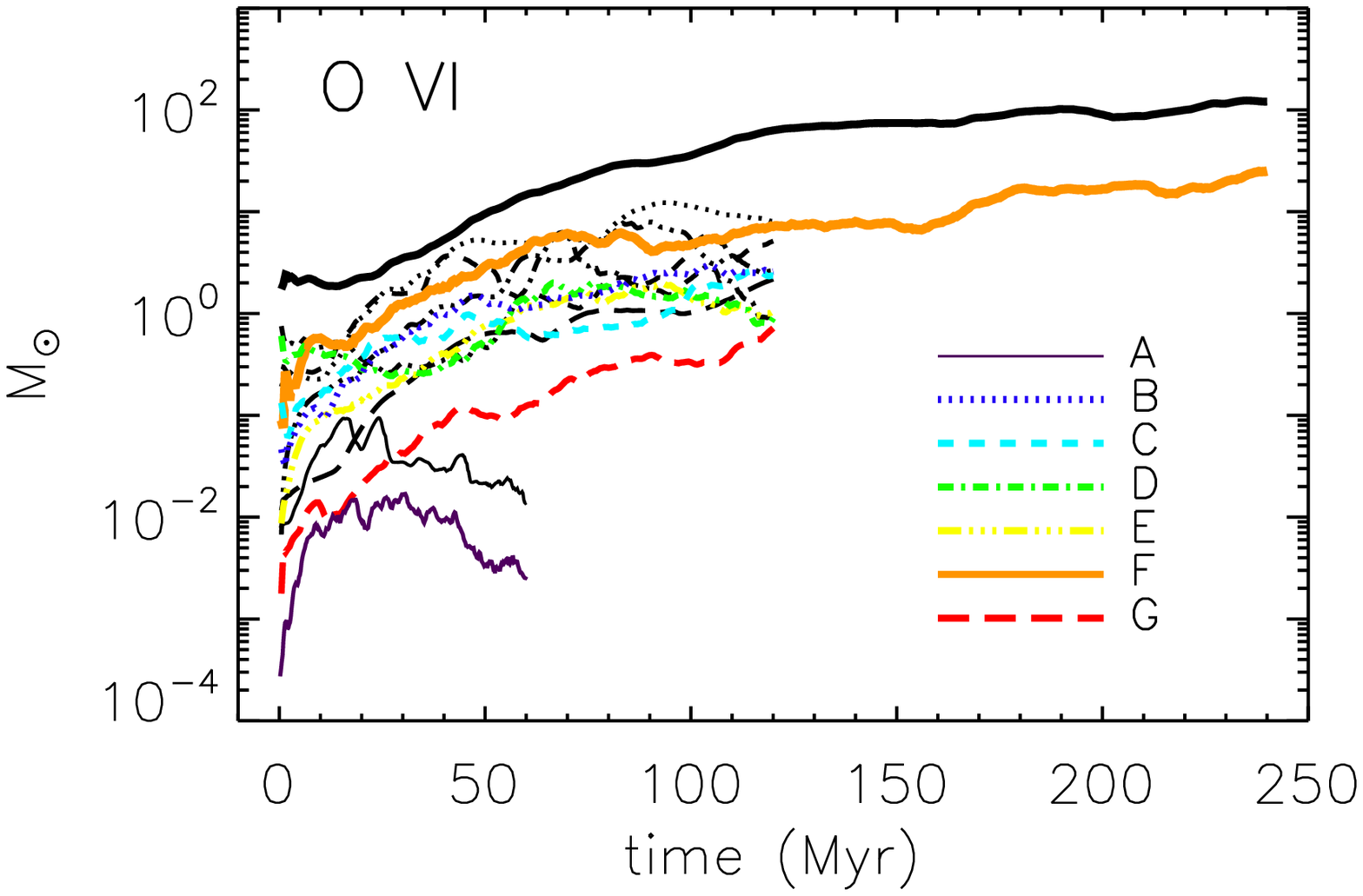}
\caption{Masses of \CIV\ (top), \NV\ (middle), and \OVI\ (bottom) as a function of time from our
  various model simulations. The black lines show the total mass of each ion in the simulational
  domain, and the colored lines show the masses of the ions moving with HVC-like velocities. The same
  line styles are used for the various models as in Figure~\ref{fig:HIloss}.
  \label{fig:IonMass}}
\end{figure}

Figure~\ref{fig:IonMass} shows the masses of the high ions in our simulational domains as
functions of time. In each case we calculate two values: the total mass of a given ion in the
simulational domain, regardless of velocity (black lines), and the mass of a given ion moving with
HVC-like velocities (colored lines). When calculating the total mass of a given ion, integrated over
all velocities, we subtract off the mass of that ion that was contained in the hot ambient ISM at
$t=0$. Our values do not include the ions that have escaped from the domain. This means that the
total masses of the high ions are actually lower limits. However, as with the amount of
HVC-like \HI\ (Section~\ref{subsec:LossOfHI}), the masses of the HVC-like ions are not significantly
affected by our neglecting material that has flowed off the domain, because such material has a low
density and a low velocity in the observer's frame. Note from Figure~\ref{fig:IonMass} that the
masses of the high ions with HVC-like velocities are smaller than the corresponding total
masses integrated over all velocities. This is because the gas that ablates from the cloud
decelerates as it mixes with the ambient gas. Eventually, the mixed gas slows to halo-like
velocities, but still contains high ions.

In Models~B through G, the masses of the high ions generally increase with time, although
there are fluctuations on small timescales. Because the high ions are abundant in the material
ablated from the cloud, the time evolution of their masses is similar to that of the lost \HI\ in
two distinct ways. Firstly, similar amounts of high ions are produced regardless of the
cloud's initial velocity (compare Models~B, C, and D; cf.\ Section~\ref{subsubsec:EffectOfVelocity}) or
density profile (compare Models~C and E; cf.\ Section~\ref{subsubsec:EffectOfDensity}). Typically, the
amounts of high ions produced in Models~B through E agree within a factor of 2. Secondly, more
high ions are produced from the larger cloud than from the smaller cloud (compare Models~A, B, and F;
cf.\ Section~\ref{subsubsec:EffectOfSize}). The density of the Model~G cloud is 1/10 that of the
Model~B cloud, and so the masses of the high ions produced in Model~G are commensurately
lower.  Unlike the other models, Model~A shows a decrease in the mass of high ions at later
times. This occurs because Model~A is sampling a much later phase of the cloud evolution (see
Section~\ref{subsubsec:EffectOfSize}), such that by the end of the Model~A simulation, the cloud is
mostly destroyed and so provides fresh cool gas for mixing with the ISM more slowly.

\begin{figure*}
\centering
\plottwo{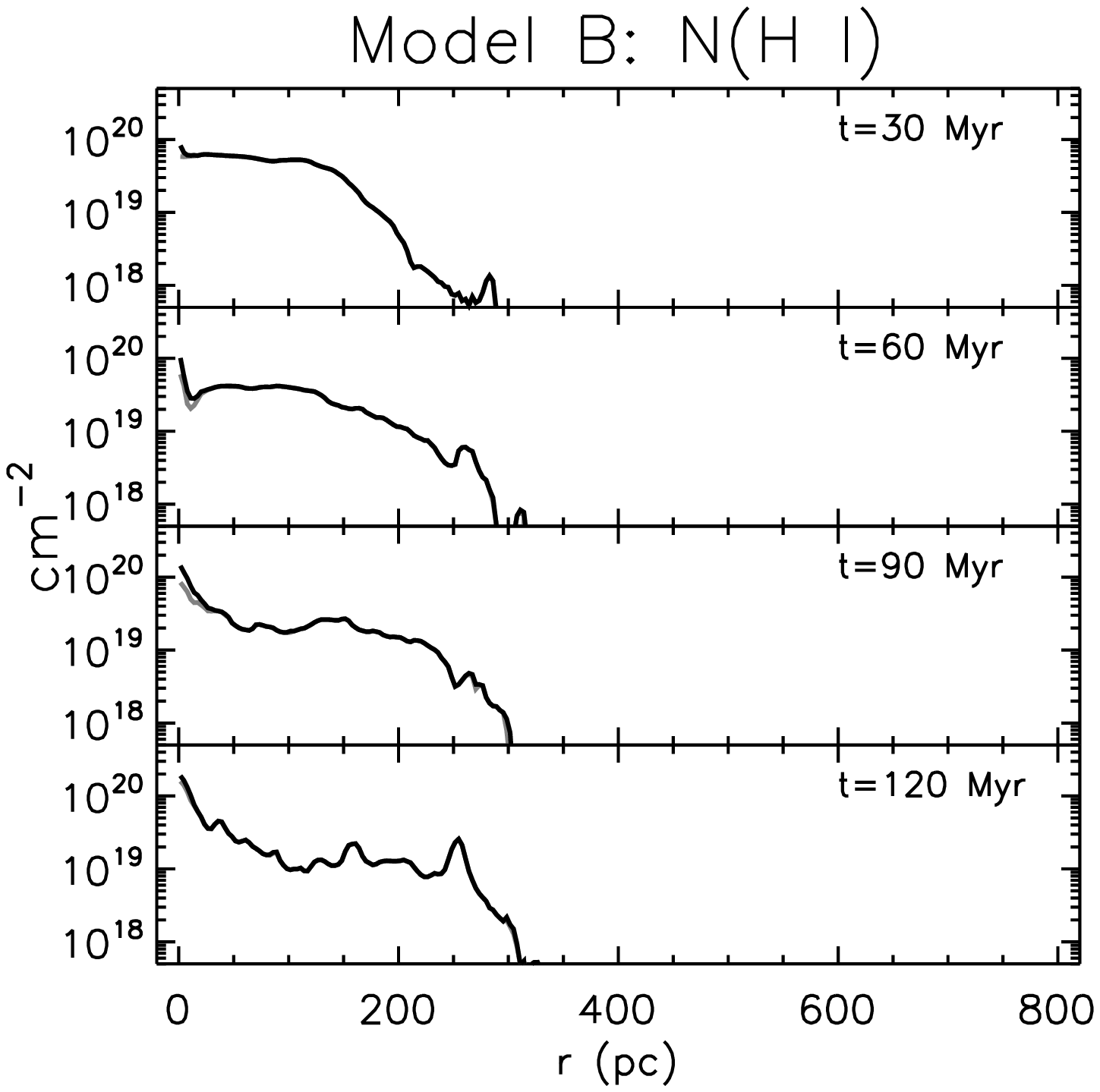}{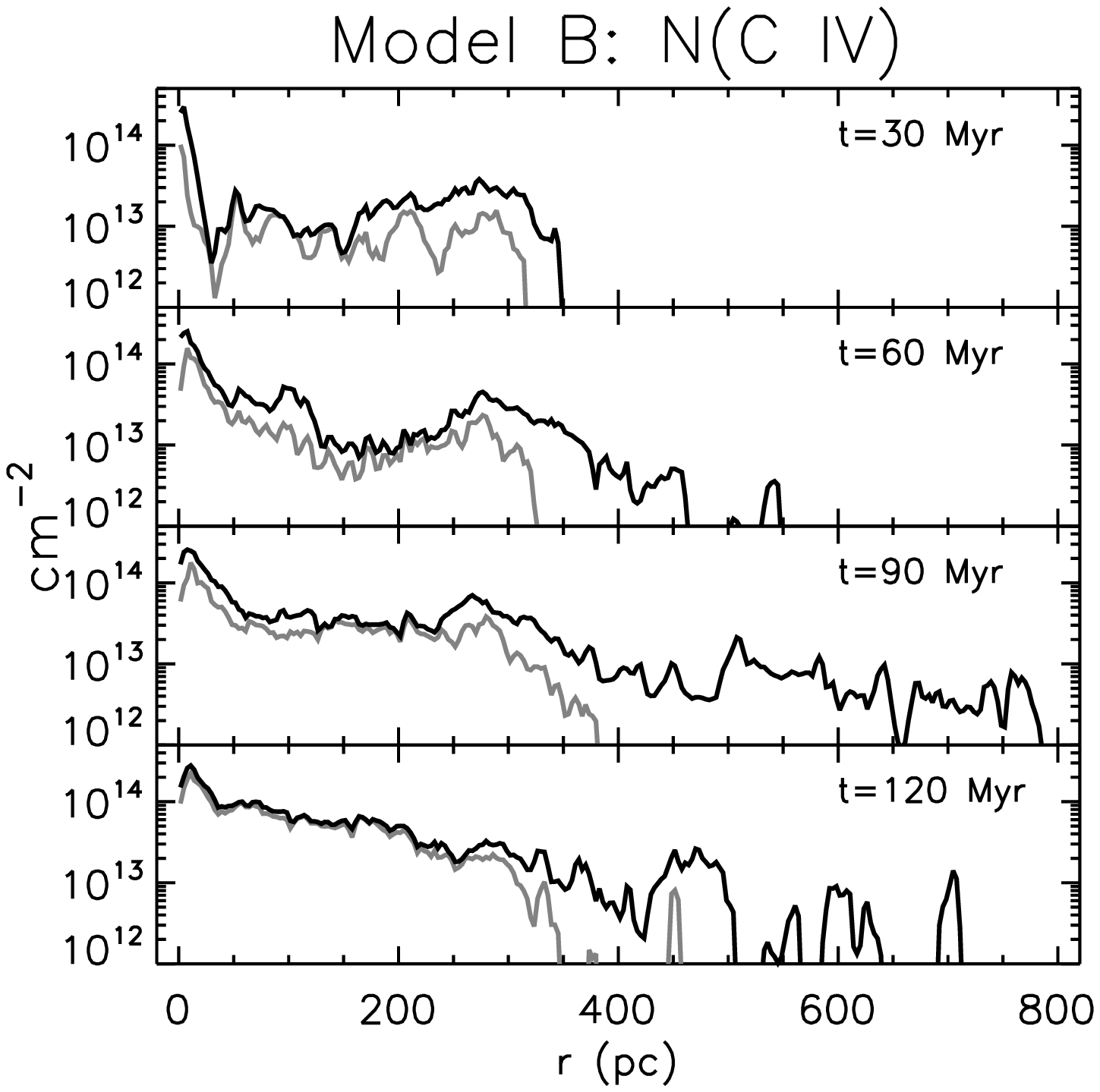} \\
\plottwo{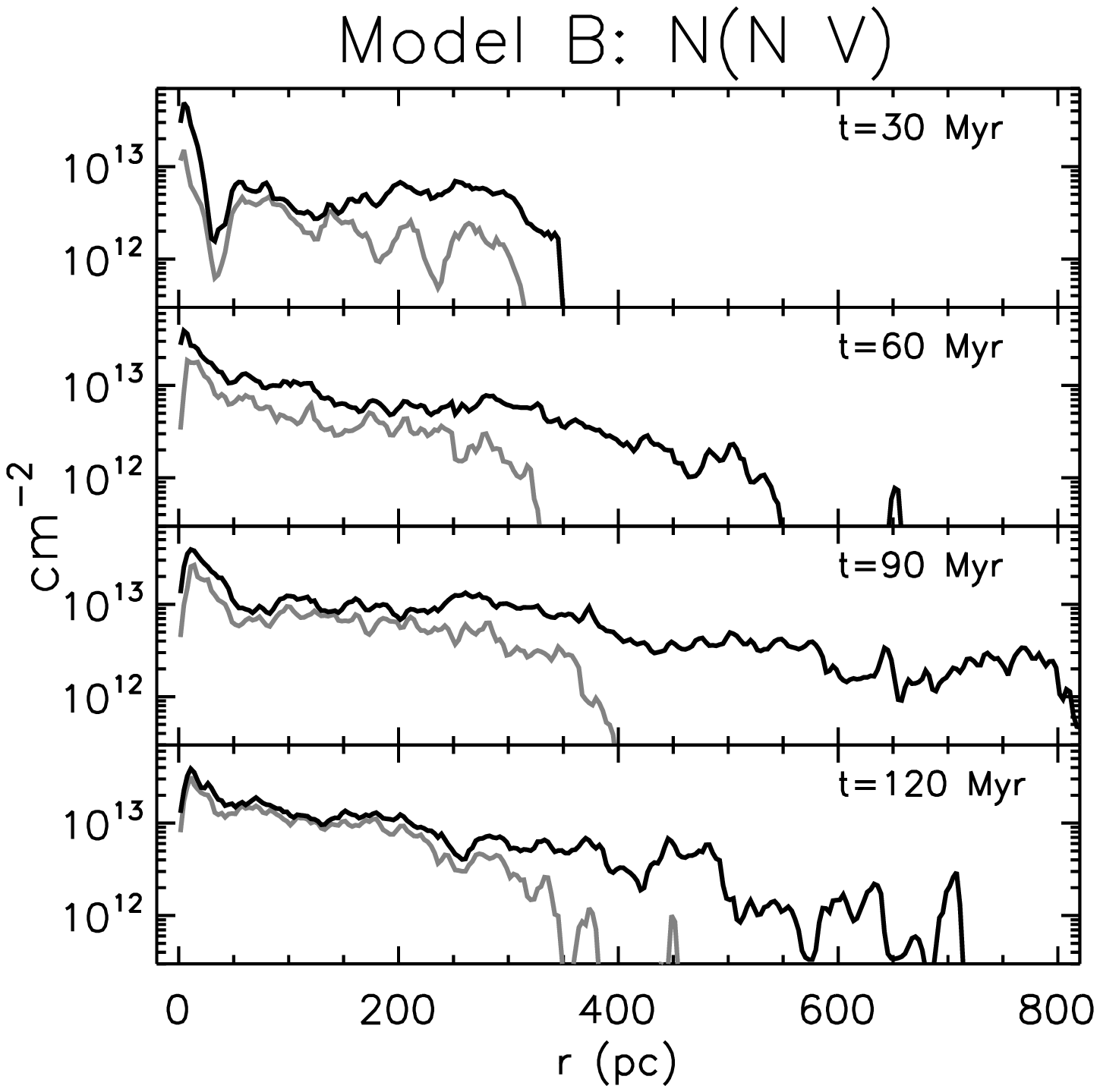}{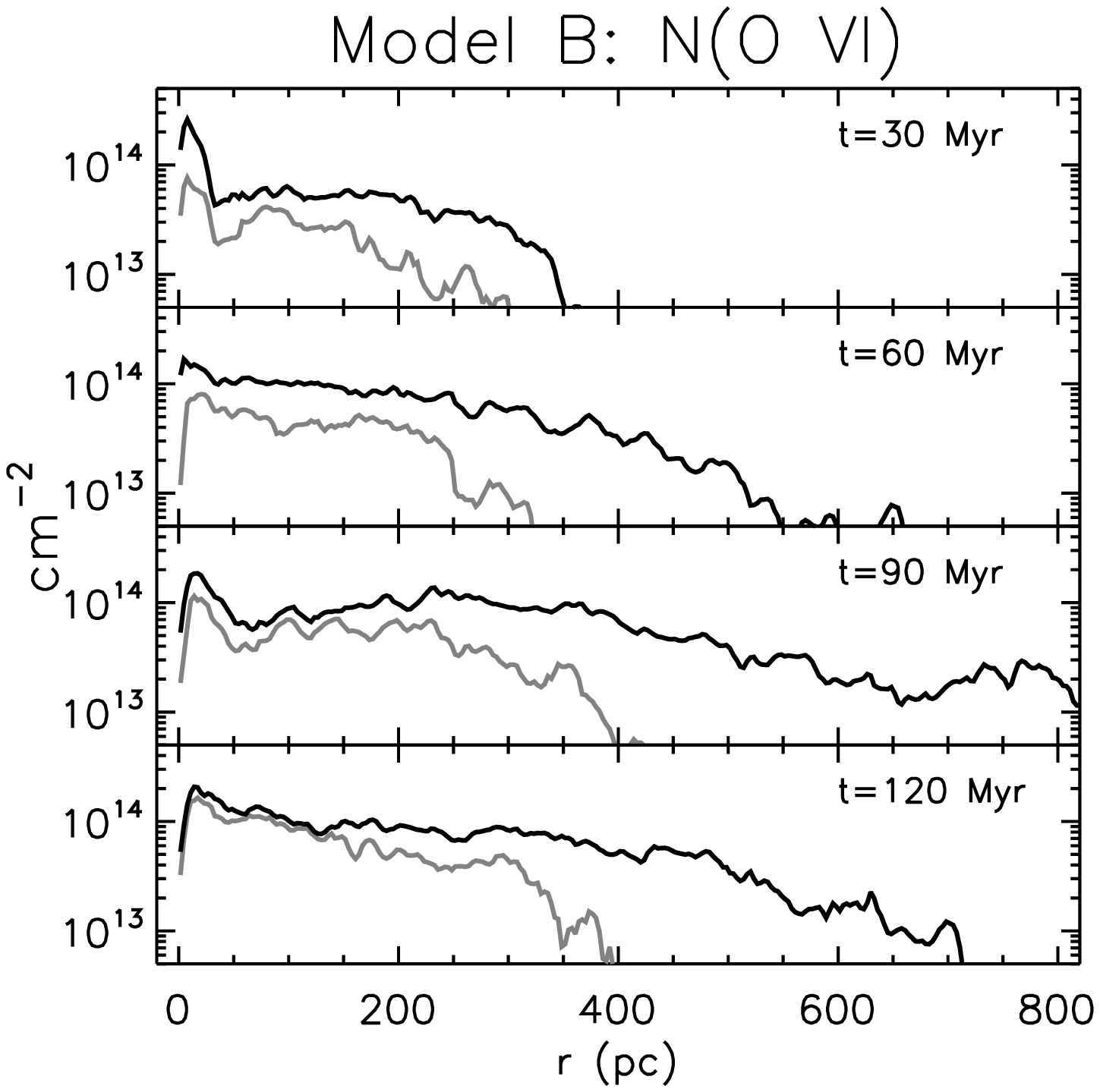}
\caption{Column density as a function of radius for \HI\ (top left), \CIV\ (top right), \NV\ (bottom
  left), and \OVI\ (bottom right), from Model~B.  In each plot, the panels correspond to $t=30$, 60,
  90, and 120~\Myr\ (top to bottom). The black lines are the column densities integrated over all
  velocities, and the gray lines for the column densities for material with HVC-like
  velocities. Note that for \HI\ (top left), these two lines are almost identical.
  \label{fig:ModelBColumn}}
\end{figure*}

\begin{figure}
\centering
\plotone{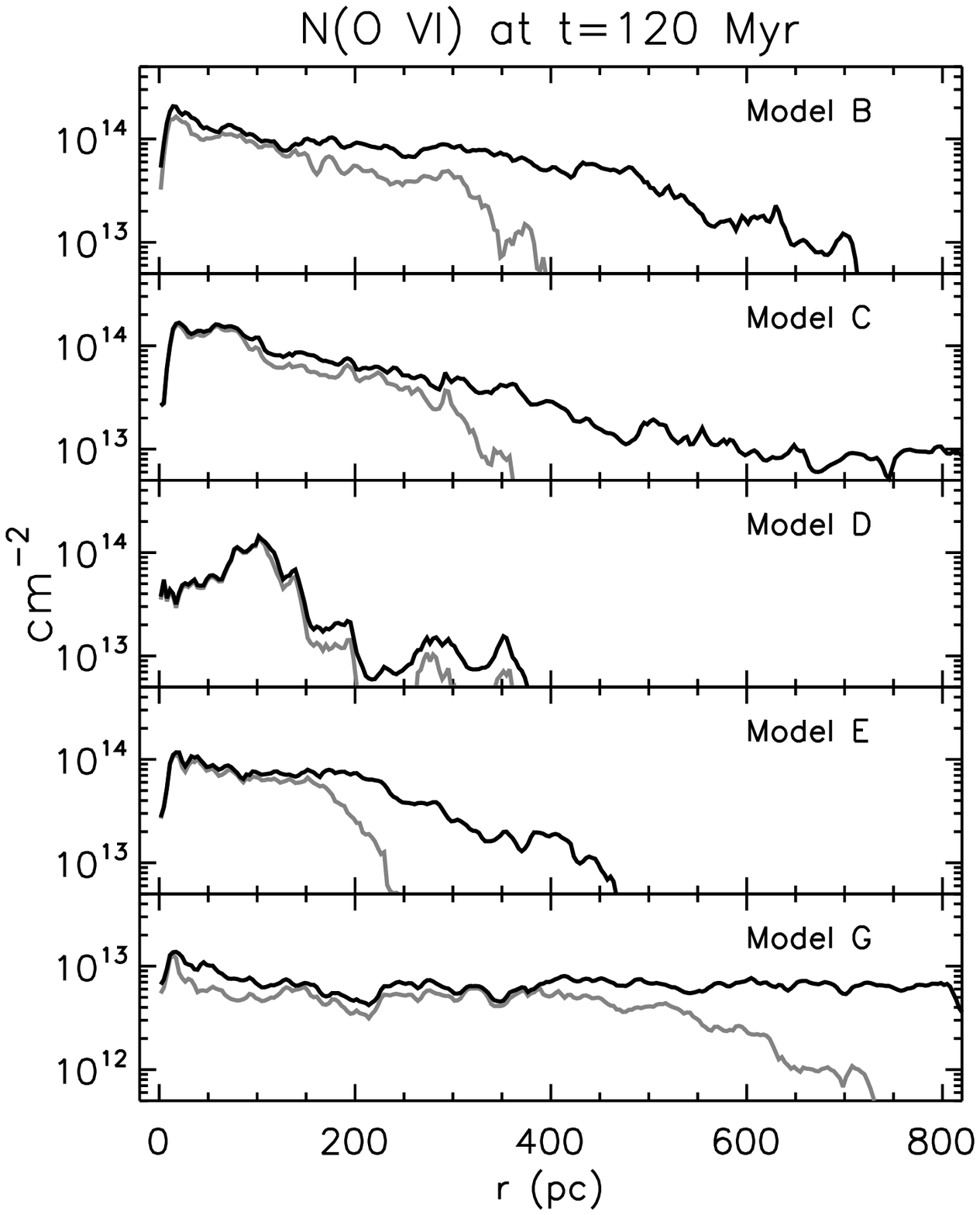}\\
\plotone{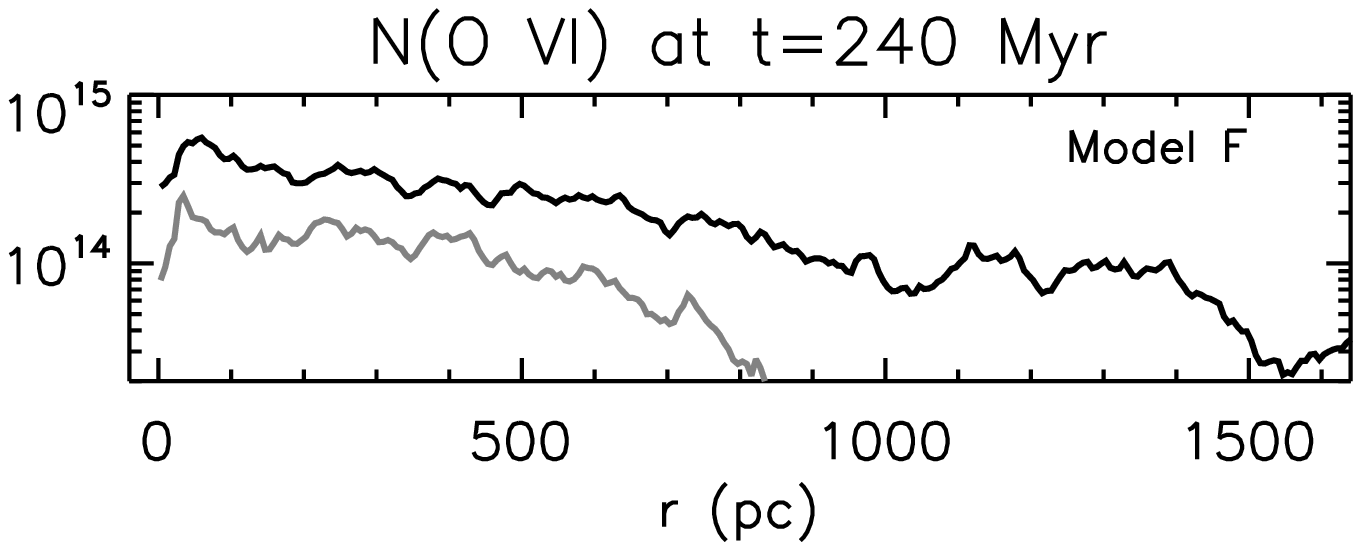}
\caption{\OVI\ column density as a function of radius from Models~B, C, D, E, G, and F (top to bottom).
  All plots are for $t=120~\Myr$, except for the Model~F plot, which is for $t=240~\Myr$.
  Note the larger $r$ axis range for Model~F. As in Figure~\ref{fig:ModelBColumn},
  the black lines show the total column densities, regardless of velocity, and the gray lines show
  the column densities of HVC-like material.
  \label{fig:ModelC-GColumn}}
\end{figure}

\begin{figure}
\centering
\plotone{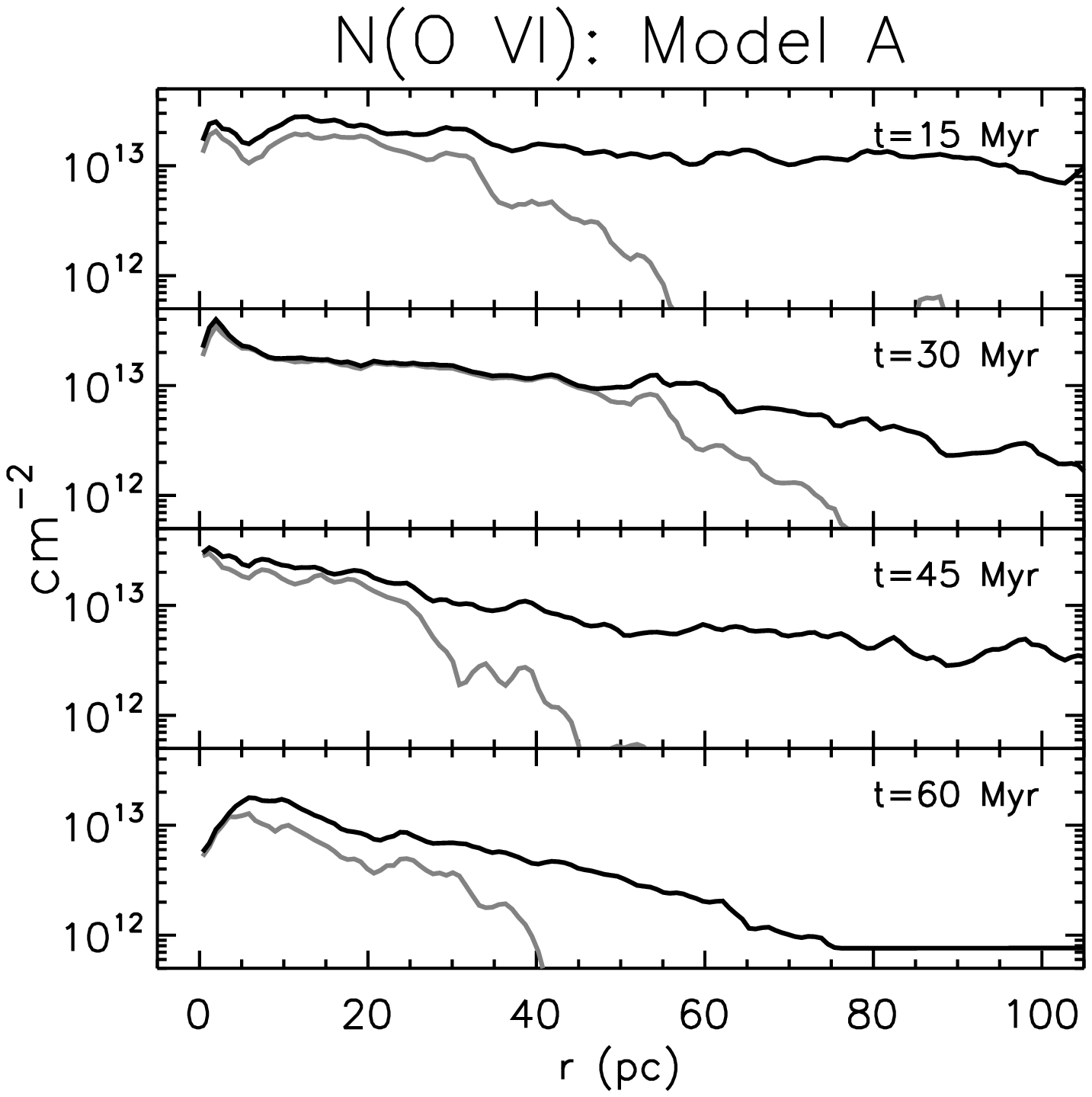}
\caption{\OVI\ column density as a function of radius from Model~A at $t=15$, 30, 45, and 60~Myr
  (from top to bottom; these are the same times as the Model~A panels in
  Figure~\ref{fig:CloudEvolutionDifferences}.)  As in Figure~\ref{fig:ModelBColumn}, the black lines
  show the total column densities, regardless of velocity, and the gray lines show the column
  densities of HVC-like material.
  \label{fig:ModelAColumn}}
\end{figure}

To investigate the spatial distributions of the high ions, we calculated column densities for
sightlines running vertically through the Model~B domain. Figure~\ref{fig:ModelBColumn} shows \HI,
\CIV, \NV, and \OVI\ column densities as functions of radius for $t=30$, 60, 90, and 120~\Myr. The
footprints of the high ion distributions increase with time, especially for the ions with slow
velocities. The column densities of the high ions generally increase with time while the \HI\ column
density decreases, because the cool \HI-rich gas is ionized and converted to high ion-bearing
gas. Most of the material containing high ions with HVC-like velocities was recently ablated from
the cloud and ionized to this level, which is why the footprints of the HVC-like high ions (gray
lines in Figure~\ref{fig:ModelBColumn}) are similar to the footprint of the \HI\ cloud.

The column densities of \HI\ calculated throughout the Model~B simulation are above the current
21-cm detection limit (a few times $10^{18}~\pcmsq$) over most of the cloud, so if the Model~B cloud
were a real cloud, it could be identified as an HVC throughout the whole simulation period. We find
that simulations with similar clouds (i.e., Models~C, D, and E) have similar \HI\ column densities
to Model~B. The \HI\ column densities for larger (Model~F), smaller (Model~A), or less dense
(Model~G) clouds than the Model~B cloud vary according to their size or density. Even the smaller
column densities (from Models~A and G) are above the current detection limit over much of the clouds
up to the ends of the simulations.

At the edges of the clouds, however, the \HI\ can become undetectable, while the high ions remain
detectable. For example, in Model~B at $t = 120~\Myr$, the \HI\ column density falls below the 21-cm
detection limit beyond $r \approx 300~\pc$, but there are substantial amounts of \OVI\ (up to
$\sim$$10^{13.5}~\pcmsq$) out to larger radii (see Figure~\ref{fig:ModelBColumn}).  Also, it should
be noted that the column densities plotted in Figure~\ref{fig:ModelBColumn} are for vertical
sightlines through our model domains. The high-velocity high ions reside mainly behind the
main body of the cloud, in material that has ablated from the cloud, mixed with the ambient medium,
and fallen behind the cloud. Therefore, diagonal sightlines through the model domain could intersect
significant column densities of high ions, while missing most of the \HI. Our simulation results
could therefore partially explain the presence of highly ionized high-velocity gas on sightlines
without corresponding high-velocity \HI\ \citep{sembach03}. Our results offer only a partial explanation
because they account only for sightlines with high-velocity high ions that have high-velocity \HI\
nearby; some of the high-velocity \OVI\ detections in \citet{sembach03} are on sightlines without any
high-velocity \HI\ within several degrees.

We noted above that similar amounts of high ions are produced in Models~C, D, and E as in
Model~B (Figure~\ref{fig:IonMass}). The top four panels in Figure~\ref{fig:ModelC-GColumn} show that
the \OVI\ column densities from Models~C, D, and E are similar to those from Model~B at $t = 120~\Myr$,
although the footprints vary from model to model. In contrast, the Model~G \OVI\ column densities (5th panel
in Figure~\ref{fig:ModelC-GColumn}) are about an order of magnitude smaller than those from Model~B,
which is commensurate with the density difference between these two models.

As was noted in Section~\ref{subsubsec:EffectOfSize}, Model~A is mostly in a later evolutionary
phase than the other models. From Figure~\ref{fig:ModelAColumn}, we see that the \OVI\ column
densities in this model peak around $t = 30~\Myr$ and subsequently decrease (see also the bottom
panel in Figure~\ref{fig:IonMass}).  When we compare Models~A, B, and F at similar stages of their
evolution ($t=15$, 120, and 240~\Myr, respectively; see Section~\ref{subsubsec:EffectOfSize}), we
find that the \OVI\ column densities with HVC-like velocities from Models~A and F are factors of
$\sim$5 smaller and $\sim$1.5 larger than those from Model~B, respectively ($\sim$$2 \times
10^{13}$, $\sim$$1 \times 10^{14}$, and $\sim$$1.5 \times 10^{14}~\pcmsq$ for Models~A, B, and F,
respectively). The ratio of the \OVI\ column densities from Models~A, B, and F (approximately
0.2:1:1.5) is similar to but not equal to the ratio of initial radii, 0.13:1:2.

In the following section, we compare our column density predictions with observations of
Complex~C. In particular, we will consider the ratios of the high ions to \HI\ and to each
other. Such ratios predicted by different models are frequently compared with the observed ratios in
order to identify the physical process(es) by which the high ions are produced.

\section{COMPARISON WITH OBSERVATIONS OF COMPLEX~C}
\label{sec:ComparisonWithObservations}

Among the several HVC complexes, Complex~C has been studied most extensively, perhaps because of its
location in the northern hemisphere and its large size on the sky ($\sim$1800~deg$^2$;
\citealt{wakker91}). Its distance is reasonably well constrained ($10\pm2.5~\kpc$;
\citealt{thom08}), and it is of rather low metallicity ($\sim$0.13 solar;
\citealt{collins07}). Observations of high-velocity high ions along several sight lines passing
through the Complex~C region have been reported by \citet{sembach03}, \citet{fox04}, and
\citet{collins07}. In addition, Complex~C is thought to be colliding with the Milky Way's thick disk
\citep{tripp03}. In this section, we will compare our model predictions with observations of
Complex~C in several different ways. First, in Section~\ref{subsec:ComplexCColumns}, we compare the
ion column densities, $N$, predicted by our models with those measured for Complex~C
\citep{sembach03,fox04,collins07}. \citet{sembach03} looked for but did not find a correlation
between the Complex~C \OVI\ and \HI\ column densities. We discuss this lack of correlation in terms
of our model predictions in Section~\ref{subsec:NOVIversusNHI}. Finally, in
Section~\ref{subsec:ColumnOverview} we calculate the ratios of column densities averaged over the
whole of the model cloud, and we compare these ratios with observations in
Sections~\ref{subsec:ColumnIonToHI} and \ref{subsec:ColumnIonToIon}.

\subsection{Ion Column Densities}
\label{subsec:ComplexCColumns}

\citet{sembach03} searched for high-velocity \OVI\ in the survey of high-latitude \textit{Far
  Ultraviolet Spectroscopic Explorer} (\fuse) observations \citep{wakker03}. They detected a total
of 84 high-velocity \OVI\ absorption features on 59 out of the 102 sightlines in the survey, spread
over the high-latitude sky.  The median \OVI\ column density of these 84 detected features is $9.3 \times
10^{13}~\pcmsq$, although there is significant sightline-to-sightline variation (between
$\sim$$10^{13}$ and $\sim$$3 \times 10^{14}~\pcmsq$; \citealt{sembach03}). For the 9 sightlines
toward Complex~C with detections, the median high-velocity \OVI\ column density is $7.6 \times
10^{13}~\pcmsq$ (range: $(\mbox{4.7--16.6}) \times 10^{13}~\pcmsq$; \citealt{sembach03}).  In
Model~B, the column densities of HVC-like \OVI\ approach the median of the observed column densities
at some times for some impact parameters, but are generally lower than the observed
median. Furthermore, along some sight lines toward Complex~C, there are measurements of \CIV\ and
\NV\ column densities \citep{fox04,collins07}. The observed high-velocity \CIV\ column density
ranges from $<$$6.9 \times 10^{12}$ to $6.5 \times 10^{13}~\pcmsq$ while that of \NV\ ranges from
$<$$6.9 \times 10^{12}$ to $<$$3.1 \times 10^{13}~\pcmsq$ \citep{collins07}. For certain choices of
age and impact parameter, our Model~B results match the column densities at the higher end of the
observed ranges (e.g., $r \sim 150~\pc$ at $t=120~\Myr$ for \CIV\ and $r \sim 0~\pc$ at $t=120~\Myr$
for \NV) and at the lower end (e.g., $r \sim 150~\pc$ at $t=60~\Myr$ for \CIV\ and $r \ga 100~\pc$
at $t=60~\Myr$ for \NV).

\subsection{\NOVI\ versus \NHI}
\label{subsec:NOVIversusNHI}

\begin{figure}
\centering
\plotone{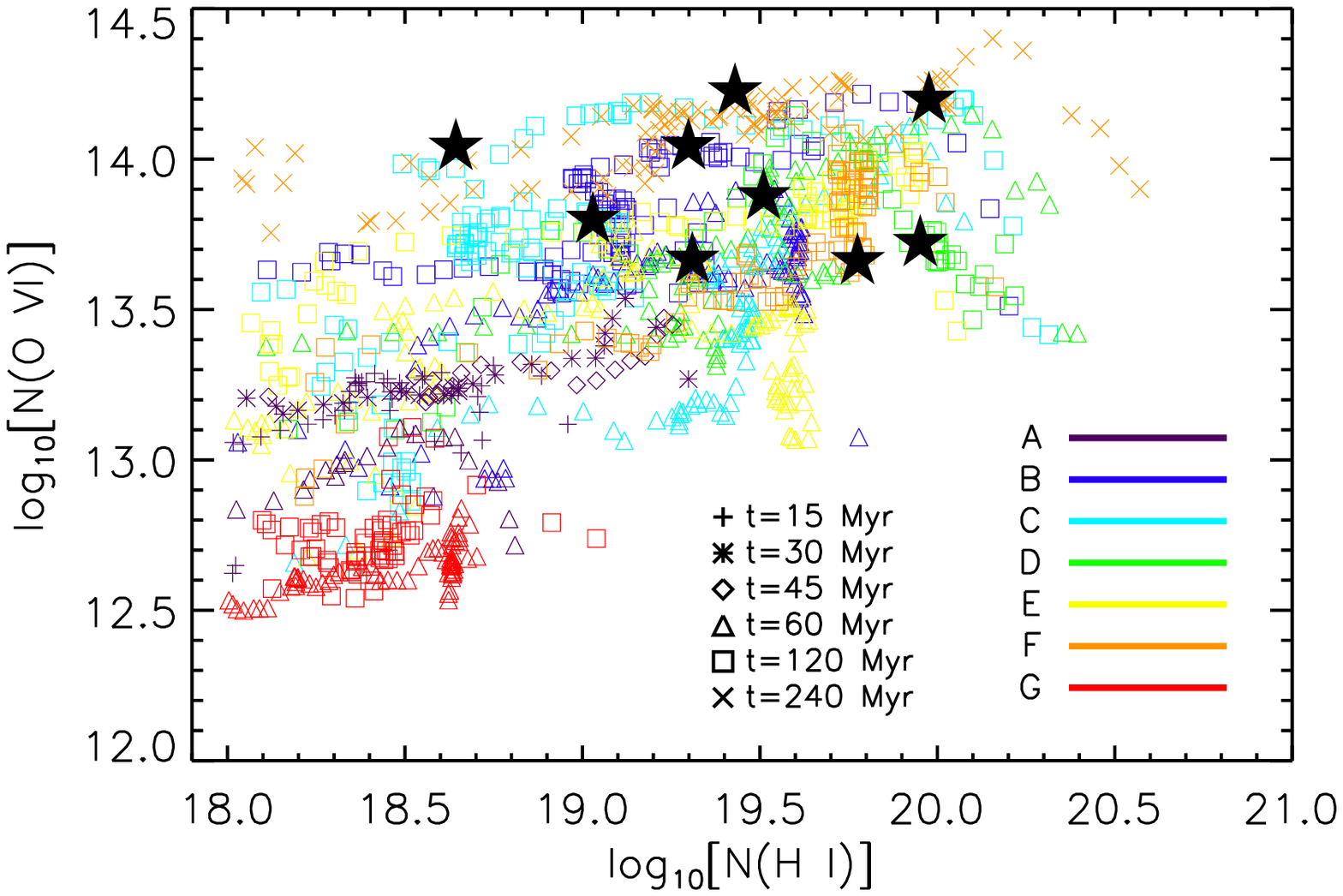}
\caption{Observed and predicted \OVI\ versus \HI\ column densities (plotted on logarithmic scales).
  The filled stars show the Complex~C observations (Table~8 from \citealt{sembach03}). The other
  symbols show predictions from our simulations for material with HVC-like velocities. The colors
  and symbols respectively denote the different models and different times (e.g., orange diagonal
  crosses denote Model~F at $t = 240~\Myr$). For each model and for each time, column densities were
  extracted for multiple vertical sightlines; the results for each sightline are plotted. Note that
  not all the times in the key were used for each model. We considered column densities of \HI\ and
  \OVI\ above $10^{18}$ and $10^{12}~\pcmsq$, respectively.
  \label{fig:NOVIversusNH}}
\end{figure}

The filled stars in Figure~\ref{fig:NOVIversusNH} show \NOVI\ against \NHI\ for 9 sightlines through
Complex~C (\citealt{sembach03}, Table~8; we followed \citeauthor{sembach03} and did not include the
\HI\ components in brackets in their table). As noted by \citeauthor{sembach03}, there is no
correlation between \NOVI\ and \NHI; \NOVI\ varies by $\approx$0.5~dex over Complex~C while
\NHI\ varies by $\approx$1.5~dex.  The other symbols in Figure~\ref{fig:NOVIversusNH} show the
predictions from our various models, extracted for multiple vertical sightlines through our model
domains at several different times. For any given model at a given time, the predictions for
different sightlines exhibit a similar trend to the observations; i.e., \NOVI\ is approximately
constant over a large range of \NHI. The trend in the predictions is because, in our simulations,
the \HI\ column density decreases significantly toward larger radii, whereas the \OVI\ column
densities remain more or less constant out to larger radii (see Figure~\ref{fig:ModelBColumn}). Not
only do the predictions exhibit the same trend as the observations, but also some of the predicted
values are in good agreement with the observations. However, not all of the observations can be
matched by a single model at a specific time. For example, 5 of the 9 observed data points (the four
with the largest values of \NOVI\ and the one with $\NHI \approx 10^{19.0}~\pcmsq$, $\NOVI \approx
10^{13.8}~\pcmsq$) are similar to values from Model~C at $t = 120~\Myr$ (cyan squares) and from
Model~F at $t = 240~\Myr$ (orange diagonal crosses). The remaining data points are similar to
predicted values at other times and/or from other models.

The predicted values from Models~A and G (small and low-density clouds, respectively) have \HI\ and
\OVI\ column densities that are smaller than the observed values for Complex~C, regardless of
simulation time and sightline. However, it is possible that Complex~C is composed of many small or
low-density clouds, each of which individually resembles a Model~A or G cloud. If this is the case,
the total column densities would be obtained by multiplying the Model~A or G predictions (which are
for a single cloud) by the number of clouds along the line of sight, resulting in a shift toward
larger values of \NHI\ and \NOVI. This means that the predictions for multiple Model~A or G clouds
would belong upwards and to the right of the Model~A or G points in Figure~\ref{fig:NOVIversusNH},
and therefore could become consistent with the observations. Note that, in order to carry out this
shift, we must assume that each cloud contributes the same amount of \OVI\ and \HI\ at the same
period in time.

The consistency between the observed and predicted values of \NOVI\ against \NHI\ supports the idea
that Complex~C has interacted with the hot ISM to produce \OVI. In particular, this consistency
confirms that the loss of cloud material due to shear instabilities and its subsequent mixing with
the hot ISM, subject to radiative cooling, is a viable physical process for the production of
\OVI. However, the timescale for Complex~C's interaction with the hot ISM is uncertain, as it is
dependent on the size and density of the constituent clouds -- a set of smaller clouds could produce
similar values of \NOVI\ and \NHI\ to a single larger cloud that has been evolving for a longer
time.

In the following subsections, we will look at the evolution of the column density ratios for each
model, and compare them with the Complex~C observations. This will enable us to place a rough
constraint on the timescale for Complex~C's interaction with the hot ISM.

\subsection{Column Density Ratios}
\label{subsec:ColumnOverview}

For comparison with the observed column density ratios, we calculated ratios of column densities
that had been averaged over the model clouds. We considered only sightlines for which the column
density was above some cut-off, \Ncutoff, so that material not significantly affected by the
cloud-ISM interaction is not included in the averages. For a given ion, the average column density,
$\bar{N}(t)$, at time $t$ is given by
\begin{equation}
  \bar{N}(t) = \frac { \int N(t,r) \alpha(t,r) r\,dr }{\int \alpha(t,r) r\,dr } ,
\end{equation}
where $N(t,r)$ is the column density for a sightline at radius $r$, and
\begin{equation}
  \alpha = \left\{
  \begin{array}{rl}
    1 & \mbox{if $N > \Ncutoff$;} \\
    0 & \mbox{otherwise.}
  \end{array}
  \right.
\end{equation}
For \HI, $\Ncutoff = 10^{16}~\pcmsq$ for Models~B through F and $10^{15}~\pcmsq$ for Models~A and
G. For the high ions, $\Ncutoff = 10^{11}~\pcmsq$ for Models~B through F and $10^{10}~\pcmsq$
for Models~A and G. We found that the choices of \Ncutoff\ did not significantly affect the average
column densities and the resulting ratios.\footnote{It is also possible to calculate averaged column
  density ratios by calculating the column density ratios for each sightline, and then averaging
  these ratios; i.e., the ratio for ions X and Y is given by $\overline{N(\mathrm{X}) /
    N(\mathrm{Y})}$, as opposed to $\bar{N}(\mathrm{X}) / \bar{N}(\mathrm{Y})$. We found that the
  two methods gave similar results.} We did not subtract off a background column density due to the
hot ambient medium: because the high ion fractions are small in this medium, the background
column density is negligible.

\begin{figure}
\centering
\plotone{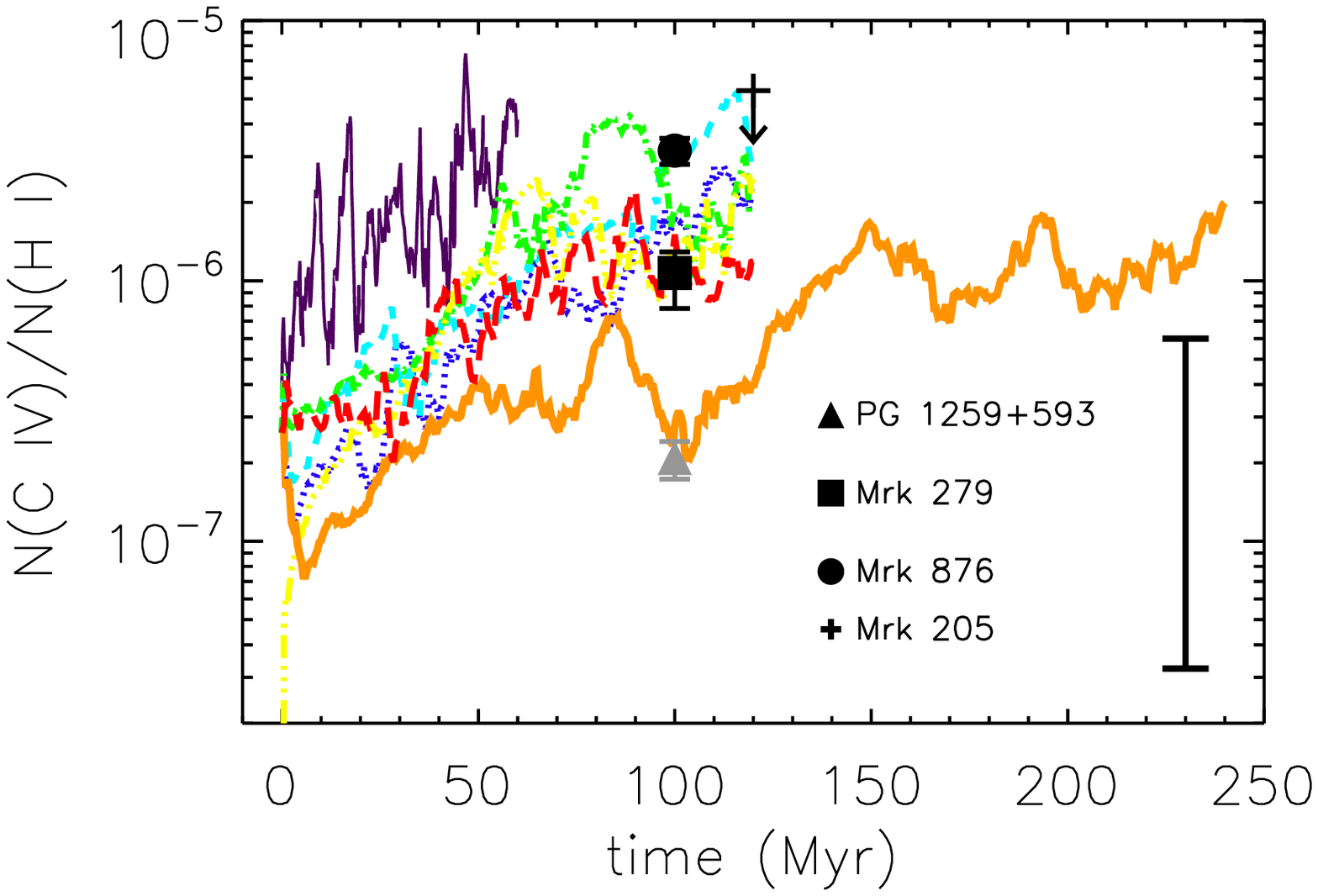} \\
\plotone{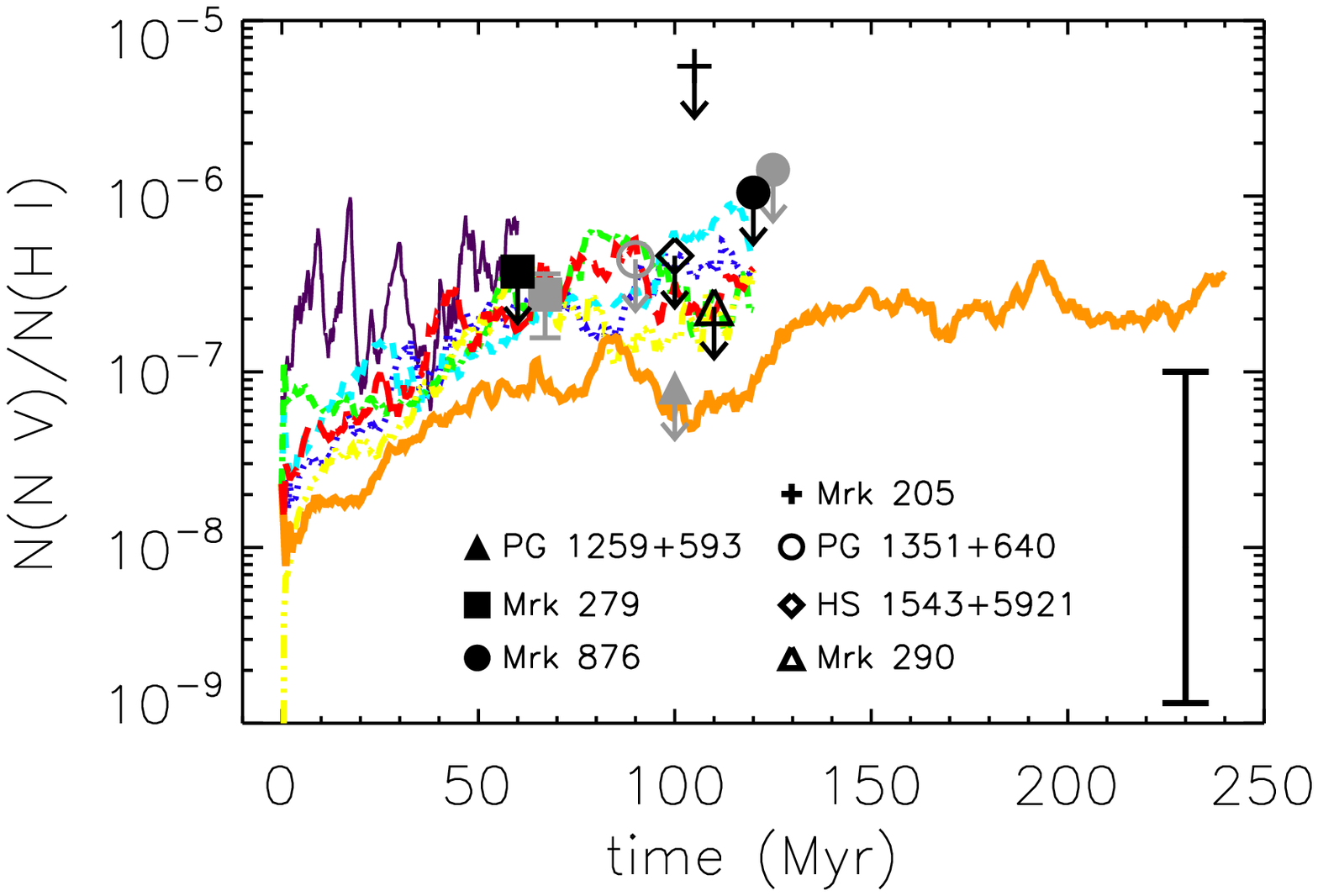} \\
\plotone{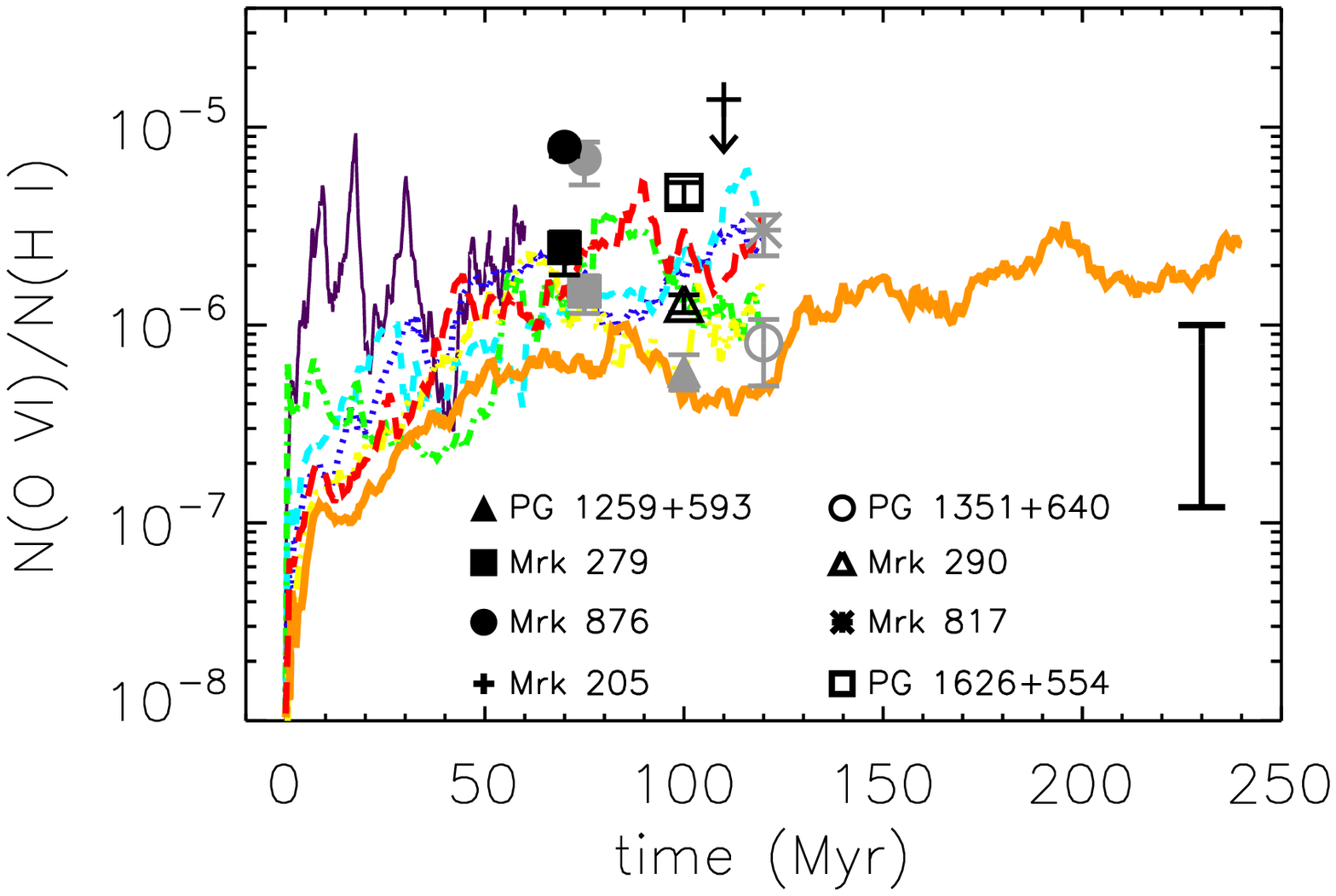}
\caption{Comparison of predicted high-ion-to-\HI\ column density ratios with Complex~C
  observations (top to bottom: \CIV, \NV, \OVI\ to \HI, respectively). The predicted ratios are
  calculated for material moving with HVC-like velocities using solar abundances \citep{allen73};
  the different line styles and colors correspond to the different models shown in
  Figures~\ref{fig:HIloss} and \ref{fig:IonMass}. The black vertical lines denote the amount the
  curves should be shifted downward to go from solar abundances to the Complex~C abundances measured
  along the PG~1259+593 sightline \citep{fox04}. The observed ratios for different sightlines are
  also plotted (note that these values are independent of time, and so their positions along the
  time axis are arbitrary). The black data points are from \citet{collins07}, and the gray data
  points from \citet{fox04}. We also plot the errors on the measurements, although in most cases the
  error bars are smaller than the plotting symbols. Upper limits are denoted by the
  downward-pointing arrows.
  \label{fig:IonToHIComplexC}}
\end{figure}

\begin{figure}
\centering
\plotone{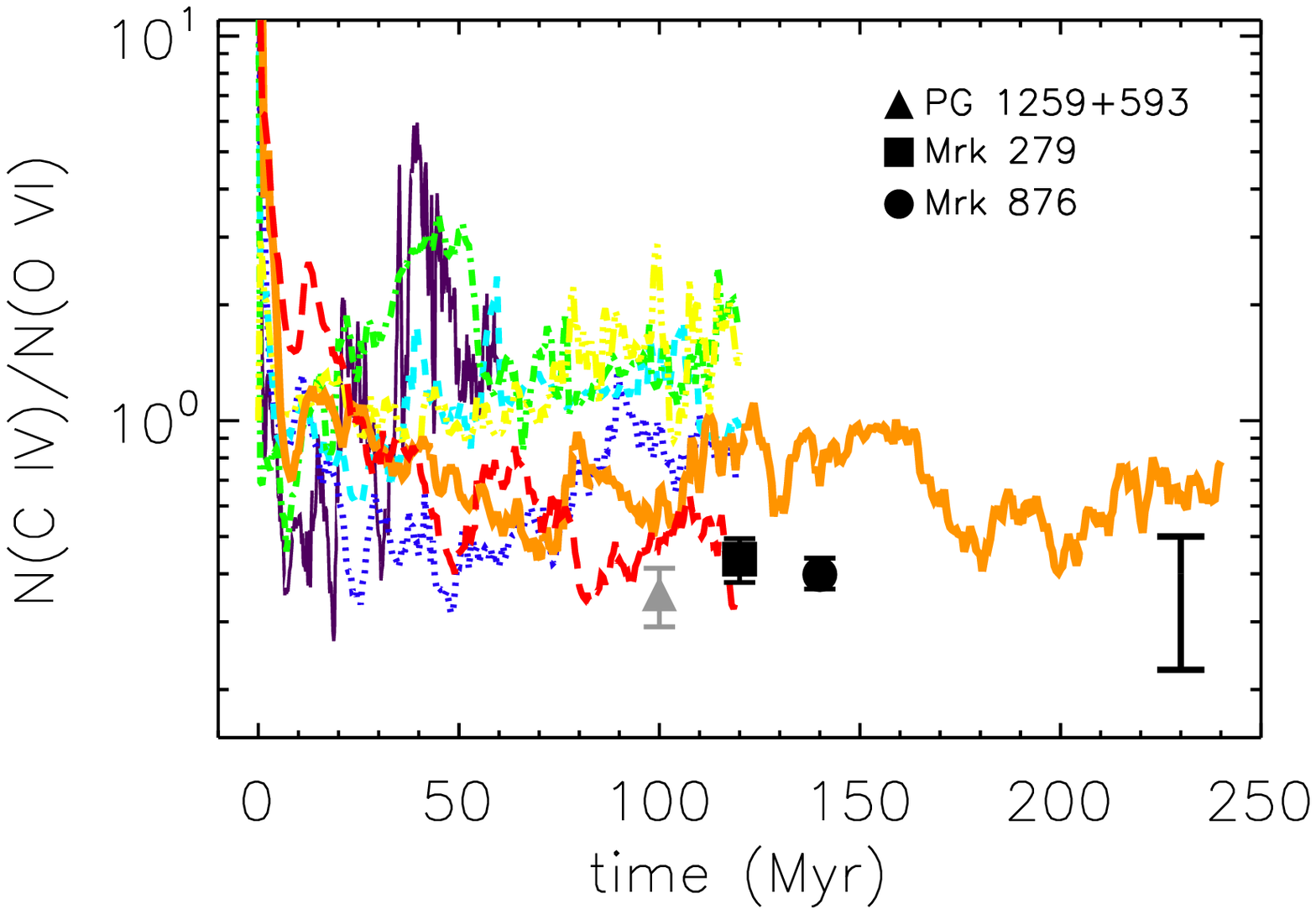}
\plotone{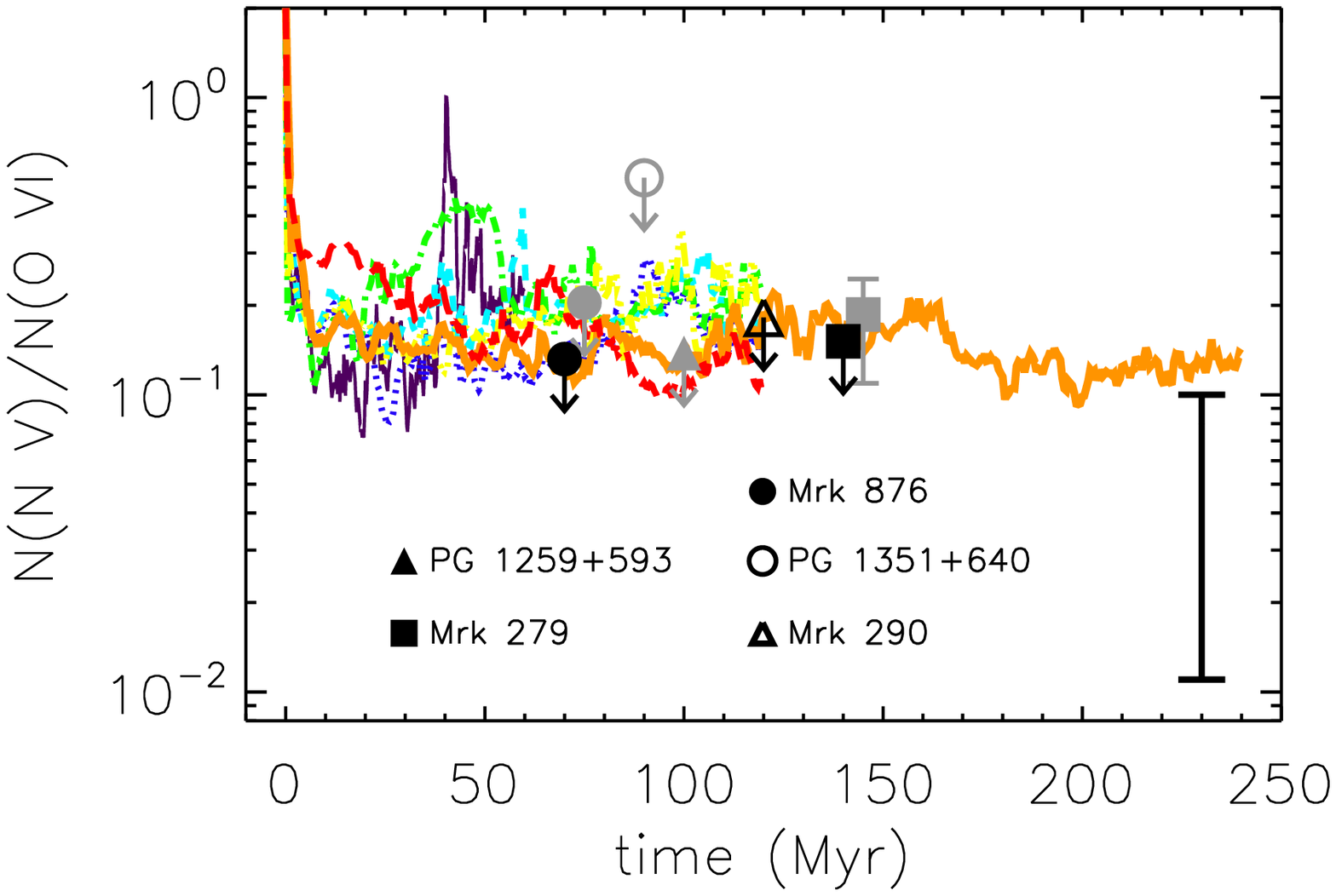}
\caption{As Figure~\ref{fig:IonToHIComplexC}, but showing \CIV\ to \OVI\ (top) and \NV\ to \OVI\ (bottom).
  \label{fig:IonToIonComplexC}}
\end{figure}

The evolution of the column density ratios calculated above is presented in
Figures~\ref{fig:IonToHIComplexC} (ion to \HI) and \ref{fig:IonToIonComplexC} (ion to ion). Because
we will compare the models with Complex~C observations, we plot only the ratios for HVC-like
material. These curves were calculated using solar abundances \citep{allen73}, but subsolar
abundances have been reported for Complex~C \citep{fox04,collins07}. We can adjust our predicted
ratios by shifting the curves in Figures~\ref{fig:IonToHIComplexC} and \ref{fig:IonToIonComplexC}
according to the ratios of the Complex~C abundances to solar abundances. The black vertical lines in
Figures~\ref{fig:IonToHIComplexC} and \ref{fig:IonToIonComplexC} indicate the amounts by which the
curves should be shifted downward to go from solar abundances to the abundances measured along the
PG~1259+593 sightline (\citealt{fox04}; see Table~8 in KS10 for the relevant multiplication
factors). However, it is important to note that the abundances measured along the PG~1259+593
sightline are the lowest among the Complex~C measurements \citep{collins07}, and so the shifts
indicated in Figures~\ref{fig:IonToHIComplexC} and \ref{fig:IonToIonComplexC} would be the most
extreme cases. It is also important to note that the Complex~C abundance measurements use low
ions, and thus may only be sampling cloud material, rather than the turbulently mixed material in
which the high ions are produced. The turbulently mixed material includes a contribution from
the ambient medium, which may have a different metallicity from the cloud. Note that nitrogen is
more depleted along the PG~1259+593 sightline than carbon and oxygen.

Figures~\ref{fig:IonToHIComplexC} and \ref{fig:IonToIonComplexC} also show the observed ratios for
Complex~C from \citet{fox04} and \citet{collins07}.  In some cases, these two studies give slightly
different column densities for the same sightline, due to differences in their measurement methods
(e.g., the determination of the continuum and the velocity range used for the column density
integration).  In such cases, we plot the measurements from both studies.  In the following two
subsections, we discuss the evolution of the ion-to-\HI\ and ion-to-ion ratios that we have plotted,
and compare these ratios with the Complex~C observations.

\subsection{Ion-to-\HI\ Column Density Ratios}
\label{subsec:ColumnIonToHI}

The predicted ratios of high ions to \HI\ generally increase over time, although large fluctuations
occur in some models at earlier times (Figure~\ref{fig:IonToHIComplexC}). The ratios are largest for
the smallest cloud (Model~A) and smallest for the largest cloud (Model~F) at a given time, and so an
observed ratio can correspond to a smaller cloud that has evolved for a short period of time, or a
larger cloud that has evolved for a longer period of time
(cf.\ Section~\ref{subsec:NOVIversusNHI}). Either interpretation would be possible without knowing
Complex~C's three-dimensional structure (i.e., whether Complex~C is composed of many small clumps,
fewer larger clumps, or a few large clumps mingled with many small clumps).

The observed ratios vary by more than an order of magnitude from sightline to sightline. The
variations in the predicted ratios, due to differences in the age or size of a cloud, are as large
as the observed variations.  The range of ratios predicted with solar abundances covers the observed
range. However, when we shift the predicted ratios according to the subsolar Complex~C abundances,
the models can only match the lower observed ratios.

Although there may be a few ways to interpret the observed ratios in terms of our models, we provide
the following interpretation as a possible example. The observed $\NNV/\NHI$ ratios are almost all
upper limits, and do not help to constrain the models. The $\NCIV/\NHI$ and $\NOVI/\NHI$ ratios
observed toward PG~1259+593 are similar to those predicted by Model~A at $t \sim \mbox{few} \times
10~\Myr$ or by Models~B through E at $t \sim 100~\Myr$, after the abundance shifts indicated by the
vertical bars in Figure~\ref{fig:IonToHIComplexC} have been applied. Nearly all the other sightlines
have larger observed $\NCIV/\NHI$ and $\NOVI/\NHI$ ratios than the PG~1259+593 sightline. As the
predicted ratios rise with time, the model predictions (after the abundance shift) will match these
other sightlines at even later times. As noted above, the abundance shifts indicated in
Figure~\ref{fig:IonToHIComplexC} are the most extreme cases, as the PG~1259+593 sightline has the
lowest abundances of the Complex~C sightlines. However, even the solar-abundance predictions from
some models (e.g., B through E) do not match the ratios observed along some sightlines until $t \sim
100~\Myr$. These results suggest that the models require at least $\sim$100~\Myr\ of evolution to
match the observed ratios, if the radii of the composite clouds were initially
$\ga$150~\pc\ (Models~B through E). As the model clouds' initial velocities were
100--300~\kmps\ (see Table~\ref{tab:ModelParameters}), this implies that Complex~C has moved
$\ga$10--30~\kpc\ from its initial location to its current location, suggesting that
Complex~C likely originated as extragalactic material. Note, however, that the measured metallicity
of Complex~C is too high for a purely extragalactic origin \citep{tripp03,collins03} and requires
a metal-enrichment process. In preliminary calculations, we have found that mixing with the Galactic
gas noticeably enhances the metallicity of the \HI\ HVC gas, but a full discussion is beyond the scope
of our current study.

As noted above, there is significant sightline-to-sightline variation in the observed ratios.
\citet{sembach03} reported a weak trend for the ratio of \OVI\ to \HI\ to increase toward lower
Galactic longitude for 9 sightlines through Complex~C. Complex~C is oriented diagonally on the
Galactic coordinate grid, running from low latitudes and longitudes to high latitudes and longitudes
\citep[e.g.,][]{wakker91}, with the higher-latitude, higher-longitude region of the complex being
higher above the disk \citep{thom08}. Hence, the trend reported by \citet{sembach03} means that the
ratio of \OVI\ to \HI\ increases toward lower Galactic latitude and lower height above the disk. The
ratio of \CIV\ to \HI\ also increases toward lower latitudes, although there are fewer data points for
this trend (the \NV\ measurements are mostly upper limits).

This variation of $\NOVI/\NHI$ with longitude was discussed by \citet{tripp03}, who suggested that
the lower-longitude, lower-latitude region of Complex~C is closer to the disk, where the ambient
medium is denser. They suggested that this greater density leads to a more vigorous interaction
between the ambient medium and the HVC, leading to greater production of \OVI\ and greater
ionization of \HI. Our simulations are unsuitable for directly examining this suggestion, as we have
not investigated the effect of an increasing ambient density on high ion production. An
increasing ambient density will lead to both greater ram pressure stripping of material, and
faster-growing shear instabilities, because of the lower density contrast between the cloud and
ambient medium \citep[Section~101]{chandrasekhar61}. Both of these effects should lead to greater
production of high ions.

An alternative explanation for the observed $\NOVI/\NHI$ trend is that the region of Complex~C
nearer the disk is at a later stage in its evolution, having passed through more ambient medium, and
so has a higher high-ion-to-\HI\ ratio (see Figure~\ref{fig:IonToHIComplexC}). However, the range of
heights spanned by Complex~C ($\sim$2~\kpc, using data from \citealt{thom08}) divided by a velocity
of $\sim$100~\kmps\ corresponds to a timescale of only $\sim$20~\Myr. From
Figure~\ref{fig:IonToHIComplexC}, we can see that this timescale is insufficient to explain the
range of observed ratios, which vary by more than an order of magnitude from sightline to sightline.
The trend could also be explained by the region of Complex~C toward lower longitudes and latitudes
being initially composed of smaller clouds than the higher-longitude, higher-latitude region, which
give rise to larger high-ion-to-\HI\ ratios (see Figure~\ref{fig:IonToHIComplexC}). However, there
is no obvious reason for the size of the clouds of which Complex~C was composed when it began
interacting with the ISM to vary systematically with position.

Another possible explanation for the observed $\NOVI/\NHI$ trend is that the turbulently mixed
material toward lower longitudes and latitudes has a higher metallicity. Such a difference in
metallicity could arise from the mixing of metal-poor Complex~C gas with the metal-rich ISM. If the
metallicity of the halo decreases with height and/or Galactocentric distance,\footnote{Abundances in
  the disk ISM decrease with Galactocentric distance \citep{shaver83,rudolph06}, but in the halo the
  situation is unclear. However, because fountains of material from the disk into the halo likely
  dominate the heating of the halo, especially the lower halo \citep[e.g.,][and references
    therein]{henley10b}, it seems not unlikely that the metallicity of the halo decreases with
  height and Galactocentric distance.} then the lower-longitude, lower-latitude region of Complex~C
will be interacting with higher-metallicity gas than the higher-longitude, higher-latitude
region. As a result, the turbulently mixed material toward lower longitudes and latitudes would have
higher abundances and thus higher ratios of high ions to \HI.  If this speculation is correct,
then variation in elemental abundances with latitude may be apparent from other ions. \OI, for
example, does \textit{not} show an anticorrelation between abundance and latitude (using data from
\citealt{collins07}), but these abundance measurements are probably probing undisturbed cloud
material rather than turbulently mixed material. More measurements of elemental (as opposed to
ionic) abundances for Complex~C are needed to test whether or not the abundances vary with latitude.

\subsection{Ion-to-Ion Column Density Ratios}
\label{subsec:ColumnIonToIon}

Unlike the ion-to-\HI\ ratios, the observed ratios of \CIV\ and \NV\ to \OVI\ do not vary greatly
from sightline to sightline, nor do the predicted ratios vary systematically with cloud size or
time. The observed \CIV-to-\OVI\ ratios are in good agreement with the predicted values; the
agreement is even better when the predicted ratios are shifted according to the Complex~C
abundances. For \NV\ to \OVI, all but the Mrk~279 measurement from \citet[filled gray square]{fox04}
are upper limits, and so do not strongly constrain the models. If the ratio of gas phase nitrogen to
oxygen is equal to the solar ratio, then our results are in good agreement with the observed column
density ratio. However, if the ratio of gas phase nitrogen to oxygen is about 1/10~solar, as it is
on the PG~1259+593 sight line \citep{fox04}, then our results are overabundant in \NV\ compared to
\OVI\ by about a factor of 10.  Note that PG~1259+593 is more than 10 degrees higher in Galactic
latitude than Mrk~279, so it is not implausible that the relative abundances along these sightlines
could differ.

\begin{figure*}
\centering
\plotone{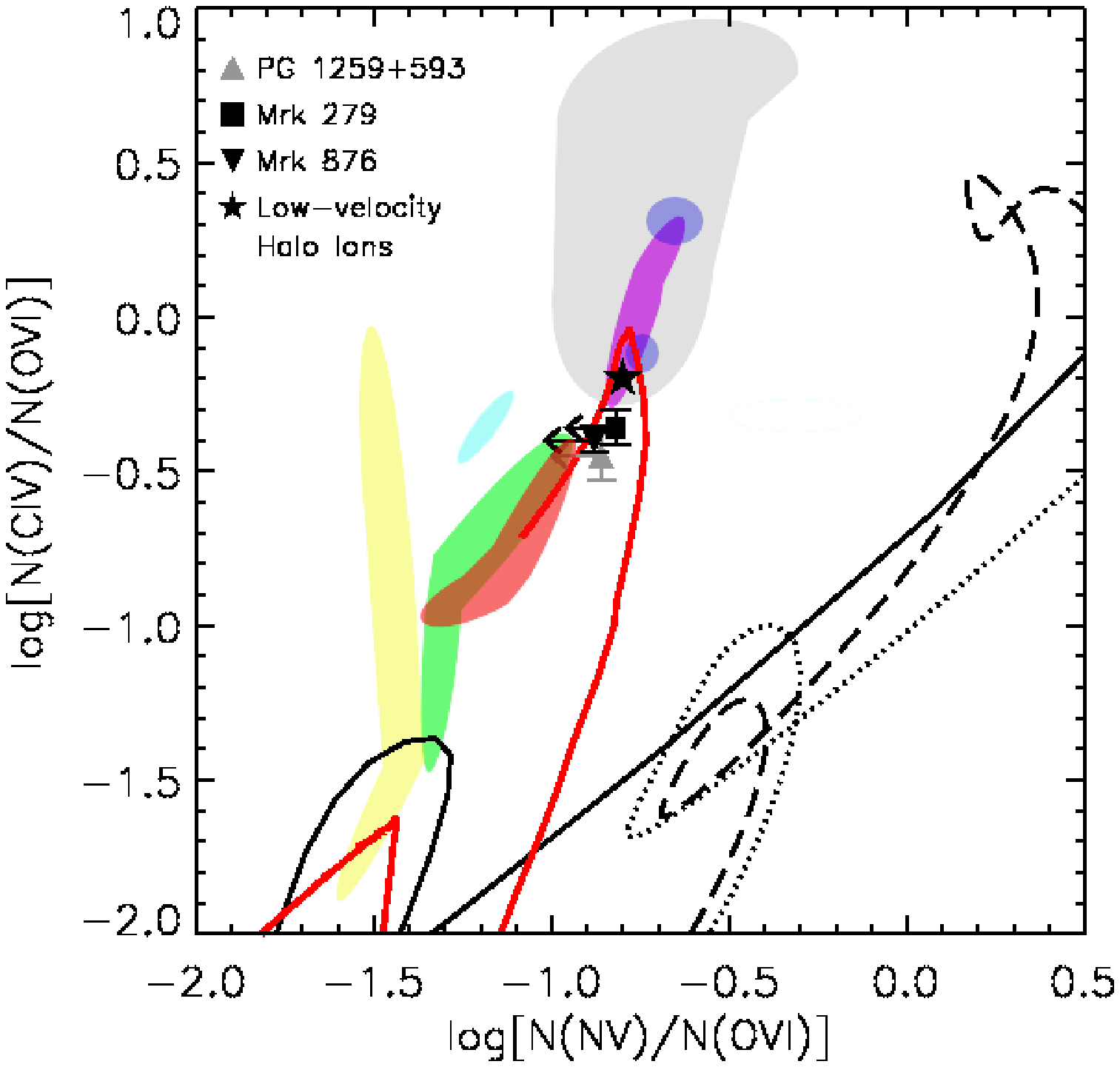}
\caption{Observed and predicted \CIV-to-\OVI\ column density ratios against
  \NV-to-\OVI\ column density ratios, plotted on logarithmic scales. The filled star near the center
  shows the average values for low-velocity high ions in the halo \citep{indebetouw04a}; these
  low-velocity ions will be discussed in Paper~II. The three solid symbols below the star show the
  observed ratios for three sightlines toward Complex~C, as indicated in the key
  (\citealt{collins07} for Mrk~876 and Mrk~279; \citealt{fox04} for PG~1259+593).  The predictions
  from various models are shown by the shaded regions and the lines, as follows.
    \textit{Gray region}: The TML models of \citet{slavin93a} and \citet{esquivel06} (the
  \citeauthor{esquivel06} predictions with and without radiative cooling lie in the lower and upper
  parts of this region, respectively, while the \citeauthor{slavin93a} predictions are concentrated
  to the top and the left of this region; see Figure~7 in KS10).
    \textit{Blue regions}: The TML model of KS10 (upper: NEI, lower: CIE).
    \textit{Purple region}: This paper, specifically, the values obtained from each
  model by time-averaging the ratios for $t \geq 10~\Myr$.
    \textit{Yellow region}: NEI calculations of a radiatively cooling shock, for a range of
  magnetic field strengths \citep{dopita96}.
    \textit{Cyan region}: Radiative cooling of Galactic fountain gas \citep{shapiro93,benjamin93}.
    \textit{Green region}: Radiatively cooling supernova remnant shells \citep{slavin93b,shelton98},
  and steady-state cooling flows \citep{edgar86} (the predicted ratios from these models are similar
  to each other).
    \textit{Red region}: Conductive heating and evaporation of spherical \citep{bohringer87} and
  planar \citep{borkowski90} clouds (the predicted ratios from these two models are similar to each
  other).
    \textit{Red curve}: Conductively evaporating cloud with photoionization included
  \citep[specifically, the model in their Figure 11f]{gnat10}. Different points along the curve
  correspond to different impact parameters; the end of the line at $\approx$$(-1.1,-0.7)$
  corresponds to a sightline through the cloud center.
    \textit{Black curves}: Gas cooling radiatively from a single temperature: \textit{solid line}:
  CIE cooling \citep{sutherland93}; \textit{dotted line}: CIE cooling with solar abundances
  \citep[Figure~5a]{gnat07}; \textit{dashed line}: NEI isochoric cooling with solar abundances
  \citep[Figure~5e]{gnat07}. As the gas cools, the ions ratios will move along the curves, from the
  lower left to the upper right.
  \label{fig:IonRatio}}
\end{figure*}

Another way to look at the ratios of \CIV\ and \NV\ to \OVI\ is to plot them in
$\NNV/\NOVI$--$\NCIV/\NOVI$ space (Figure~\ref{fig:IonRatio}). Although the ratios for ions moving
with HVC-like velocities predicted by the simulations in this paper overlap the ion ratios predicted
by our previous numerical models of TMLs (KS10), the ratios from some our models are lower than
KS10's, and better approach the ratios observed along three sightlines toward Complex~C.  However,
as mentioned earlier, if our results are rescaled according to the Complex~C abundances measured
along the PG~1259+593 sightline \citep{fox04}, our predicted data points are shifted by $-0.97$ and
$-0.35$ along the $x$ and $y$ axes, respectively. After such a shift, some of our models'
predictions would match the observed ratios (note that the observed $\NNV/\NOVI$ ratios are upper
limits, so our model predictions would be consistent with the observations, even after a shift of
$\approx$1~dex to the left). Figure~\ref{fig:IonRatio} confirms that ablation of cloud material
followed by mixing with the hot ambient gas is a viable physical process for the production of
high-velocity high ions. Note that some other models shown in Figure~\ref{fig:IonRatio} are in
agreement with the observed ion ratios toward Complex~C. However, these models do not distinguish
between high- and low-velocity ions; such a distinction is needed for an accurate comparison with
observations of HVCs.

\section{Neglected Physical Processes and Their Effects}
\label{sec:NeglectedPhysics}

Our simulations used a 2D geometry because the memory requirements for tracking the ionization
evolution of carbon, nitrogen, and oxygen in three dimensions are prohibitive. This assumed geometry
prevented us from modeling realistic magnetic field configurations, and so magnetohydrodynamic
effects were ignored. We also neglected thermal conduction, external heating, such as
photoelectric heating by Galactic or extragalactic UV radiation, and photoionization.

KS10 discuss the effect of assuming a 2D geometry on turbulent mixing, compared with a 3D geometry,
as well as the effect of a magnetic field. Generally, turbulent mixing is more severe in 3D
simulations than in 2D, which would lead to greater production of high ions, while a magnetic
field is known to suppress turbulence (see references in KS10). It is unclear whether the
combination of a magnetic field and a 3D geometry would result in more or fewer high ions,
compared with our 2D simulations. In addition, our previous 3D MHD simulations showed that the
dynamical effect of the magnetic field depends strongly on the orientation of the field
\citep{kwak09}.  As a result, the formation and size of the TMLs, and thus the production of
high ions, would also be affected by the orientation of the magnetic field. However, it should
be noted that, for a plane-parallel geometry, the 3D MHD TML simulations of \citet{esquivel06}
predicted similar high ion column density ratios as the 2D NEI simulations of KS10.
Finally, it should be noted that the magnetic field strength is likely to be small in the upper
halo, which is the region of interest of this study.

Our simulations did not include thermal conduction. \citet{vieser07a,vieser07b} studied numerically
the effect of thermal conduction on the evaporation of a cool cloud embedded in a hot ambient
medium. They found that thermal conduction suppresses shear instabilities by smoothing out the steep
temperature and density gradients at the interface between the hot and cool gas
\citep{vieser07a}. However, in our simulations, except for Model~E, there are not steep gradients at
the initial interface between the cool cloud and the hot ambient medium (see
Figure~\ref{fig:CloudProfile}). Also, in our simulations, heat diffusion due to thermal conduction
is unlikely to be important relative to heat diffusion by turbulence (\citealt{esquivel06};
KS10). For the mixed gas in our simulations ($T \sim 10^5~\K$, $n \sim 10^{-2}~\pcmsq$), the Spitzer
thermal diffusion coefficient\footnote{Derived from the \citet{spitzer56b} thermal conductivity,
  $6.1 \times 10^{-7} (T/\K)^{5/2}$~\erg\ \pcmsq\ \ps\ $\K^{-1}$.} is $\Ksp \sim
10^{24}$~\cmsq\ \ps. In contrast, the turbulent diffusion coefficient $\Kturb \sim \onethird \vturb
\Linj$ \citep{cho03} is $\ga$$10^{25}$~\cmsq\ \ps, where the turbulent velocity $\vturb \sim
100~\kmps$ and the energy injection scale $\Linj \ga 1~\pc$ (note that \Linj\ corresponds to the size
of the largest eddies; \citealt{cho03}). These
diffusion coefficients mean that turbulent diffusion of heat would dominate over diffusion by
thermal conduction, even if thermal conduction were included in our simulations.

We included radiative cooling in our simulations (albeit under the assumption of CIE), but not
photoelectric heating due to UV irradiation of dust grains. Including such heating would, in
principle, reduce the effect of radiative cooling. However, in practice, the photoelectric heating
rate is unlikely to be important for our simulations of HVCs in the upper halo, because of the
decrease in the UV radiation field and the dust density with height. For example, the photoelectric
heating rate for the upper halo ($|z| \ga 3~\kpc$) used by \citet{joung06} in their ISM models
would be $\la$$10^{-32}$~\erg\ \pcc\ \ps\ in our mixed gas, whereas the radiative cooling rate is
$\sim$$10^{-25}$~\erg\ \pcc\ \ps\ (using the 1993 version of the \citealt{raymond77} code).

A more important cooling-related effect is that the lower metal abundances seen on some sightlines
through Complex~C \citep{fox04,collins07} would reduce the cooling rate, as the cooling is dominated
by metal line emission. Radiative cooling tends to stabilize a cool cloud against disruption as it
moves through hot ambient gas \citep{vietri97}, and so reduced cooling rates would tend to
accelerate the disruption of the cloud and so may lead to the production of more high ions
than are currently predicted by our simulations. We defer investigation of this effect to a later
study.

Finally, we did not consider radiative transfer, and the possibility that some of the ion
populations may be affected by photoionization. Including photoionization is beyond the scope of
this study. However, we note that, if it is important, photoionization by escaping Galactic
radiation would tend to increase the amount of \CIV\ relative to \OVI.

\section{SUMMARY}
\label{sec:Summary}

We have presented the results of hydrodynamical simulations of HVCs that self-consistently trace the
non-equilibrium ionization evolution of carbon, nitrogen, and oxygen. This is the first time such
simulations have been carried out with an initially spherical cloud moving through hot, tenuous
ambient gas. We have concentrated particularly on the production of high ions (\CIV, \NV, and \OVI)
in the turbulent mixing layers that form between the cold clouds and the hot ambient medium. In contrast
to the previous plane-parallel TML models (e.g., KS10), the new simulations enable us to investigate
how the cold gas ablates from an initially spherical cloud and subsequently mixes with the hot ambient
gas.

Material is constantly being ablated from the clouds (Section~\ref{subsec:CloudEvolution}). This
initially cold material mixes with the hot ambient medium, and becomes ionized, so the clouds are
constantly losing their \HI\ content (Section~\ref{subsec:LossOfHI}). However, despite this loss of
material, we find that HVCs can survive for long times: at least a few hundred megayears for a cloud
with initial mass $\sim$$4 \times 10^5~\Msol$. Note that clouds that are currently
$\sim$10~\kpc\ from the disk (e.g., Complex~C with $z \sim 8~\kpc$; \citealt{thom08}) and traveling
toward the disk at $\sim$100~\kmps\ would require only $\sim$100~\Myr\ to reach the disk. It is
therefore possible that the more massive regions of HVC complexes, if not composed of many small
cloudlets, may reach the disk at least partially as neutral hydrogen
(Section~\ref{subsec:FateOfHVCs}).

High ions are produced in the mixed gas, both by ionization of the initially cool cloud material and
recombinations in the initially hot ambient material (Sections~\ref{subsec:CloudEvolution} and
\ref{subsec:HighStageIons}). The high ions initially appear while the mixed gas is still traveling
at HVC-like velocities. The material stripped from the clouds slows while remaining abundant in high
ions, and ultimately ends up with halo-like velocities (Paper~II), ceasing to be recognizable as HVC
material.

The simulations may also explain why some high-velocity highly ionized gas is seen on sightlines
that lack high-velocity \HI\ \citep{sembach03}, at least for sightlines that have high-velocity
\HI\ nearby (Section~\ref{subsec:HighStageIons}). At the edges of the clouds, the high-velocity
\HI\ column density can fall below the 21-cm detection limit ($\NHI \la 10^{18}~\pcmsq$), while the
column densities of high-velocity high ions remain detectable (e.g., $\NOVI\ \ga
10^{13}~\pcmsq$). In addition, the high-velocity high ions exist mainly in material that has
ablated from the main body of the cloud and mixed with the ambient gas. This material continues to
travel at HVC-like velocities, although it has a tendency to fall behind the cloud. As a result,
sightlines that cross the ablated material but not the cloud (such sight lines would cross our
domain diagonally) would intersect large numbers of high ions but small numbers of neutral hydrogen
atoms.

We investigated a suite of models with a range of model parameters. We found that the cloud's
initial velocity does not affect the rate of \HI\ loss (Section~\ref{subsubsec:EffectOfVelocity}) or
high ion production (Section~\ref{subsec:HighStageIons}). The rate of \HI\ loss also does not
strongly depend on the initial density of the cloud or its density profile
(Section~\ref{subsubsec:EffectOfDensity}).  However, the cloud's initial size does affect the rate
of \HI\ loss -- a smaller cloud loses its \HI\ mass relative to its initial mass more rapidly than a
larger cloud. This difference is at least partially due to the fact that a smaller cloud has a
larger surface-area-to-volume ratio (Section~\ref{subsubsec:EffectOfSize}).

Significant column densities of high ions are produced in our simulations. In some models, at some
times, these column densities are large enough to be observed, and are in reasonable agreement with
the ion column densities observed for Complex~C (Section~\ref{subsec:ComplexCColumns}). As the ion
column densities generally rise throughout our simulations, it is likely that the predicted column
densities would be in better agreement with the observed values if our simulations were continued
beyond their current end times. Our models also compare well with the Complex~C observations in
terms of the trend between \NOVI\ and \NHI\ (Section~\ref{subsec:NOVIversusNHI}), the
ion-to-\HI\ ratios (Section~\ref{subsec:ColumnIonToHI}), and the ion-to-ion ratios
(Section~\ref{subsec:ColumnIonToIon}). We therefore conclude that the ablation of cloud material and
its subsequent turbulent mixing with the hot ambient medium is a viable mechanism for the production
of high-velocity high ions.

The column density ratios of high ions to \HI\ depend on the initial cloud size. A smaller cloud
reaches a given ion-to-\HI\ ratio at an earlier time than a larger cloud. If Complex~C were
initially composed of clouds with $r \ga 150~\pc$ (which cannot be ruled out, given the lack of
knowledge of Complex C's three-dimensional structure), then the observed ion-to-\HI\ ratios would
imply an age for Complex~C of $\ga$100~\Myr. This in turn implies that Complex~C has traveled
$\ga$10--30~\kpc\ from its initial location, suggesting that Complex~C formed from extragalactic
material. Furthermore, if parts of Complex~C are composed of clouds as large as our Model~F cloud
($M \sim 4 \times 10^5~\Msol$), these clouds may reach the disk partially in the form of neutral
hydrogen, as noted above. HVCs could therefore provide new material to the disk for star formation.
We also noted that there is a trend in Complex~C for the \OVI-to-\HI\ ratio to increase as one gets
closer to the disk, and speculated on possible reasons for this trend in
Section~\ref{subsec:ColumnIonToHI}.

In this paper, we have concentrated particularly on high-velocity ions in the Galactic
halo. However, low-velocity high ions are also observed in the halo, and, as noted above,
low-velocity ions are produced abundantly in our simulations. These low-velocity ions will be
discussed in Paper~II.

\acknowledgements

We thank Orly Gnat for providing us with the data from \citet{gnat10} plotted in our
Figure~\ref{fig:IonRatio}.  We also thank the anonymous referee, whose comments have helped improve
this paper. The software used in this work was in part developed by the DOE-supported ASC/Alliance
Center for Astrophysical Thermonuclear Flashes at the University of Chicago. The simulations were
performed at the Research Computing Center (RCC) of the University of Georgia. This work was
supported by NASA grant NNX09AD13G, awarded through the Astrophysics Theory and Fundamental Physics
Program. DBH acknowledges funding from NASA grant NNX08AJ47G, awarded through the Astrophysics Data
Analysis Program.

\bibliography{references}

\begin{thebibliography}{75}
\expandafter\ifx\csname natexlab\endcsname\relax\def\natexlab#1{#1}\fi

\bibitem[{Allen(1973)}]{allen73}
Allen, C.~W. 1973, {Astrophysical Quantities}, 3rd edn. (London: Athlone)

\bibitem[{Begelman \& Fabian(1990)}]{begelman90}
Begelman, M.~C., \& Fabian, A.~C. 1990, MNRAS, 244, 26P

\bibitem[{Benjamin \& Shapiro(1993)}]{benjamin93}
Benjamin, R.~A., \& Shapiro, P.~R. 1993, in UV and X-Ray Spectroscopy of
  Astrophysical and Laboratory Plasmas, ed. E.~H. Silver \& S.~M. Kahn, 275

\bibitem[{Bland-Hawthorn(2009)}]{blandhawthorn09}
Bland-Hawthorn, J. 2009, in Proc. IAU Symp. 254, The Galaxy Disk in
  Cosmological Context, ed. J.~Andersen, J.~Bland-Hawthorn, \&
  B.~Nordstr{\"o}m, 241

\bibitem[{Blitz {et~al.}(1999)Blitz, Spergel, Teuben, Hartmann, \&
  Burton}]{blitz99}
Blitz, L., Spergel, D.~N., Teuben, P.~J., Hartmann, D., \& Burton, W.~B. 1999,
  ApJ, 514, 818

\bibitem[{B{\"o}hringer \& Hartquist(1987)}]{bohringer87}
B{\"o}hringer, H., \& Hartquist, T.~W. 1987, MNRAS, 228, 915

\bibitem[{Borkowski {et~al.}(1990)Borkowski, Balbus, \& Fristrom}]{borkowski90}
Borkowski, K.~J., Balbus, S.~A., \& Fristrom, C.~C. 1990, ApJ, 355, 501

\bibitem[{Bregman \& Lloyd-Davies(2007)}]{bregman07}
Bregman, J.~N., \& Lloyd-Davies, E.~J. 2007, ApJ, 669, 990

\bibitem[{Burrows \& Mendenhall(1991)}]{burrows91}
Burrows, D.~N., \& Mendenhall, J.~A. 1991, Nature, 351, 629

\bibitem[{Chandrasekhar(1961)}]{chandrasekhar61}
Chandrasekhar, S. 1961, {Hydrodynamic and Hydromagnetic Stability} (Oxford:
  Clarendon Press; Dover reprint, 1981)

\bibitem[{Cho {et~al.}(2003)Cho, Lazarian, Honein, Knaepen, Kassinos, \&
  Moin}]{cho03}
Cho, J., Lazarian, A., Honein, A., Knaepen, B., Kassinos, S., \& Moin, P. 2003,
  ApJL, 589, L77

\bibitem[{Collins {et~al.}(2003)Collins, Shull, \& Giroux}]{collins03}
Collins, J.~A., Shull, J.~M., \& Giroux, M.~L. 2003, ApJ, 585, 336

\bibitem[{Collins {et~al.}(2007)Collins, Shull, \& Giroux}]{collins07}
Collins, J.~A., Shull, J.~M., \& Giroux, M.~L. 2007, ApJ, 657, 271

\bibitem[{Danly {et~al.}(1993)Danly, Albert, \& Kuntz}]{danly93}
Danly, L., Albert, C.~E., \& Kuntz, K.~D. 1993, ApJ, 416, L29

\bibitem[{Dopita \& Sutherland(1996)}]{dopita96}
Dopita, M.~A., \& Sutherland, R.~S. 1996, ApJS, 102, 161

\bibitem[{Edgar \& Chevalier(1986)}]{edgar86}
Edgar, R.~J., \& Chevalier, R.~A. 1986, ApJ, 310, L27

\bibitem[{Esquivel {et~al.}(2006)Esquivel, Benjamin, Lazarian, Cho, \&
  Leitner}]{esquivel06}
Esquivel, A., Benjamin, R.~A., Lazarian, A., Cho, J., \& Leitner, S.~N. 2006,
  ApJ, 648, 1043

\bibitem[{Fang {et~al.}(2006)Fang, McKee, Canizares, \& Wolfire}]{fang06}
Fang, T., McKee, C.~F., Canizares, C.~R., \& Wolfire, M. 2006, ApJ, 644, 174

\bibitem[{Fox {et~al.}(2006)Fox, Savage, \& Wakker}]{fox06}
Fox, A.~J., Savage, B.~D., \& Wakker, B.~P. 2006, ApJS, 165, 229

\bibitem[{Fox {et~al.}(2004)Fox, Savage, Wakker, Richter, Sembach, \&
  Tripp}]{fox04}
Fox, A.~J., Savage, B.~D., Wakker, B.~P., Richter, P., Sembach, K.~R., \&
  Tripp, T.~M. 2004, ApJ, 602, 738

\bibitem[{Fox {et~al.}(2005)Fox, Wakker, Savage, Tripp, Sembach, \&
  Bland-Hawthorn}]{fox05}
Fox, A.~J., Wakker, B.~P., Savage, B.~D., Tripp, T.~M., Sembach, K.~R., \&
  Bland-Hawthorn, J. 2005, ApJ, 630, 332

\bibitem[{Fryxell {et~al.}(2000)Fryxell, Olson, Ricker, Timmes, Zingale, Lamb,
  MacNeice, Rosner, Truran, \& Tufo}]{fryxell00}
Fryxell, B., {et~al.} 2000, ApJS, 131, 273

\bibitem[{Gaensler {et~al.}(2008)Gaensler, Madsen, Chatterjee, \&
  Mao}]{gaensler08}
Gaensler, B.~M., Madsen, G.~J., Chatterjee, S., \& Mao, S.~A. 2008, PASA, 25,
  184

\bibitem[{Galeazzi {et~al.}(2007)Galeazzi, Gupta, Covey, \&
  Ursino}]{galeazzi07}
Galeazzi, M., Gupta, A., Covey, K., \& Ursino, E. 2007, ApJ, 658, 1081

\bibitem[{Gardiner \& Noguchi(1996)}]{gardiner96}
Gardiner, L.~T., \& Noguchi, M. 1996, MNRAS, 278, 191

\bibitem[{Gnat \& Sternberg(2007)}]{gnat07}
Gnat, O., \& Sternberg, A. 2007, ApJS, 168, 213

\bibitem[{Gnat {et~al.}(2010)Gnat, Sternberg, \& McKee}]{gnat10}
Gnat, O., Sternberg, A., \& McKee, C.~F. 2010, ApJ, 718, 1315

\bibitem[{Grcevich \& Putman(2009)}]{grcevich09}
Grcevich, J., \& Putman, M.~E. 2009, ApJ, 696, 385

\bibitem[{Heitsch \& Putman(2009)}]{heitsch09}
Heitsch, F., \& Putman, M.~E. 2009, ApJ, 698, 1485

\bibitem[{Henley \& Shelton(2008)}]{henley08a}
Henley, D.~B., \& Shelton, R.~L. 2008, ApJ, 676, 335

\bibitem[{Henley {et~al.}(2010)Henley, Shelton, Kwak, Joung, \&
  Mac~Low}]{henley10b}
Henley, D.~B., Shelton, R.~L., Kwak, K., Joung, M.~R., \& Mac~Low, M.-M. 2010,
  ApJ, 723, 935

\bibitem[{Hulsbosch \& Wakker(1988)}]{hulsbosch88}
Hulsbosch, A.~N.~M., \& Wakker, B.~P. 1988, A\&AS, 75, 191

\bibitem[{Indebetouw \& Shull(2004)}]{indebetouw04a}
Indebetouw, R., \& Shull, J.~M. 2004, ApJ, 605, 205

\bibitem[{Joung \& Mac~Low(2006)}]{joung06}
Joung, M.~K.~R., \& Mac~Low, M.-M. 2006, ApJ, 653, 1266

\bibitem[{Keenan {et~al.}(1995)Keenan, Shaw, Bates, Dufton, \& Kemp}]{keenan95}
Keenan, F.~P., Shaw, C.~R., Bates, B., Dufton, P.~L., \& Kemp, S.~N. 1995,
  MNRAS, 272, 599

\bibitem[{Kuntz \& Snowden(2000)}]{kuntz00}
Kuntz, K.~D., \& Snowden, S.~L. 2000, ApJ, 543, 195

\bibitem[{Kwak \& Shelton(2010)}]{kwak10}
Kwak, K., \& Shelton, R.~L. 2010, ApJ, 719, 523 (KS10)

\bibitem[{Kwak {et~al.}(2009)Kwak, Shelton, \& Raley}]{kwak09}
Kwak, K., Shelton, R.~L., \& Raley, E.~A. 2009, ApJ, 699, 1775

\bibitem[{Lei {et~al.}(2009)Lei, Shelton, \& Henley}]{lei09}
Lei, S., Shelton, R.~L., \& Henley, D.~B. 2009, ApJ, 699, 1891

\bibitem[{Lockman(2003)}]{lockman03}
Lockman, F.~J. 2003, ApJ, 591, L33

\bibitem[{Lockman {et~al.}(2002)Lockman, Murphy, Petty-Powell, \&
  Urick}]{lockman02}
Lockman, F.~J., Murphy, E.~M., Petty-Powell, S., \& Urick, V.~J. 2002, ApJS,
  140, 331

\bibitem[{Maller \& Bullock(2004)}]{maller04}
Maller, A.~H., \& Bullock, J.~S. 2004, MNRAS, 355, 694

\bibitem[{Muller {et~al.}(1963)Muller, Oort, \& Raimond}]{muller63}
Muller, C.~A., Oort, J.~H., \& Raimond, E. 1963, C. R. Acad. Sci. Paris, 257,
  1661

\bibitem[{Nichols \& Bland-Hawthorn(2009)}]{nichols09}
Nichols, M., \& Bland-Hawthorn, J. 2009, ApJ, 707, 1642

\bibitem[{Oort(1966)}]{oort66}
Oort, J.~H. 1966, Bull. Astron. Inst. Netherlands, 18, 421

\bibitem[{Peek {et~al.}(2007)Peek, Putman, McKee, Heiles, \&
  Stanimirovi{\'c}}]{peek07}
Peek, J.~E.~G., Putman, M.~E., McKee, C.~F., Heiles, C., \& Stanimirovi{\'c},
  S. 2007, ApJ, 656, 907

\bibitem[{Putman {et~al.}(2004)Putman, Thom, Gibson, \&
  Staveley-Smith}]{putman04}
Putman, M.~E., Thom, C., Gibson, B.~K., \& Staveley-Smith, L. 2004, ApJ, 603,
  L77

\bibitem[{Raymond \& Smith(1977)}]{raymond77}
Raymond, J.~C., \& Smith, B.~W. 1977, ApJS, 35, 419

\bibitem[{Rudolph {et~al.}(2006)Rudolph, Fich, Bell, Norsen, Simpson, Haas, \&
  Erickson}]{rudolph06}
Rudolph, A.~L., Fich, M., Bell, G.~R., Norsen, T., Simpson, J.~P., Haas, M.~R.,
  \& Erickson, E.~F. 2006, ApJS, 162, 346

\bibitem[{Sembach {et~al.}(2003)Sembach, Wakker, Savage, Richter, Meade, Shull,
  Jenkins, Sonneborn, \& Moos}]{sembach03}
Sembach, K.~R., {et~al.} 2003, ApJS, 146, 165

\bibitem[{Shapiro \& Benjamin(1993)}]{shapiro93}
Shapiro, P.~R., \& Benjamin, R.~A. 1993, in Star Formation, Galaxies, and the
  Interstellar Medium, ed. J.~Franco, F.~Ferrini, \& G.~Tenorio-Tagle, 275

\bibitem[{Shaver {et~al.}(1983)Shaver, McGee, Newton, Danks, \&
  Pottasch}]{shaver83}
Shaver, P.~A., McGee, R.~X., Newton, L.~M., Danks, A.~C., \& Pottasch, S.~R.
  1983, MNRAS, 204, 53

\bibitem[{Shelton(1998)}]{shelton98}
Shelton, R.~L. 1998, ApJ, 504, 785

\bibitem[{Shull {et~al.}(2009)Shull, Jones, Danforth, \& Collins}]{shull09}
Shull, J.~M., Jones, J.~R., Danforth, C.~W., \& Collins, J.~A. 2009, ApJ, 699,
  754

\bibitem[{Slavin \& Cox(1993)}]{slavin93b}
Slavin, J.~D., \& Cox, D.~P. 1993, ApJ, 417, 187

\bibitem[{Slavin {et~al.}(1993)Slavin, Shull, \& Begelman}]{slavin93a}
Slavin, J.~D., Shull, J.~M., \& Begelman, M.~C. 1993, ApJ, 407, 83

\bibitem[{Smith {et~al.}(2007)Smith, Bautz, Edgar, Fujimoto, Hamaguchi, Hughes,
  Ishida, Kelley, Kilbourne, Kuntz, McCammon, Miller, Mitsuda, Mukai,
  Plucinsky, Porter, Snowden, Takei, Terada, Tsuboi, \& Yamasaki}]{smith07a}
Smith, R.~K., {et~al.} 2007, PASJ, 59, S141

\bibitem[{Snowden {et~al.}(1998)Snowden, Egger, Finkbeiner, Freyberg, \&
  Plucinsky}]{snowden98}
Snowden, S.~L., Egger, R., Finkbeiner, D.~P., Freyberg, M.~J., \& Plucinsky,
  P.~P. 1998, ApJ, 493, 715

\bibitem[{Snowden {et~al.}(1991)Snowden, Mebold, Hirth, Herbstmeier, \&
  Schmitt}]{snowden91}
Snowden, S.~L., Mebold, U., Hirth, W., Herbstmeier, U., \& Schmitt, J.~H.~M.~M.
  1991, Science, 252, 1529

\bibitem[{Spitzer(1956)}]{spitzer56b}
Spitzer, L. 1956, ApJ, 124, 20

\bibitem[{Sutherland \& Dopita(1993)}]{sutherland93}
Sutherland, R.~S., \& Dopita, M.~A. 1993, ApJS, 88, 253

\bibitem[{Thom {et~al.}(2008)Thom, Peek, Putman, Heiles, Peek, \&
  Wilhelm}]{thom08}
Thom, C., Peek, J.~E.~G., Putman, M.~E., Heiles, C., Peek, K.~M.~G., \&
  Wilhelm, R. 2008, ApJ, 684, 364

\bibitem[{Tripp {et~al.}(2003)Tripp, Wakker, Jenkins, Bowers, Danks, Green,
  Heap, Joseph, Kaiser, Linsky, \& Woodgate}]{tripp03}
Tripp, T.~M., {et~al.} 2003, AJ, 125, 3122

\bibitem[{Vieser \& Hensler(2007{\natexlab{a}})}]{vieser07b}
Vieser, W., \& Hensler, G. 2007{\natexlab{a}}, A\&A, 475, 251

\bibitem[{Vieser \& Hensler(2007{\natexlab{b}})}]{vieser07a}
Vieser, W., \& Hensler, G. 2007{\natexlab{b}}, A\&A, 472, 141

\bibitem[{Vietri {et~al.}(1997)Vietri, Ferrara, \& Miniati}]{vietri97}
Vietri, M., Ferrara, A., \& Miniati, F. 1997, ApJ, 483, 262

\bibitem[{Wakker \& van Woerden(1991)}]{wakker91}
Wakker, B.~P., \& van Woerden, H. 1991, A\&A, 250, 509

\bibitem[{Wakker \& van Woerden(1997)}]{wakker97}
Wakker, B.~P., \& van Woerden, H. 1997, ARA\&A, 35, 217

\bibitem[{Wakker {et~al.}(2003)Wakker, Savage, Sembach, Richter, Meade,
  Jenkins, Shull, Ake, Blair, Dixon, Friedman, Green, Green, Kruk, Moos,
  Murphy, Oegerle, Sahnow, Sonneborn, Wilkinson, \& York}]{wakker03}
Wakker, B.~P., {et~al.} 2003, ApJS, 146, 1

\bibitem[{Wakker {et~al.}(2007)Wakker, York, Howk, Barentine, Wilhelm,
  Peletier, van Woerden, Beers, Ivezi{\'c}, Richter, \& Schwarz}]{wakker07}
Wakker, B.~P., {et~al.} 2007, ApJ, 670, L113

\bibitem[{Weiner \& Williams(1996)}]{weiner96}
Weiner, B.~J., \& Williams, T.~B. 1996, AJ, 111, 1156

\bibitem[{Yao \& Wang(2005)}]{yao05}
Yao, Y., \& Wang, Q.~D. 2005, ApJ, 624, 751

\bibitem[{Yao \& Wang(2007)}]{yao07a}
Yao, Y., \& Wang, Q.~D. 2007, ApJ, 658, 1088

\bibitem[{Yao {et~al.}(2009)Yao, Wang, Hagihara, Mitsuda, McCammon, \&
  Yamasaki}]{yao09}
Yao, Y., Wang, Q.~D., Hagihara, T., Mitsuda, K., McCammon, D., \& Yamasaki,
  N.~Y. 2009, ApJ, 690, 143

\bibitem[{Yoshino {et~al.}(2009)Yoshino, Mitsuda, Yamasaki, Takei, Hagihara,
  Masui, Bauer, McCammon, Fujimoto, Wang, \& Yao}]{yoshino09}
Yoshino, T., {et~al.} 2009, PASJ, 61, 805

\end{thebibliography}

\end{document}